\documentstyle[aps,eqsecnum,twoside]{revtex}
\begin{document}

%\draft

\twocolumn[ % this optional argument is put on top of the page before the
            % \twocolumn formatting begins

\title{\vspace*{5mm}
       Exact Hypersurface-Homogeneous Solutions in Cosmology and Astrophysics}

\author{Claes Uggla\cite{thanks}}
\address{
Department of Physics, University of Stockholm,
Box 6730, S-113 85 Stockholm, Sweden}

\author{Robert T. Jantzen}
\address{
  Department of Mathematical Physics, Villanova University,
  Villanova, PA 19085, USA and\\
  International Center for Relativistic Astrophysics,
  University of Rome, I--00185 Roma, Italy}

\author{Kjell Rosquist\cite{thanks}\cite{thanksEC}}
\address{
Department of Physics, University of Stockholm,
Box 6730, S--113 85 Stockholm, Sweden and\\
RGGR group, Chimie-Physique, Universit\'e Libre de Bruxelles, Brussels,
Belgium}

%\date{\today}

%%%%%%%%%%%%%%%%%%%%%%%%%%%%%%%%%%%%%%%%%%%%%%%%%%%%%%%%%%%%%%%%%%%%%%%%%%%%%%
\maketitle

\widetext
\begin{abstract}
A framework is introduced which explains
the existence and similarities of most exact solutions of the Einstein
equations with a wide range of sources
for the class of hypersurface-homogeneous spacetimes
which admit a Hamiltonian formulation.
This class includes
the spatially homogeneous cosmological models and the astrophysically
interesting static spherically symmetric models as well as the
stationary cylindrically symmetric models.
The framework involves methods for finding and exploiting hidden
symmetries and invariant submanifolds of the Hamiltonian formulation of the
field equations. It unifies, simplifies and extends most known work on
hypersurface-homogeneous exact solutions.
It is shown that the same framework is also relevant
to gravitational theories with a similar structure, like
Brans-Dicke or higher-dimensional theories.
\end{abstract}

] % end of optional argument for \twocolumn

%%%%%%%%%%%%%%%%%%%%%%%%%%%%%%%%%%%%%%%%%%%%%%%%%%%%%%%%%%%%%%%%%%%%%%%%%%%%%%
% Defining page header (a la The REVTeX Input Guide)
%%%%%%%%%%%%%%%%%%%%%%%%%%%%%%%%%%%%%%%%%%%%%%%%%%%%%%%%%%%%%%%%%%%%%%%%%%%%%%
\makeatletter % temporarily putting @ in the `letter' category

\global\@specialpagefalse
\def\@oddhead{To appear in Phys.\ Rev.\ D
              \hfill Preprint, March 1995, gr-qc/9503061}
\let\@evenhead\@oddhead

\makeatother % returning @ to the `other' category
%%%%%%%%%%%%%%%%%%%%%%%%%%%%%%%%%%%%%%%%%%%%%%%%%%%%%%%%%%%%%%%%%%%%%%%%%%%%%%

%%%%%%%%%%%%%%%%%%%%%%%%%%%%%%%%%%%%%%%%%%%%%%%%%%%%%%%%%%%%%%%%%%%%%%%%%%%%%%
% section 1
%%%%%%%%%%%%%%%%%%%%%%%%%%%%%%%%%%%%%%%%%%%%%%%%%%%%%%%%%%%%%%%%%%%%%%%%%%%%%%

\narrowtext
\section{Introduction}

Exact solutions have always played a central role in the investigation
of physical theories whose content is encoded in a complicated set of
differential equations. The theory of general relativity is indeed an
example of this. A number of exact solutions of Einstein's field equations
have been of key importance in the discussion of physical problems.
Solutions have been found which
describe black holes, stellar interiors,
gravitational waves, and even the large scale structure of the universe
itself. Exact solutions have also served as a guide to point out mathematical
features of the theory.
The Taub-NUT-M (Taub-Newman-Unti-Tamburino-Misner \cite{tnutm,rs})
solution, for example, has been of crucial importance for the very definitions
one uses in describing the singularities of the full theory.
Thus exact solutions may point out features which are not just special to
themselves but characterize in some way properties of a wider class of
solutions. They may also play a role as ``building blocks" for more general
solutions. For example, in certain ways the general spatially homogeneous
cosmological model near an initial singularity can be understood in terms of
very special exact solutions, notably the Kasner and the vacuum Bianchi type
II solutions, which to some extent also describe aspects of general
cosmological singularities \cite{blk}.
Sometimes exactly solvable problems are even used as a guide in developing
ideas for the construction of more general theories, e.g., quantum gravity.
For example, solvable problems in spatially homogeneous (SH) cosmology have
been used to implement a number of different quantization schemes.

Thus there is ample motivation to try to find exact solutions.
Indeed, the book by Kramer et al \cite{kraetal} is largely dedicated
to the listing of exact solutions.
Several chapters of that book deal with hypersurface-homogeneous (HH)
solutions, a class for which the Einstein equations reduce to
more manageable ordinary differential equations.
Within the class of HH solutions there are several subclasses of
considerable physical interest, the cosmological SH models and the
astrophysical static spherically symmetric spacetimes being the
most prominent ones.

Since the birth of general relativity nearly 80 years
ago, an overwhelming number of HH solutions have been produced.
A look at physics abstracts shows that this production continues even
today at a considerable pace. However, often this search
is undertaken as an end in itself without attempting
to understand how particular successes fit into a larger scheme and without
employing any systematic method of attack revealing possible
underlying mechanisms. Exceptions do exist though, as illustrated very nicely
by the numerous approaches to the problem of finding vacuum solutions
for spacetimes with one or two commuting Killing vector
fields \cite{kraetal,ver}.
However, these techniques have not contributed much to the
problem of finding HH solutions.

It is the purpose of this article
\begin{itemize}
\item to give a basic underlying explanation of why exact
      solutions arise for models with nonnull homogeneous hypersurfaces,
\item to provide a set of
      techniques which makes it possible
      to obtain these solutions in as simple a form as possible,
\item to show how these techniques are applicable to a wide set of different
      physical problems, and
\item to unify an otherwise seemingly unrelated zoo of particular
      results.
\end{itemize}

Rather than exhaustively treating all possible cases,
a wide variety of examples will be discussed which illustrate the
utility of the
approach presented here.
The choice of examples is a subjective one reflecting our particular tastes.
Among the many models and source types considered,
the exact SH perfect fluid solutions will be more exhaustively
surveyed, updating the work of Kramer et al.
Scalar fields, not considered in that catalog but now currently fashionable
and physically interesting,
will also be examined. Special attention will also be given to the
static spherically symmetric perfect fluid models which are important in
astrophysical applications.

The present approach will use a Hamil\-to\-ni\-an/La\-gran\-gian formulation
of the field equations, using the Hamiltonian constraint together with
the Lagrangian equations. This may come as a surprise
considering the statement by Kramer et al \cite{kraetal} (p.131)
that this formulation
``is not well-adapted to searching for exact solutions$\ldots$",
but upon second thought, this is quite natural since the Hamiltonian
function contains all the dynamical content of the Einstein equations.
This enables one to study a single function instead of a whole system
of equations, armed with many powerful techniques from classical
mechanics.
The economy of the Hamiltonian approach also reveals
the close mathematical relationships among different types of models.
These relationships
are often obscured by the particular way in which a particular
physical problem suggests expressing the field equations.
For example, the usual approaches to static spherically symmetric
models and SH cosmological models are quite different, but the
Hamiltonian formulation shows their mathematical similarities.
Furthermore,
the Hamiltonian approach used in this article
will also show how most exact solutions are either associated with
essentially 1-dimensional problems (in terms of degrees of freedom)
or with problems which admit a sufficient number of
a certain type of Hamiltonian symmetry.

The article will proceed as follows.
Since the Hamiltonian is essential for this
discussion, Hamiltonians will be derived for a variety
of different problems in section 2.
This section starts with an outline of the way in which the
Hamiltonian function is obtained for the HH models.
This is followed by the explicit evaluation of this function,
considering separately the diagonal and nondiagonal models.
For the more numerous diagonal models, the gravitational Hamiltonian
is discussed first and then a number of source contributions
to the Hamiltonian are examined.
For the fewer nondiagonal examples,
each case is studied individually, and only the vacuum and
perfect fluid Hamiltonians are derived.
In all cases the Hamiltonian is put into a ``canonical" form
that makes evident the mathematical similarities  among different HH models.

In section 3 the generalized Friedmann equation is reviewed.
This equation often arises
in the context of 1-dimensional invariant submanifolds and
as a part of solvable higher-dimensional problems.
It therefore frequently plays an important role
as a fundamental building block when it comes
to finding exact solutions.

To solve a problem with two or more degrees of freedom it is necessary to
find a sufficient number of symmetries which makes it possible to decompose
the problem into smaller solvable parts. In section 4 a particular
type of Hamiltonian
symmetry is discussed, a so-called Killing tensor symmetry. It turns
out that this symmetry, which usually is ``hidden,'' is responsible for all
known solvable Hamiltonian problems with two or more degrees of
freedom that we are aware of.
This symmetry is a generalization of symmetries related to the existence
of cyclic variables and Hamilton-Jacobi separability.
To exploit the Killing tensor symmetries
in order to find explicit solutions,
one has to find symmetry compatible dependent and
independent variables. A way of doing this is presented.

Section 5 presents one method
for finding invariant Hamiltonian submanifolds,
i.e., consistent subsystems of the Hamiltonian equations.
This in turn leads to particular solutions.
The existence of invariant submanifolds is an important issue in the
search for exact solutions since it is usually impossible to solve the
most general problems.

Section 6 lists problems leading to
exact solutions and indicates how to solve them by referencing
the relevant parts of this article.
The models in this section
are again divided into diagonal and nondiagonal models.
However, the diagonal models previously arranged in section 2
according to how their Hamiltonians are evaluated are now instead
collected together according to the dimension of their
intrinsic symmetry group, i.e., the group of symmetries of the intrinsic
geometry of the individual homogeneous hypersurfaces.
This categorization
reveals an underlying mathematical unity of entire classes of
physically distinct models and allows them to be treated collectively.

The methods developed in this article are also applicable in
contexts beyond 4-dimensional classical general relativity.
This is discussed in the concluding section 7.
As examples some remarks are made about higher-dimensional theories,
nonminimally coupled scalar field theories,
and quantum cosmology.

In section 8 the present approach is compared with some other
exact solution techniques, followed by a concluding discussion
which addresses a variety of other issues.

%%%%%%%%%%%%%%%%%%%%%%%%%%%%%%%%%%%%%%%%%%%%%%%%%%%%%%%%%%%%%%%%%%%%%%%%%%%%%%
% section 2
%%%%%%%%%%%%%%%%%%%%%%%%%%%%%%%%%%%%%%%%%%%%%%%%%%%%%%%%%%%%%%%%%%%%%%%%%%%%%%

\section{A Hamiltonian Approach to the Field Equations} \label{sec:HA}

In this section the Hamiltonian will be evaluated for a wide
variety of HH models. However, the goal is not just to produce a
Hamiltonian, but to obtain it in
a certain ``canonical" form  which reveals the mathematical
similarities between different HH models.
This is not only useful for the present purpose of finding
exact solutions, but may also serve as a starting point for
qualitative analysis of the many problems which cannot be
solved exactly.
Furthermore, one can study a single function rather than be
overwhelmed by a whole system of
equations which also hides the mathematical connections between different
kinds of models.
This section concludes with a discussion of the general
form of the Hamiltonians which have been obtained in the individual cases.

\subsection{The line element}

To explicitly obtain a Hamiltonian for a given model,
it is necessary to introduce a line element.
In this article we will consider
the Taub-NUT-M model, to be discussed below, and certain models
with  nonnull homogeneous hypersurfaces.
For the latter models the
line element can be expressed in the form
\begin{equation}
     ds^2 = \epsilon N(\lambda)^2 d\lambda^2
               + g_{ab}(\lambda) \omega^a\omega^b \,,
\end{equation}
where the 1-forms
$\omega^a$ ($a = 1,2,3$)
are associated with the symmetry group acting on the homogeneous
hypersurfaces.
The quantity $\epsilon = n_\alpha n^\alpha$ is the sign of the norm
of the unit normal $n^\alpha$ to the homogeneous hypersurfaces,
having the value $-1$ for the SH models and $1$ for the static ones
($\alpha$ assumes the values
$0, 1, 2, 3$ with $0$ referring to the component associated with the $\lambda$
direction). For the SH models the single independent variable $\lambda$
is a time variable $t$ and $N$ is the familiar lapse function.
In the static case $\lambda$ is instead
a spatial variable.
The gravitational degrees of freedom are associated with
the component functions $g_{ab}$.

For later purposes
it is convenient to introduce the function $x$ defined by
\begin{equation}
   x = N_{\rm(taub)}/N\,,\quad N_{\rm(taub)} = 12|g|^{1/2}\,,
\end{equation}
where $g =\det (g_{ab}) = -\epsilon |g|$.
The function $x$ is the reciprocal of the
relative ``slicing gauge function" ${\cal N} = N/N_{\rm(taub)}$
which is sometimes used in $3 + 1$ canonical gravity as a relative lapse.
One must fix $N$ or $x$ to determine the
parametrization of the family of homogeneous hypersurfaces by the
independent variable $\lambda$; this will be referred to as a choice
of slicing gauge.
The notation $N_{\rm(taub)}$ comes from the very
useful slicing gauge introduced by Taub \cite{tau} in the context of
SH cosmology.

The HH models admit simply transitive
or multiply transitive (MT) homogeneity groups.
The models which admit a simply transitive
three-dimensional homogeneity group are called Bianchi models, for which
the 1-forms may be chosen to satisfy
\begin{equation}
    d\omega^a=-\case12 C^a{}_{bc} \omega^b\wedge\omega^c\,,
\end{equation}
where $C^a{}_{bc}$ are the components of the structure constant tensor
of the Lie algebra of the homogeneity group.
The Bianchi models are divided into two classes, class A
and class B, according to the vanishing or nonvanishing of the trace
$C^b{}_{ab}$.
For the class A models one can choose
$C^a{}_{bc} = n^{(a)} \epsilon_{abc}$, where the parameters
$n^{(a)}$ characterizing the various symmetry types can be chosen
to have the values in Table 1; explicit coordinate representations
of the 1-forms can be found in \cite{uni}.
Class B models which admit a Hamiltonian formulation will be discussed
below.

%\typeout{**********************************Table 1:}

Among the Bianchi models there is a special family
of Bianchi type I, II, III, V, VII$_0$, VII$_h$, VIII, and IX models
which admits MT symmetry groups.
There are also MT models which do not admit a simply transitive
subgroup (i.e., they are not Bianchi models);
the most prominent ones are the SH Kantowski-Sachs models and
the static spherically symmetric models.

\subsection{How to find the Hamiltonian function}

Not all HH models admit a Hamiltonian formulation of the field equations.
However, of those that do,
the Lagrangian and Hamiltonian have the usual simple forms
in terms of kinetic and potential terms
\begin{equation}
          L = T - U\,, \quad H = T + U\,.
\end{equation}
The Hamiltonian must vanish as a consequence of one of the field
equations; this is the ``Hamiltonian constraint" $H=0$.
 From the expression for $H$ in terms of the velocities, one can read off
the Lagrangian as well by a simple change of sign.
The Hamiltonian constraint and the
Lagrangian equations are
both  needed since
a mixture of first and second order
differential equations is often required to find explicit solutions.
Since the Hamiltonian constraint is
of crucial importance in obtaining solutions and
is the most economical way of describing the field equations,
our method will be referred to as a Hamiltonian approach.

The starting point for
obtaining the above Hamiltonian is
the ADM Hamiltonian of the
$3+1$ or ADM approach described in Misner, Thorne and Wheeler \cite{mtw}.
This approach
reformulates the Einstein equations expressed with respect to a spacelike
slicing as a parametrized Hamiltonian system with constraints.
Since all their formulas involve explicitly the sign $\epsilon$ of the
norm of the
unit normal to the slicing, they continue to hold for a timelike slicing.
The ADM Hamiltonian is a linear combination of the super-Hamiltonian and
supermomentum constraint functions
(whose vanishing is equivalent to the $00$ and $0a$ components of
the Einstein equations respectively),
with the lapse and shift as the respective
Lagrange multiplier coefficients.
However, in the
present application the supermomentum constraints
can be solved and the
shift set to zero, reducing the ADM Hamiltonian to the lapse times the
super-Hamiltonian. Allowing the lapse function in this Hamiltonian
to depend explicitly
on the gravitational and possible source variables
leads to Hamiltonian equations which
differ from each other by terms involving the super-Hamiltonian,
required to vanish as the remaining unsolved constraint.

This specialized ADM Hamiltonian is the first step
towards obtaining the desired Hamiltonian for
HH models and will be referred to simply as the Hamiltonian function.
It has the following form
\cite{mtw}
\begin{eqnarray}
\label{eq:H}
   H &=& - 2 N|g|^{1/2}n^\alpha n^\beta
        [ G_{\alpha \beta} - \kappa T_{\alpha \beta}]
\nonumber\\
     &=& - 24 x^{-1} |g| n^\alpha n^\beta
        [ G_{\alpha \beta} - \kappa T_{\alpha \beta}]
\nonumber\\
     &=& H_{\rm(G)} + H_{\rm(source)} \,,
\end{eqnarray}
where $ G_{\alpha \beta}$ is the Einstein tensor
and $T_{\alpha \beta}$ is the total energy-momentum tensor, including
a possible cosmological constant term, while $\kappa$ is the appropriate
multiple of the gravitational constant. The field equations require
$H=0$.

The gravitational part of $H$ is
\begin{eqnarray}\label{eq:Hgrav}
   H_{\rm(G)} &=& T_{\rm(G)} + U_{\rm(G)}
                 = - 2N |g|^{1/2}n^\alpha n^\beta  G_{\alpha \beta}
\nonumber\\
      &=& N|g|^{1/2} [ K^a{}_b K^b{}_a - K^a{}_a K^b{}_b
                           + \epsilon \, {}^{(3)}\kern-1pt R ]
\nonumber\\
      &=& 12 x^{-1} |g| [ K^a{}_b K^b{}_a - K^a{}_a K^b{}_b
                           + \epsilon \, {}^{(3)}\kern-1pt R ] \,,
\end{eqnarray} %
where the extrinsic curvature tensor has the usual expression
\begin{equation}
     K^a{}_b = -\case12 N^{-1} g^{ac} \dot{g}_{cb}
              = -  \case{1}{24} x |g|^{-1/2} g^{ac} \dot{g}_{cb}
\end{equation}
(a dot indicates the $\lambda$ derivative)
and ${}^{(3)}\kern-1pt
 R$ is the scalar curvature of the homogeneous hypersurfaces.

To go further one must discuss how the source enters the Hamiltonian.
Rather than attacking the general case,
we will confine our attention either to a scalar field
which contributes one additional degree of freedom, or to
sources whose additional degrees of freedom can be eliminated
by integrating the source field equations in such a way that
the source variables can be expressed entirely in terms of the
gravitational variables and constants of integration.
The second step towards obtaining the HH Hamiltonian
involves inserting those expressions for the source
variables into the Hamiltonian function.

The third step is to solve the supermomentum constraints, if nontrivial.
If these latter constraints are holonomic, i.e., can be expressed as the
vanishing of a total derivative, they can be integrated.
In this case they can be used to reduce the number of gravitational
degrees of freedom, and the above Hamiltonian function gives the
correct field equations.
On the other hand if these constraints are
nonholonomic, the reduction of the number of degrees of freedom may
not be possible or may not yield a Hamiltonian which gives the correct
field equations
because of symmetry considerations \cite{taumac}.
Because of these difficulties
one should always check that the Hamiltonian
function yields the correct field equations.

Recently the relationship between topology and the Hamiltonian formulation
of the field equations has attracted some interest \cite{ashsam,kod}.
If the symmetry type allows a closed topology for the homogeneous
hypersurfaces, then a Hamiltonian formulation exists. However, the
nonexistence of such a topology does not exclude a Hamiltonian formulation,
as exemplified by certain class B type VI$_h$ models treated below.

However, one does not just want to obtain a Hamiltonian for the problem,
but also to express it as simply as possible. This is helpful
not only in finding exact solutions but also for a qualitative analysis of
dynamics which cannot be described by exact solutions.
The fourth and final step towards obtaining the HH Hamiltonians
in a useful form
is accomplished by choosing new variables
which diagonalize the kinetic part $T$ of the Hamiltonian.
The way in which this is achieved differs for models which
are diagonal or nondiagonal, i.e., models for which the line element
can or cannot be expressed in diagonal form.
These two cases will be treated separately.

\subsection{Diagonal models}

The line element for diagonal HH models can be
written in the form
\begin{eqnarray}
  ds^2  &=&
  \epsilon N(\lambda)^2\, d\lambda^2
    + g_{11}{}(\lambda)\, (\omega^1)^2
    + g_{22}{}(\lambda)\, (\omega^2)^2
\nonumber\\ &&\quad
    + g_{33}{}(\lambda)\, (\omega^3)^2\,.
\end{eqnarray}
It is convenient to introduce the following variables
\begin{equation}
  (|g_{11}|,|g_{22}|,|g_{33}|) = (R_1{}^2,R_2{}^2,R_3{}^2)
   = (e^{2\beta^1},e^{2\beta^2},e^{2\beta^3})\,.
\end{equation}

The diagonal models all have a Hamiltonian
formulation.
The kinetic part $T_{\rm(G)}$ of the gravitational Hamiltonian
can not only be diagonalized but even made ``conformally flat"
by a linear transformation of the $\beta^a$ variables.

\subsubsection{Class A models}

For these models the momentum constraints are satisfied identically.
The parametrization
\begin{equation}\label{eq:parsA}
  \pmatrix{ \beta^1\cr \beta^2\cr \beta^3\cr} =
       \pmatrix{ \beta^0 + \beta^+ + \sqrt3 \beta^-\cr
                 \beta^0 + \beta^+ - \sqrt3 \beta^-\cr
                 \beta^0 - 2 \beta^+ \cr}
\end{equation}
introduced by Misner \cite{misnerbeta},
leads to the ``conformally flat" form
\begin{equation}
  T_{(G)} = \case12 x \eta_{AB} \dot\beta{}^A \dot\beta{}^B
\end{equation}
of the expression (\ref{eq:Hgrav}) for the
kinetic part of the gravitational Hamiltonian.
Here $\eta_{AB} = \eta^{AB}$ ($A,B=0,+,-$)
are the orthonormal components of the Minkowski metric
\begin{equation}
  - \eta_{00} = \eta_{++} = \eta_{--} = 1
\end{equation}
and the quantity $x$ is the conformal factor, making it more convenient
to use than $N$.
Equation (\ref{eq:Hgrav}) also yields the gravitational potential
\begin{eqnarray}\label{eq:gpotA1}
   U_{\rm(G)} &=& \epsilon N |g|^{1/2} \, {}^{(3)}\kern-1pt R
              = 12 \epsilon x^{-1} e^{6\beta^0} {}^{(3)}\kern-1pt R
\nonumber\\
              &=& 6 x^{-1}
     [ \sum_{a=1}^3 (n^{(a)}g_{aa}){}^2
         - 2 n^{(1)} n^{(2)} g_{11} g_{22}
\nonumber\\ & &\quad
         - 2 n^{(2)} n^{(3)} g_{22} g_{33}
         - 2 n^{(3)} n^{(1)} g_{33} g_{11} ] \,.
\end{eqnarray} %
In addition to being mathematically convenient, the $\beta^A$
variables are also adapted to physical quantities.
The quantity $|g|^{1/2} = e^{\beta^1+\beta^2+\beta^3}=e^{3\beta^0}$
is the volume-element density. The $\lambda$ derivatives of the
$\beta^0$, $\beta^\pm$ variables are respectively
proportional to the expansion and the nonzero shear components
of the congruence normal to the homogeneous hypersurfaces.

The possibility of choosing the timelike direction along different
axes in the static case yields an abundance of models.
To avoid being lost in details, the
signature freedom will be restricted
by requiring $\omega^1$ and $\omega^2$ to be spacelike,
a condition which
will be assumed in the rest of this article.
This leads to the line element
\begin{eqnarray}\label{eq:lineA}
   ds^2  &=&
        \epsilon [ N(\lambda)^2\, d\lambda^2
          - R_3(\lambda){}^2\, (\omega^3)^2 ]
          + R_1{}(\lambda)^2\, (\omega^1)^2
\nonumber\\ &&\quad
          + R_2(\lambda){}^2\, (\omega^2)^2 \,.
\end{eqnarray}
With this choice of timelike direction, the gravitational Hamiltonian
takes the form
\begin{eqnarray}
   H_{\rm(G)}
     &=& \case12 x \eta_{AB} \dot\beta{}^A \dot\beta{}^B + U_{\rm(G)}
\nonumber\\
     &=& \case12 x \eta_{AB} \dot\beta{}^A \dot\beta{}^B
                  + 12 x^{-1} e^{4\beta^0} V^\ast(\beta^\pm) \,,
\end{eqnarray}
where
\begin{eqnarray}\label{eq:vstar}
   V^\ast &=& \case12 e^{4\beta^+} h_-{}^2
                    + \epsilon n^{(3)} e^{-2\beta^+} h_+
                      + \case12(n^{(3)})^2e^{-8\beta^+} \,,
\nonumber\\
    h_{\pm} &=& n^{(1)} e^{2\sqrt3 \beta^-}
                         \pm n^{(2)} e^{-2\sqrt3 \beta^-} \,.
\end{eqnarray} %

An interesting subclass of these models is obtained by setting
$\beta^-=0$ (equivalent to $R_1=R_2$). For the Bianchi types I, II,
VII$_0$, VIII, and IX, these models admit multiply transitive symmetry groups
which correspond to local rotational symmetry
for the choice made for the timelike axis in (\ref{eq:lineA}).
The Bianchi type I and VII$_0$ models of this type coincide and are
plane-symmetric.
Only the type I models will be referred to in what follows.

\subsubsection{Bianchi type V and VI$_h$ models} \label{sec:classBm}

Most class B models do not admit a Hamiltonian formulation, but
some particular nondiagonal and all diagonal models (consistent with
the field equations and sources studied here) do.
In the following we will only consider SH class B models since the
static class B models seem to lack physical interest.
The SH class B Bianchi types V and VI$_h$
admit diagonal cases provided that
the structure constants have the form \cite{uni,ecs,symvac}
\begin{equation}
     C^1{}_{31} = a + q\,, \quad
     C^2{}_{32} = a - q \,, \quad
     a^2= -h q^2 \,.
\end{equation}
The line element has the same form (\ref{eq:lineA})
as in the class A case, but
the variables $R_a$ are related by an algebraic constraint
obtained from the single nontrivial holonomic momentum constraint.
Canonical choices for the structure constants are
\begin{eqnarray}
      {\rm V}:  \quad   && q=0\,, \quad a=1\,,
\nonumber\\
     {\rm VI}_h: \quad  && q=1\,, \quad a>0\,.
\end{eqnarray} %
Explicit coordinate representations of the 1-forms may be found in
\cite{uni,symvac}.

The following parametrization solves the momentum constraint
and diagonalizes $T_{\rm(G)}$
\begin{equation}\label{eq:parsB}
\pmatrix{\beta^1\cr \beta^2\cr \beta^3\cr} =
       \pmatrix{ \beta^0 - c(q-3a)\beta^\times \cr
                 \beta^0 - c(q+3a)\beta^\times \cr
                 \beta^0 + 2cq\beta^\times \cr } \,,
\end{equation}
where
$c=(q^2 + 3a^2)^{-1/2}$. Note that for the Bianchi type V case,
this reduces to the class A parametrization with $\beta^+=0$ and
$\beta^- = \beta^\times$. The parameter combination
$cq$ takes values in the interval
$[0,1)$ with the endpoint value 1 corresponding to the
class A limit of type VI$_0$ which arises from $a\to0$, with $q\ne0$.
The parametrization then reduces to the
corresponding class A case with $\beta^- =0$
and $\beta^+ = -\beta^\times$.
It is sometimes convenient to choose $a=1$ in Bianchi type V, leading
to $c^{-2}=3$, since this choice is the one usually used for the open
isotropic FRW models which are obtained by setting $\beta^\times = 0$ in
type V.

With the above parametrization the gravitational Hamil\-to\-ni\-an has the
expression
\FL
\begin{equation} \label{eq:Bvac}
    H_{\rm(G)}
       = \case12 x ( - \dot{\beta}{}^{0\,2} + \dot{\beta}{}^{\times\,2} )
              + 24 x^{-1} c^{-2} e^{4(\beta^0 - cq \beta^\times)} \,.
\end{equation}

\subsubsection{Multiply transitive models} \label{sec:mtm}

The MT models we have encountered so far are those belonging to class A
and the isotropic open type V model. There are some additional ones
described by the same form (\ref{eq:lineA}) of the line element as in
the class A case with the Misner parametrization restricted by the condition
$\beta^-=0 $ corresponding to $R_1 = R_2$.
For this case,
the $( \omega^1)^2 + (\omega^2 )^2 $ part of the line element represents
a 2-space of constant curvature $\sigma\in\{1,0,-1\}$, while
$\omega^3=dx^3$ is an exact differential. Explicit coordinate representations
of the other two 1-forms may be found in \cite{kraetal}.

For this class of models
the case $\sigma=0$ coincides with the locally rotationally symmetric (LRS)
Bianchi type I models already considered.
The case $\sigma=-1$ is an LRS class B
Bianchi type III case. The case $\sigma=1$ is the exceptional case
where no three-dimensional simply transitive subgroup exists, not
falling under the Bianchi classification.
The SH models of this latter case are called Kantowski-Sachs (KS)
models, while the static models are the spherically symmetric models.

The 3-curvature ${}^{(3)}\kern-1pt
 R$ reduces to the curvature of the
constant curvature 2-spaces
${}^{(3)}\kern-1pt
 R = R_1{}^{-2} \sigma = e^{-2\beta^1} \sigma$,
leading to the gravitational Hamiltonian
\begin{equation}\label{eq:MTHam}
      H_{\rm(G)} = \case12 x ( -\dot{\beta}{}^{0\,2} + \dot{\beta}{}^{+\,2} )
                 + 24 \epsilon \sigma x^{-1} e^{4\beta^0-2\beta^+} \,.
\end{equation}

The type III models can be obtained from the previously discussed type
VI$_h$ models. The values $a=1=q$, for which $c=\case12$, correspond to
the usual Bianchi type III = VI$_{h=-1}$ structure constants.
However, the equivalent choice $a=\case12=q$
leads to $c=1$ and the same Hamiltonian as
(\ref{eq:MTHam}) (with $\sigma=-1$)
for this case. In both cases one has $cq=\case12$.
The different values of $c$ are related by a
translation in the $\beta^A$ variables associated with a scaling of the
corresponding 1-forms.

When not otherwise stated a reference to MT models will refer to
the LRS type I, III models and the KS and static spherically symmetric
models described by the Hamiltonian (\ref{eq:MTHam}).

\subsubsection{Source terms}

Any combination of potential terms of the following form may be considered,
representing different arrangements of sources.

\bigskip\paragraph{Cosmological Constant.}

A cosmological constant term in Einstein's equations will be considered
as an additional term $ (-\Lambda/\kappa) g_{\alpha\beta}$
in the total energy-momentum tensor $T_{\alpha\beta}$.
Inserting this term in Eq.~(\ref{eq:H}) gives rise  to the potential
\begin{equation}
  U_{\rm(\Lambda)} = -2\epsilon N |g|^{1/2} \Lambda
           = -24\epsilon x^{-1} e^{6\beta^0} \Lambda \,.
\end{equation}

\bigskip\paragraph{Scalar Field.}

A scalar field adds another dependent variable and
contributes both to the kinetic part of the Hamiltonian
and to the potential \cite{wal}
\begin{equation}
   T_{\rm(sc)}  = \case12 x {\dot {\beta^\dagger}}{}^2\,,\quad
   U_{\rm(sc)}  = -24\kappa \epsilon x^{-1}e^{6\beta^0}
                               V_{\rm(sc)} ({\beta^\dagger})\,,
\end{equation}
where $V_{\rm(sc)} ({\beta^\dagger})$ is the scalar field potential and where
${\beta^\dagger} = \sqrt{\kappa/6} \phi$
is the relation between the scaled scalar
field variable ${\beta^\dagger}$ and the usual scalar field $\phi$.
It will be convenient to introduce $\beta^4 = \beta^\dagger$ and
an exponential scalar field variable
analogous to the gravitational variables $R_a$ by defining
\begin{equation}
            R_4 = e^{\beta^\dagger} \,.
\end{equation}
When convenient the index $a$ will take the values in the set $\{1,2,3,4\}$
and the range of the index $A$ may include the value ``$\dagger$."

\bigskip\paragraph{Perfect Fluids.}

For a perfect fluid source with energy density $\rho$, pressure $p$, and
4-velocity $u^\alpha$, the energy-momentum tensor is
\begin{equation}
  T^\alpha{}_\beta = (\rho +p) u^\alpha u_\beta + p \delta^\alpha{}_\beta \,.
\end{equation}

For the cosmological case $\epsilon=-1$, the 4-velocity for a diagonal
source must
be $u^\alpha= n^\alpha$.
For the static case $\epsilon=1$,
the 4-velocity must be along the timelike third direction
$u_\alpha = u_3 \delta^3{}_\alpha$ (recall Eq.~(\ref{eq:lineA})).
According to Eq.~(\ref{eq:H}), the fluid contributes the
potential
\begin{eqnarray}
 U_{\rm(fluid)}
    &=& 2\kappa N|g|^{1/2}n^\alpha n^\beta T_{\alpha \beta}
\nonumber\\
    &=&\cases{
     2\kappa N |g|^{1/2} \rho & (SH case)  \cr
     2\kappa N |g|^{1/2} p & (static case) \cr}
\end{eqnarray}
to the Hamiltonian but an equation of state is needed to express this
entirely in terms of the gravitational variables using the conservation
equations.

In the SH case an equation of state $p = p(\rho)$ is imposed.
Note that dust models do not exist in
the static diagonal case without additional source terms (to prevent the
dust from collapsing).
Since the fluid potential is proportional to $p$ in this case,
an equation of state $\rho = \rho(p)$ is therefore imposed instead.

Following Misner, Thorne and Wheeler \cite{mtw},
it is convenient to introduce the baryon
number density $n$ and
the chemical potential $\mu = (\rho + p)/n$ which satisfy
\begin{equation}\label{eq:dlnnmu}
   d \ln n  = (\rho +p )^{-1} d\rho\,,\quad
   d \ln \mu  = (\rho +p )^{-1} dp\,.
\end{equation}
The conservation equations then imply
\begin{eqnarray}
   n^\alpha T_\alpha{}^\beta{}_{;\beta}
  &=&
  (\rho + p) \{ u_\alpha [ u^\beta{}_{;\beta} + (\ln n\mu)_{;\beta} u^\beta]
\nonumber\\ &&\quad
   + u_{\alpha;\beta}u^\beta + (\ln \mu)_{;\alpha} \} n^\alpha = 0\,.
\end{eqnarray}

In the SH case the conservation equation reduces to
\FL
\begin{equation}\label{eq:ell}
   ( n u^\beta )_{;\beta} = 0 \quad \rightarrow \quad
   n g^{1/2} = ne^{3\beta^0} = \ell = const \,.
\end{equation}
This equation together with (\ref{eq:dlnnmu}) leads to $\rho= \rho(\beta^0)$
which gives the fluid potential
\begin{equation}
   U_{\rm(fluid)} = 24 \kappa x^{-1} e^{6\beta^0}\rho(\beta^0) \,.
\end{equation}

For the static case
the conservation equation reduces to the simple relation
\begin{equation}\label{eq:mu}
   d \ln \mu = \frac{dp}{\rho + p} = - d\beta^3 \,,
\end{equation}
which may be integrated to yield
\begin{equation}
   \mu R_3 = \mu e^{\beta^3}  = const \,, \quad
    \mu \propto e^{-\beta^3} \,,
\end{equation}
which means that $p= p(\beta^3) = p(\beta^0-2\beta^+)$.
This yields the the potential
\begin{equation}
   U_{\rm(fluid)} = 24 \kappa x^{-1} e^{6\beta^0}p(\beta^0-2\beta^+) \,.
\end{equation}

For a fluid with the equation of state $p = (\gamma-1) \rho$,
implying $\mu = \gamma \rho /n$,
the above relations (\ref{eq:dlnnmu}) can be integrated to yield
\begin{equation}
   \rho/\rho_{(0)} = (n/n_{(0)})^{\gamma} \,, \quad
     \mu/\mu_{(0)} = (n/n_{(0)})^{\gamma-1}  \,,
\end{equation}
and without loss of generality one can set $\rho_{(0)} = n_{(0)}{}^\gamma$
leading to
\begin{equation}\label{eq:rhomun}
   \rho = n^{\gamma} \,,\quad
     \mu = \gamma n^{\gamma-1}  \,.
\end{equation}
The values $\gamma= 1, \case43, 2$ describe respectively dust, radiation,
and stiff perfect fluids.

For the SH case, this leads to
\begin{equation}
   \rho = \rho_{(0)} e^{-3\gamma \beta^0}\,,
\end{equation}
where the choice $\rho_{(0)} = \ell^\gamma$
leads to the fluid contribution
\begin{equation}
   U_{\rm(fluid)} = 24 \kappa x^{-1} \ell^\gamma e^{3(2-\gamma)\beta^0}
\end{equation}
to the Hamiltonian.

In the static case one often assumes
$ \rho= \rho_{(0)} + (\eta-1) p $
as an equation of state, where $\eta$ and $\rho_{(0)}$ are constants.
This includes both $p=(\gamma-1)\rho$ when $\rho_{(0)} = 0 $, so that
$\eta= \gamma/(\gamma-1)$, as well as the case of constant energy density
obtained by setting $\eta=1$.
Inserting this into Eq.~(\ref{eq:mu}) yields
\begin{equation}
    p = (p_{(0)} + \rho_{(0)}) e^{-\eta\beta^3} - \eta^{-1} \rho_{(0)} \,,
\end{equation}
where $p_{(0)}$ is a constant of integration.
This leads to the fluid potential
\begin{equation}
    U_{\rm(fluid)} = 24\kappa x^{-1}
         [ (\rho_{(0)} + p_{(0)}) e^{ (6-\eta)\beta^0 + 2\eta \beta^+}
                  - \rho_{(0)} e^{6\beta^0} ] \,.
\end{equation}

Occasionally, particularly in the SH case,
one considers
several noninteracting perfect fluids with a common 4-velocity.
In this situation one has a perfect fluid potential for each such component
fluid.

\bigskip\paragraph{Electromagnetic Fields.}

An electromagnetic field has the energy-momentum tensor \cite{mtw}
\begin{equation}
    T^\alpha{}_\beta = - \case{1}{4\pi}
     \left( F^\alpha{}_\gamma F^\gamma{}_\beta
           - \case{1}{4} \delta^\alpha{}_\beta
           F^\gamma{}_\delta F^\delta{}_\gamma \right) \,.
\end{equation}
The simplest electromagnetic fields are aligned with one of the axes,
leading to a diagonal energy-momentum tensor.
Examining Maxwell's equations
for an electric or magnetic field or some combination of the two
along a single axis, one finds which directions are
allowed (if any). For the class A and MT models
the third axis is such a direction, leading to an energy-momentum
tensor with nonzero components
\begin{equation}
    T^0{}_0 = T^3{}_3 = - T^1{}_1 = - T^2{}_2 = - \rho\em \,.
\end{equation}
An examination of the conservation equations similar to that of the
perfect fluid case shows that the positive quantity
\begin{equation}
     g_{11} g_{22} \rho\em  = \kappa^{-1} e^2
\end{equation}
is a constant.
The
electromagnetic contribution to the Hamiltonian then takes the simple
form
\begin{equation}
  U\em =  -24\epsilon x^{-1} {\rm e}^2 e^{2\beta^3}  \,.
\end{equation}

In the static cylindrically symmetric case the
inhomogeneity together with the singular axis of symmetry breaks the duality
symmetry, and a magnetic field must lie along one of the spatial directions
not associated with the spatial independent variable $\lambda$.
However, the potential is identical with the previous case once a
suitable permutation of the dependent variables $R_a$ is made.

%%%%%%%%%%%%%%%%%%%%%%%%%%%%%%%%%%%%%%%%%%%%%%%%%%%%%%%%%%%%%%%%%%%%%%%%%%%
%%%%%%%%%%%%%%%%%%%%%%%%%%%%%%%%%%%%%%%%%%%%%%%%%%%%%%%%%%%%%%%%%%%%%%%%%%

\subsection{Some nondiagonal models}

Similar methods can be used to treat some of
the nondiagonalizable
cases. Some of the simplest nondiagonal cases with a Hamiltonian
formulation will be considered to illustrate this procedure.
Solving the momentum constraints leads to
a problem with an additional nondiagonal gravitational degree of freedom
which is associated with a cyclic variable.
Using the associated constant momentum, one obtains a reduced Hamiltonian for
the diagonal degrees of freedom with
an effective potential left behind in the
kinetic part of the Hamiltonian.
This is analogous to the centrifugal
potential which appears in the central force problem.

\subsubsection{The Taub-NUT-M model}

The Taub-NUT-M model is an excellent example to study
in this context for two reasons.
First it has a homogeneous slicing whose causal character changes from
spacelike to timelike and back to spacelike again.
Second it is a nondiagonal model which pieces together
diagonal models of each causality type, illustrating the way in which
some nondiagonalizable models behave.

Its line element can be put in the form \cite{rs}
\FL
\begin{equation}
    ds^2  = - 2 z^{-1} d\lambda \, \omega^1
           + g (\omega^1){}^2 + e^{2 w} [ (\omega^2){}^2 + (\omega^3){}^2 ]
       \,,
\end{equation}
where the $\omega^a$ are the same 1-forms as for the diagonal type IX
models and $z$, $g$, and $w$ are functions of $\lambda$.
The function $z$ is a slicing gauge function.
The function $g$ is positive in the SH Taub region, negative in
the static NUT region,
and zero at the bridging null hypersurfaces between them.
The $\lambda$ coordinate lines are null in contrast with the usual
orthogonal coordinate lines for which the line element is diagonalized in the
Taub and NUT regions.

Due to the existence of null hypersurfaces, it is perhaps most
straightforward to specialize the full curvature scalar Lagrangian
to this case, removing a total $\lambda$ derivative to obtain a
Lagrangian valid for the entire spacetime. This is equivalent
to the ADM Lagrangian in the separate Taub and NUT regions with the
shift and lapse freedom
fixed by the null condition on the $\lambda$ coordinate lines,
modulo a conformal factor $z$
representing the freedom remaining in the
parametrization of the slicing.
The corresponding Hamiltonian is given by
\FL
\begin{equation} \label{eq:tnutm}
                + z^{-1} ( \case12 g e^{-2 w} - 2 ) = 0 \,.
\end{equation}
The kinetic part can be easily diagonalized for $g\neq0$ and $g=0$
separately, but not for all values of $g$ simultaneously
which is needed to describe the full Taub-NUT-M spacetime.

\subsubsection{Stationary cylindrically symmetric models}

The stationary cylindrically symmetric models have a line element
which can be written as \cite{kraetal}
\begin{equation}
    ds^2  = N^2 d\lambda^2 - e^{2\beta^1} (dt + C d\phi)^2
           + e^{2\beta^2} d\phi^2 + e^{2\beta^3} dz^2 \,,
\end{equation}
where $\beta^a$, $N$ and $C$ are functions of the
independent variable $\lambda$ which is interpreted here as a radial
coordinate ordinarily denoted by the symbol $\rho$.
As for the diagonal cases,
the Hamiltonian is given by
\begin{equation}
   H = - 2 N |g|^{1/2} n^\alpha n^\beta
            ( G_{\alpha\beta} - \kappa T_{\alpha\beta} ) = 0 \,.
\end{equation}
Expressing the variables $\beta^a$ in terms of the Misner parametrization,
the vacuum Hamiltonian assumes the explicit form
\begin{equation}
  H = \case12 x  ( \eta_{AB} \dot\beta{}^A \dot\beta{}^B
                    + \case{1}{12} e^{4\sqrt3 \beta^-} \dot C{}^2 ) = 0\,.
\end{equation}
The momentum $p_C$ associated with the cyclic variable $C$
is constant, leading to
\begin{equation}
   \dot C = 12 e^{-\sqrt3 \beta^-} p_C
\end{equation}
and the reduced Hamiltonian
\begin{equation} \label{eq:Hcyl}
  H = \case12 x \eta_{AB} \dot\beta{}^A \dot\beta{}^B
          + 12 x^{-1} e^{-4\sqrt3 \beta^-}p_C{}^2  = 0\,.
\end{equation}

\subsubsection{Spatially homogeneous Bianchi type VI$_{-1/9}$ models}
\label{sec:sixn}

The orthogonal perfect fluid models ($u^\alpha = n^\alpha$)
of this type permit a nondiagonalizable line element of the form
\begin{equation}
     ds^2  = -N(t)^{2}dt^{2}+g_{ab}(t)\omega^{a}\omega^{b} \,,
\end{equation}
where $\omega^{a}$ are the invariant 1-forms for this symmetry type,
with structure constants
\begin{eqnarray}
   && C^a{}_{bc} = n^{(a)} \epsilon_{abc} + a \delta^a{}_3
              \quad (\hbox{no sum on $a$}) \,,
\nonumber\\
   && n^{(1)} = - n^{(2)} = 1\,, \quad n^{(3)} = 0 \,,\quad a= \case13\,.
\end{eqnarray} %
The 3-metric $g_{ab}$ can be conveniently parametrized
by $ g_{ab} = S^{c}{}_{a} S^{d}{}_{b} g^\prime_{cd} $,
where
$g'_{ab} = {\rm diag}(e^{2\beta ^{1} } ,e^{2\beta ^{2} } ,
     e^{2\beta ^{3} } )$
and
the matrix $S^{a}{}_{b}$ lies in the 3-dimensional special
automorphism group of the type VI$_{-1/9}$ Lie algebra
\begin{equation}
  (S^{a}{}_{b}) = \left( \begin{array}{ccc}
      \cosh \theta^{3}  &  -\sinh \theta^{3}  & \sqrt{2} \theta^{2} \\
     -\sinh \theta^{3}  &   \cosh \theta^{3}  & \sqrt{2} \theta^{1} \\
                    0   &                 0   &                  1
  \end{array} \right) \,.
\end{equation}

The momentum constraints remain to be satisfied in this
nondiagonal case. Using the standard Misner parametrization for the
$\beta^a$ variables and the above $\theta^a$ variables in the constraint
equations leads to \cite{kjell??}
\begin{equation}
        \theta ^{2} = \theta ^{1} \,, \quad
        \theta ^{3} = - \beta ^{+} \,.
\end{equation}
These conditions together with the transformation
\begin{equation}
         \beta ^{0} = \tilde{\beta} ^{0}\,, \quad
         \beta ^{+} = - \case{\sqrt3}{2}   \tilde{\beta} ^{+} \,, \quad
         \theta ^{1} = \varphi e^{ \beta ^{+}} \,,
\end{equation}
leads to the Hamiltonian
\begin{eqnarray}\label{eq:Hsixn}
     H &=& \case12 x [-(\dot{\tilde{\beta}}{}^{0})^{2}
       + (\dot{\tilde{\beta}}{}^{+})^{2}
       + \case{2}{3} e^{- 4\sqrt{3}\tilde{\beta}^{+}} \dot{\varphi}{} ^{2} ]
\nonumber\\ &&\quad
       + x^{-1} [
         32 e^{4\tilde{\beta} {}^{0} - 2\sqrt{3} \tilde{\beta} ^{+}}
       + 24 \ell ^{\gamma} e^{3(2-\gamma) \tilde{\beta}^{0}}] \,,
\end{eqnarray}
where $\ell$ is the same constant as in the diagonal case.
Since the fluid is orthogonal, it follows from Eq.~(\ref{eq:H})
that it has the same potential as in the diagonal case.

The momentum $p_\varphi$ associated with the cyclic coordinate $\varphi$ is
constant, leading to
\begin{equation}
    \dot\varphi = \case{3}{2} x^{-1} e^{ 4\sqrt3 \tilde{\beta} ^+ } p_\varphi
\end{equation}
and the reduced Hamiltonian
\begin{eqnarray}
     H &=& \case12 x [-(\dot{\tilde{\beta}}{}^{0})^{2}
       + (\dot{\tilde{\beta}}{}^{+})^{2}]
       + x^{-1} [
        \case{3}{2}e^{ 4\sqrt{3}\tilde{\beta}^{+}} p_{\varphi}{} ^{2}
\nonumber\\ &&\quad
       + 32 e^{4\tilde{\beta} ^{0} - 2\sqrt{3} \tilde{\beta} ^{+}}
       + 24 \ell ^{\gamma} e^{3(2-\gamma) \tilde{\beta}^{0}}] = 0 \,.
\end{eqnarray}
Note that $\varphi=0=p_{\varphi}$ reduces this case to the corresponding
diagonal case.

%%%%%%%%%%%%%%%%%%%%%%%%%%%%%%%%%%%%%%%%%%%%%%%%%%%%%%%%%%%%%%
\subsubsection{Spatially homogeneous class A models belonging to the
            symmetric case}

The class A perfect fluid models with an equation of state
$p= (\gamma - 1)\rho$
admit the special case where
the fluid 4-velocity has a single nonzero spatial component $u_3$
and where $g_{ab}$ in the line element
\begin{equation}
     ds^2  = -N(t)^{2}dt^{2}+g_{ab}(t)\omega^{a}\omega^{b}
\end{equation}
has only one nonzero offdiagonal coefficient $g_{12}=g_{21}$.
This is referred to as a ``symmetric case" \cite{uni}.
For Bianchi type II it is now more
convenient to choose the alternative structure constant values
$(n^{(1)}, n^{(2)}, n^{(3)}) = (1,0,0)$
in place of those of Table 1.

As in the previous case,
$g_{ab}$ can be conveniently parametrized
by $ g_{ab} = S^{c}{}_{a} S^{d}{}_{b} g'_{cd} $,
where $g'_{ab} $ is the same but
the special automorphism matrix $S^{a}{}_{b}$ is now
\begin{equation}
(S^{a}{}_{b}) = \left( \begin{array}{ccc}
                  c_{3}  &  - \hat n{}^{(1)} s_3  & 0 \\
   \hat n{}^{(2)} s_3    &   c_3                  & 0 \\
                     0   &                 0      & 1
  \end{array} \right) \,,
\end{equation}
where
\FL
\begin{eqnarray}
 && (\hat n{}^{(1)}, \hat n{}^{(2)} ) =
     e^{-\alpha^3} (n{}^{(1)}, n{}^{(2)} ) \,,\quad
      \hat m{}^{(3)} = (-\hat n{}^{(1)} \hat n^{(2)} )^{1/2} \,,
\nonumber\\
 &&  e^{\alpha^3} = 2^{-1/2} [ (n^{(1)}){}^2 + (n^{(2)}){}^2 ]^{1/2}\,,
\nonumber\\
 &&  c_3 = \cosh \hat m{}^{(3)} \theta^3 \,, \quad
      s_3 = (\hat m{}^{(3)} )^{-1/2} \sinh \hat m{}^{(3)} \theta^3 \,.
\end{eqnarray}
For Bianchi types I and II where
some of these expressions are undefined,
one defines them as the limit in which $n^{(2)}\to 0$ and in
Bianchi type I one then lets $n^{(1)}\to0$ \cite{uni}.

The Hamiltonian constraint (\ref{eq:H}) is then
\begin{equation}
   H = T_{\rm(G)} + U_{\rm(G)} + U_{\rm(fluid)} = 0 \,,
\end{equation}
where the kinetic part is given by \cite{uni}
\begin{equation}
   T_{\rm(G)} =   \case12 x [ \eta_{AB} \dot\beta{}^A \dot\beta{}^B
              + \case{1}{12} e^{-2\alpha^3} (h_-)^2 (\dot\theta{}^3)^2 ]
         \,,
\end{equation}
while the gravitational potential $U_{\rm(G)}$ is the same as for the
diagonal case. The fluid potential is given by
\begin{eqnarray}
    U_{\rm(fluid)}
      &=& 2 \kappa N g^{1/2} n^\alpha n^\beta T_{\alpha \beta}
\nonumber\\
      &=& 2 \kappa N g^{1/2} [(\rho+p)(n^\alpha u_\alpha)^2-p]
\nonumber\\
      &=& 24\kappa x^{-1} e^{6\beta^0} [(\rho+p)Y-p] \,,
\end{eqnarray}
where $Y=(n^\alpha u_\alpha)^2$. As for the diagonal case for an
equation of state $p=(\gamma-1)\rho$ one can introduce $\rho=n^\gamma$.
For these models there exists a constant of the motion \cite{uni,pfs}
\begin{equation}
    \ell = (-n^\alpha u_\alpha)n g^{1/2} = Y^{1/2}\rho^{1/\gamma}
           e^{3\beta^0} \,.
\end{equation}
Solving for $\rho$ leads to
\begin{equation}
    \rho = \ell^\gamma e^{-3\gamma\beta^0} Y^{-\gamma/2} \,.
\end{equation}
This relation is used to eliminate $\rho$ in the fluid potential
\FL
\begin{equation}\label{eq:yfluid}
    U_{\rm(fluid)} = 24 \kappa \ell^\gamma x^{-1} e^{3(2-\gamma)\beta^0}
                 Y^{-\gamma/2} [ \gamma Y - (\gamma-1)] \,.
\end{equation}
There is an additional constant of the motion $v_3$ defined by
$v_3 = \mu u_3$ where $\mu = \gamma n^{\gamma-1} = \gamma
\rho^{(\gamma-1)/\gamma}$ \cite{uni,pfs}.
This constant of the motion can be used to
express $Y$ as a function of $\beta^0$ and $\beta^+$. Expressing
$u^\alpha u_\alpha = -1$ in terms of $Y$ and $v_3$ yields the implicit
relation
\begin{equation}
    F Y^{\gamma-1} - Y + 1 = 0  \,,
\end{equation}
where
\begin{equation}\label{eq:F}
   F = \gamma^{-2} \ell^{-2(\gamma-1)} v_3{}^2
                       e^{2[(3\gamma-4)\beta^0 + 2\beta^+]} \,.
\end{equation}
For dust ($\gamma=1$, a case discussed in \cite{ryan})
and stiff ($\gamma=2$) perfect fluids this
equation is linear and can be explicitly solved.  For some other
values of $\gamma$ (namely $\case54$, $\case43$,
$\case32$, $\case53$, $\case74$), it reduces to at most a
fourth degree polynomial equation which can be explicitly solved in
principle.  However, for certain purposes an explicit expression is
not required, as will become apparent later.
Note that the stiff perfect fluid potential is the sum of two exponentials.

The variable $\theta^3$ is cyclic and so has a constant conjugate momentum
$\tilde P_3$ and the equation of motion
\begin{equation}
    \dot\theta{}^3 = 12 x^{-1} e^{2\alpha^3} (h_-)^{-2} \tilde P_3 \,.
\end{equation}
The single nontrivial supermomentum constraint requires \cite{uni}
\begin{equation}
   e^{\alpha^3} \tilde P_3 = -2\kappa \ell v_3 \,,
\end{equation}
leading to
\begin{equation}
   H =   \case12 x  \eta_{AB} \dot\beta{}^A \dot\beta{}^B
     + U_{\rm(c)} + U_{\rm(G)} + U_{\rm(fluid)}  \,,
\end{equation}
where
\begin{equation}\label{eq:centrpot}
    U_{\rm(c)} = x^{-1} 24 \ell^2 \kappa^2 (v_3)^2 (h_-)^{-2}  \,.
\end{equation}
Note that in the type II case and in the type VI$_0$ Taublike symmetric case
(a special case defined by $\beta^-=0$), the ``centrifugal potential"
$U_{\rm(c)}$ is just an exponential and a constant respectively.

\subsection{Some remarks}

The HH symmetry types include a wide range of models describing quite
different physical situations. For example,
in the diagonal case one may consider any combination of sources
by including the corresponding potential terms in the Hamiltonian,
which together with the two possible signs of $\epsilon$
leads to numerous spacetime models that have been considered in the
literature.
Many people have attacked  such problems individually as though they were
completely unrelated to the others, each time writing
out the field equations and attempting to solve them.
However, the Hamiltonian approach reveals the close {\it mathematical\/}
relationship which exists between them.

For example, all the models considered are characterized by a Hamiltonian
which can be reduced to the form
\begin{equation}
    H = \case12 x \eta_{AB} \dot{\beta}{}^A \dot{\beta}{}^B
         + x^{-1}U_{\rm(taub)}  = 0 \,,
\end{equation}
where the Taub potential,
$U_{\rm(taub)}$, is thus just the the value of the total
potential in the Taub slicing gauge $x = 1$.
The Lorentz character of the kinetic part of the Hamiltonian \cite{dew}
suggests that Lorentz transformations of the dependent $\beta^A$ variables
play an important role in analyzing the dynamics \cite{mis:mini}.
Indeed such transformations can be used to reveal how many different
problems are mathematically equivalent.

In many cases the Taub potential is a sum of exponentials
\begin{equation}
    U_{\rm(taub)} = \sum_i A_i e^{a^i{}_A \beta^A} \,.
\end{equation}
A Hamiltonian of this type will be referred to as a SE-Hamiltonian
(for ``Sum of Exponentials").
Problems of this type are important because
the gravitational potential and many source potentials
are sums of exponentials when expressed in the Taub slicing
gauge.

By inspecting the Hamiltonian
one can see that some of these various problems are
either equivalent or closely connected.
For example, changing the sign of one of the variables $g_{aa}$ in
the class A gravitational potential (\ref{eq:gpotA1})
to go from the SH case to a
corresponding static case either leaves the Hamiltonian unchanged or
is equivalent to a change of Bianchi type,
e.g., Bianchi types VIII and IX are interchanged by this operation.
Thus there is an isomorphism between various static models and the SH
ones.
Furthermore, the MT cases (\ref{eq:MTHam})
have the same gravitational potential modulo the sign of
$\epsilon\sigma$.
When it comes to sources
the cosmological constant and electromagnetic terms only change
sign with the change in sign of $\epsilon$.
However, a perfect fluid potential differentiates between the static and
homogeneous cases in an essential way
since the 4-velocity must be along the timelike direction in each case.

The close relationship among many Hamiltonians
explains the similarity of the expressions resulting from solving the field
equations for the different models.

%%%%%%%%%%%%%%%%%%%%%%%%%%%%%%%%%%%%%%%%%%%%%%%%%%%%%%%%%%%%%%%%%%%%%%%%%%%%%%
% section 3
%%%%%%%%%%%%%%%%%%%%%%%%%%%%%%%%%%%%%%%%%%%%%%%%%%%%%%%%%%%%%%%%%%%%%%%%%%%%%%

\section{The Generalized Friedmann Equation} \label{sec:gfp}

The generalized Friedmann equation \cite{gfp}
is an equation of the form
\begin{equation}
   \dot\alpha{}^2 =  x_\alpha^{-2}  \sum_{i=1}^n a_ie^{q_i\alpha}
\end{equation}
in a single dependent variable $\alpha$, where $q_i$ are a set
of distinct constants ordered by increasing value.
This generalizes the well known Friedmann equation which has this form
for the scale factor $R=e^\alpha$.
Here the arbitrary function $ x_\alpha$ allows different choices of
the independent variable. For convenience, it will
be called the slicing gauge function.

The generalized Friedmann equation
may be converted from exponential potentials to power law potentials
by introducing the power variable
\begin{equation}
u = e^{\delta\alpha}\,,
\end{equation}
where $\delta\neq0$ is a constant parameter.
This leads to
\begin{equation}
      \dot u{}^2 =
         x_\alpha^{-2}\delta{}^2 \sum_{i=1}^n a_iu^{(2 + q_i/\delta)} \,.
\end{equation}

\subsection{The power law slicing gauge approach}

To solve the generalized Friedmann equation, it is often
convenient to introduce a power law slicing gauge function \cite{pll}
\begin{equation}
    x_\alpha^2 = e^{\Delta\alpha} = u^{\Delta/\delta} \,,
\end{equation}
where $\Delta$ is a constant parameter.
This yields
\begin{eqnarray}
   \dot\alpha{}^2 &=& \sum_{i=1}^n a_ie^{(q_i- \Delta)\alpha} \,,
\nonumber\\
   \dot u{}^2 &=& \delta{}^2 \sum_{i=1}^n a_i u^{r_i} \,, \quad
        r_i = 2 + \frac{q_i}{\delta} - \frac{\Delta}{\delta} \,.
\end{eqnarray} %
It is convenient to refer to the right hand sides of either equation as
the ``potential" for that variable.

When only one potential term is present, one can either choose
$\Delta=q_1$ or
$\Delta = 2\delta =q_1$ so that $\alpha$ or $u$ respectively is affinely
related to the independent variable.

In the case of more than one potential term one can always treat one of the
terms in the same way one deals with the single term in the
case with only one potential term.
However, since there are two parameters available $(\Delta,\delta)$,
one can vary these so that any two power exponents $(r_i,r_j)$
assume any pair of real values.  Conversely, given the values of a pair of
the original power exponents $(q_i,q_j)$, the parameters $(\Delta,\delta)$
are determined by the values of the corresponding new power exponents
$(r_i,r_j)$ in the following manner.
The scale parameter $\delta$ determines the ratio of the power exponent
increments
\begin{equation}\label{eq:darb}
  \delta = (q_j-q_i)/(r_j-r_i) \,,
\end{equation}
after which the additive parameter $\Delta$ is determined by either of the
relations
\begin{equation}\label{eq:Darb}
 \Delta = q_i - \delta(r_i - 2) = q_j - \delta(r_j - 2) \,.
\end{equation}

In the case of more than one potential term the obvious way to fix the
parameters $(\Delta,\delta)$ is to obtain the
lowest polynomial power variable potential that exists, if any. However,
even transforming the potential to a polynomial may be impossible since
this requires that the original power exponents $(q_i)$ be affinely related to
a set of nonnegative integers. When this is the case for a given
set of integers, there are two choices
of the pair $(\Delta,\delta)$ for which a polynomial potential
occurs, corresponding to a positive and negative value of the nonzero power
variable parameter $\delta$. Thus one always has two different
slicing gauges
with the same degree polynomial potential, but whose dependent variables
are related to each other in an inverse power relationship. One is an
increasing function and the other a decreasing function
of $R=e^\alpha$.

In general, for a case reducible to a polynomial potential, if $q>0$ denotes
the minimum increment between the ordered
coefficients $(q_1,\ldots,q_n)$,
then the choice
\begin{equation}\label{eq:firstdd}
  (\Delta,\delta)= (q_1+2q,q)
\end{equation}
leads to the polynomial potential
with $q_1$ corresponding to the constant term, while the choice
\begin{equation}\label{eq:lastdd}
  (\Delta,\delta)= (q_n-2q,-q)
\end{equation}
leads to a polynomial potential where the last term associated with the final
value $q_n$ is the constant term. These are the minimal degree polynomial
potentials that are possible.
The solutions of the generalized Friedmann
problem for polynomial potentials up to degree
two (four) can be expressed in terms of elementary (elliptic) functions.

In the case of two potential terms, one can reduce the potential to
polynomials of first or second degree. The parameters $(\delta,\Delta)$
determining the power variables and choices
of gauge function which accomplish this are given in Table 2,
together with the resulting powers.

%\typeout{**********************************Table 2:}

If in the case of three potential terms the coefficients $(q_1,q_2,q_3)$
are ``equally spaced", i.e., $q_2 - q_1 = q_3 - q_2$, then there are two
values of the pair $(\Delta,\delta)$ which map them onto either
$(r_i)$ equal to $(0,1,2)$ or $(2,1,0)$.
Completing the square leads to a quadratic two potential term problem
for a new variable $\bar u$ affinely related to $u$ and the same kinds
of elementary functions for $\bar u$ result that occur for $u$ in
two term quadratic case with no first power term.
For details regarding all of these cases and those involving
more than three potential terms, see \cite{gfp}.

Often the generalized Friedmann equation occurs as a first integral of
a second order equation of the general form
\begin{equation}\label{eq:diffgfe}
 \ddot{\theta} +\delta \dot{\theta}\,^2+ f(e^{\theta})=0\,,
\end{equation}
where $f(\theta)$ is a sum of exponential terms.
This equation has the first integral
\FL
\begin{equation}
  {\cal E} = e^{2\delta\theta} [ \dot{\theta}\,^2 - h(e^{\theta}) ] \,,\quad
           h(e^{\theta})= - 2 e^{-2\delta\theta}
                  \int e^{2\delta\theta} f(e^{\theta})\,d\theta\,,
\end{equation}
which may be rewritten in the form of the generalized Friedmann
equation as
\begin{equation}\label{eq:diffgfeint}
    \dot{\theta}{}^2 =  h(e^{\theta}) + {\cal E} e^{-2\delta \theta} \,,
\end{equation}
or in the power form
\begin{equation}
    \dot{u}{}^2 = H(u) + \delta^2{\cal E} \,,\quad
           H(u) = \delta^2 u^2 h(u^{1/\delta})
\end{equation}
for the power variable $u= e^{\delta\theta}$ when $\delta\neq0$.
Note that these have one
more exponential potential than the number of potential terms in the original
second order equation.
As long as $f(e^\theta)$ has no terms like $e^{-2\theta}$,
then $h(e^\theta)$
will also consist of only exponential terms and $H(u)$ of powers of $u$.

Often the generalized Friedmann equation occurs as a decoupled equation in a
larger problem. In this case the slicing gauge function may be fixed
by considerations related to the larger problem.
When
the other part of the problem consists of evaluating quadratures,
one tries to find a slicing gauge so that
both the generalized Friedmann equation and the additional
quadratures become as simple as possible.

\subsection{The intrinsic slicing gauge approach} \label{sec:gfpi}

An intrinsic slicing gauge is defined as a slicing gauge which
relates a linear combination of the $\beta^A$ variables---or an exponential
of such a linear combination (a so-called ``power variable" discussed in
section \ref{sec:power})---affinely to the independent variable,
in effect making that dependent variable the new independent variable
for the problem.
This type of slicing gauge has played a prominent role in the context of
static HH models, e.g., for the spherically symmetric models one usually
chooses the power variable $r$
as the independent variable.
In contrast the power law slicing gauges have been much more important
for the SH models.
These natural
selection effects seem a bit strange given that the static HH models
and SH models are so closely related mathematically.

Recall that the generalized Friedmann equation can either be expressed in
its ``exponential'' form
$\dot \alpha{}^2 =  x_\alpha^{-2}\sum_{i = 1}^n a_i e^{q_i\alpha}$
or its ``power'' form
$\dot u^2 =  x_\alpha^{-2}\delta^2\sum_{i = 1}^n a_i u^{2 + q_i\delta}$ where
$u = e^{\delta\alpha}$. In this context an intrinsic slicing gauge means
that one either chooses $\alpha$ or some
$u$ variable as the independent variable.
Such choices are accomplished by setting
\begin{eqnarray}
   \alpha = \lambda: & &\quad
     x_\alpha =
            (\sum_{i = 1}^n a_ie^{q_i\alpha})^{1/2}\,, \nonumber\\
u = \lambda: & & \quad
    x_\alpha =
            \delta
   ( \sum_{i = 1}^n a_i u^{2 + q_i\delta} )^{1/2}\,.
\end{eqnarray}

These choices may seem trivial. However, if the generalized Friedmann equation
is part of a larger problem, they may not be.
For example one often encounters
problems where one has a cyclic variable which is given by a quadrature
of the form
${\bar\beta}^A = \int  x_\alpha^{-1} \eta^{AB} {\bar p}_B d\lambda$, where
${\bar p}_A$ is the constant conjugate momenta associated with
${\bar\beta}^A$. Choosing a power variable $u$ leads to the integral
${\bar\beta}^A =  \eta^{AB} {\bar p}_B \delta \int
   ( \sum_{i = 1}^n a_i u^{2 + q_i\delta} )^{-1/2} \, du $
which gives the dependent variable as a function of the independent one.
If on the other hand
one chooses a power law slicing gauge, this may simplify the equation for
the ${\bar\beta}^A$ variable while transfering the difficulties to the
generalized Friedmann equation. Even if
this  latter
equation is solvable in terms of a
quadrature, it gives the independent variable as a function of the dependent
variable $u$, in contrast with
the intrinsic approach which yields the dependent
variable as a function of the independent one. Thus one is faced with the
problem of trying to invert a quadrature in the power law gauge approach,
something which often fails to lead to familiar functions. Thus the intrinsic
approach sometimes has advantages.
As an example, see \cite{fru}, where the intrinsic approach was used in the
context of SH Bianchi type V orthogonal perfect fluid models.

%%%%%%%%%%%%%%%%%%%%%%%%%%%%%%%%%%%%%%%%%%%%%%%%%%%%%%%%%%%%%%%%%%%%%%%%%%%%%%
% section 4
%%%%%%%%%%%%%%%%%%%%%%%%%%%%%%%%%%%%%%%%%%%%%%%%%%%%%%%%%%%%%%%%%%%%%%%%%%%%%%

\section{Killing Tensor Symmetries and How to Use Them} \label{sec:KT}

To solve a Hamiltonian problem, one needs to find and exploit symmetries
which lead to constants of the motion.
In the {\it search\/} for such symmetries,
a particular slicing gauge will be introduced which makes the equations
of motion equivalent to the geodesic equations associated with a
certain metric (not to be confused with the spacetime metric).
This makes it natural to look for a particular type of
symmetry, called a Killing tensor symmetry,
which generalizes symmetries corresponding to cyclic variables
and Hamilton-Jacobi separability.

However, this slicing gauge is not usually well suited to exploiting
the symmetry so that {\it explicit\/} exact solutions can be obtained.
The existence of constants
of the motion is not sufficient to obtain such solutions explicitly.
This requires the stronger condition of a decoupling of the equations of
motion (which in turn leads to constants of the motion), for which other
slicing gauges are needed.
Even within this latter class of slicing gauges
there are choices of gauge which lead to
simpler forms of the exact solutions.
The situation is similar to the case of a cyclic variable.
Any slicing gauge which doesn't involve this cyclic variable
leads to a constant of the motion and decoupling.
However, some choices lead to simpler equations and simpler expressions
for their solutions.

The present section discusses the symmetry properties of
Hamiltonians and how to explicitly find solutions of the
equations of motion in as simple a form as possible.
It then relates these results to the Hamiltonians of the type encountered
when dealing with the HH models described in section 2.

\subsection{Why Killing tensor symmetries?}

In classical mechanics one usually encounters Hamiltonians of the form
\begin{equation}
       H = \case12 g^{ab}p_a p_b + U = E \,,
\end{equation}
where the symmetric matrix $(g^{ab})$ is positive-definite. For such
problems there exists an elegant geometric reformulation of the
corresponding equations to a geodesic flow on a
certain geometry \cite{lancz,abrmar,geom}. To
accomplish such a reformulation one first introduces a new Hamiltonian
${\cal H} = H-E = 0$. A Hamiltonian system of this kind can be
reparametrized by choosing a new independent variable $\bar\lambda$ in place
of the original one $\lambda$, leading to a new
Hamiltonian
\begin{equation}
   {\cal H}_{\cal N} = {\cal N} {\cal H} = {\cal N}(H-E)\,, \quad
   {\cal N}=d\lambda/d\bar\lambda \,.
\end{equation}
The final step is to make the Hamiltonian purely
kinetic. This is accomplished by a particular choice of parametrization,
${\cal N}_J = \case12(E-U)^{-1}$, and by adding a constant to the
Hamiltonian:
\begin{equation}
      H_J = {\cal H}_{{\cal N}_J} + \case12 =
            \frac{1}{4(E-U)} g^{ab} p_a p_b = \case12 \,.
\end{equation}
The corresponding Lagrangian equations can then be reinterpreted as
those of the geodesic flow of the so-called Jacobi metric
$J_{ab} = 2(E-U) g_{ab}$,
where $g_{ab}$ is the matrix inverse of $g^{ab}$,
and $\bar\lambda$ is an affine parameter along each geodesic.

The geodesic reformulation does not rely on the positive-definiteness of
$g_{ab}$. It works locally for any nondegenerate indefinite matrix $g_{ab}$.
For the HH spacetimes of section 2,
the Hamiltonian can be put in the form
\begin{equation}
   H = \case12 x \eta^{AB} \dot\beta{}^A \dot\beta{}^B + x^{-1} U_{\rm(taub)}
           =0 \,.
\end{equation}
The Hamiltonian is already parametrized and must vanish so
there is no need for the first steps in the above procedure. The choice
\begin{equation}
    x=  x_J = 2 | U_{\rm(taub)} |
\end{equation}
leads to a Hamiltonian
with a constant value of the potential. Redefining the Hamiltonian by
this constant yields
\begin{equation}
  H_J = T_J = \case12 J_{AB}{\dot \beta^A}{\dot \beta^B} =
 -\case12 {\rm sgn}(U_{\rm(taub)})\,,
\end{equation}
where
$ J_{AB} = x_J \, \eta_{AB} $.
In contrast with the usual classical mechanical problems,
the underlying geometry here is Lorentzian rather than Riemannian.

To solve a Hamiltonian problem one needs to find enough symmetries
leading to constants of the motion, i.e., variational symmetries,
which can be used to reduce the problem sufficiently so that the reduced
problem can be solved.
The Jacobi formulation is especially suitable for finding such symmetries
since they and their associated constants of the motion
take a particularly simple form in this formulation
as discussed below.
Furthermore, the geometric framework makes available
a wide range of tools familiar from symmetry investigations on
ordinary spacetime.

{\it Variational symmetries\/} are transformations of
the phase space which can be represented as
transformations on the tangent bundle (velocity phase space)
generated by vector fields of the following form \cite{olv}
\begin{equation}
    {\bf v} = \phi^a(x,\dot x) \frac{\partial}{\partial x^a} \,.
\end{equation}
The simplest variational symmetries are the point symmetries
of the configuration space itself. In the Jacobi formulation, a generator
$\phi^a(x) {\partial}/{\partial x^a}$ of a point symmetry
corresponds to a Killing vector field of the Jacobi metric \cite{krkt}.
All other variational symmetries involve derivatives of the dependent
variables and are called generalized symmetries \cite{olv}.

The simplest generalized symmetries are the ones for which
the components of their generating vector fields
are linear and homogeneous in the
derivatives, i.e., $\phi^a = K^a{}_b(x){\dot x}^b$.
This corresponds exactly to a second rank Killing tensor
$K^a{}_b$ of the Jacobi metric \cite{krkt}.
Recall that a Killing
tensor $K_{ab}$ is a symmetric tensor for which $K_{(ab;c)} = 0$; this
includes the trivial case in which $K_{ab}$ is proportional
to $J_{ab}$ and the special
case in which the Killing
tensor is the symmetrized tensor product of Killing vectors.
Killing vector symmetries give rise to constants of the motion which are
linear and homogeneous in the momenta,
while second rank Killing tensor symmetries
correspond to homogeneous quadratic constants of the motion
($K^{ab}p_a p_b = const.$) \cite{kraetal,dietz}.
This is another way
of understanding how Killing tensor symmetries are the natural
generalizations of Killing vector symmetries.

The separability condition for the Hamilton-Jacobi equation is
equivalent to the existence of a sufficient number of second rank
Killing tensor symmetries of the Jacobi metric $J_{ab}$ \cite{dietz}.
Furthermore, in the case of an indefinite Jacobi metric,
separability requires the Killing tensors to have nonnull eigendirections.
However, it turns out that Killing tensors with null
eigendirections can also be used to solve the Hamiltonian equations.
Thus, in contrast with the positive-definite case, Killing tensor
symmetries in the indefinite case are more powerful than Hamilton-Jacobi
separability when searching for exact solutions.

The Jacobi slicing gauge is not the most convenient gauge to use when
it comes to actually producing exact solutions. There are more suitable
symmetry compatible choices of independent variable.
However, other such choices
lead to corresponding constants of the motion which are quadratic but
not homogeneous in the momenta.
Since the constant of the motion takes an especially simple form in the
Jacobi slicing gauge, entirely characterized by the Killing tensor alone,
it is this gauge which makes {\it finding\/}
the associated Killing tensor symmetries as simple as possible.
Thus it seems clear that a search for these
symmetries is a natural first step in attempting to find
exactly solvable problems.

Many of the HH field equations have long been known to admit a
Hamiltonian formulation very much like the traditional classical
mechanical problem except for the signature difference. Therefore it
is surprising that no attempt has ever been made to solve the field
equations by means of Hamilton-Jacobi separation given its success in
the Riemannian case. It is perhaps more understandable that no one has used
the more powerful but less familiar Killing tensor techniques.

\subsection{Killing tensor symmetries}

For many of the known exact HH solutions, the field equations are
ultimately expressible in terms of two nontrivial degrees of freedom
(although more variables may be involved).  It is therefore useful to
start with the case of two degrees of freedom, reviewing and extending
earlier work \cite{ru:kt}.

When dealing with 2-dimensional geometries the group of conformal
transformations plays a particularly important role since in this case
it is infinite-dimensional.  For 2-dimensional Lorentzian geometries
it is useful to use null variables since they are closely related to
this group.  A Lorentzian 2-metric can always be written in the form
\cite{wal}
\begin{equation}\label{eq:twogeom}
   ds{}^2 = -2G(w,v) dw dv
\end{equation}
in terms of the null variables $w$ and $v$.
A general conformal transformation to new null variables $\tilde V$ and
$\tilde W$ is of the form
$v = F({\tilde V}), w = {\tilde F}({\tilde W})$, and results
in a new conformal factor
${\tilde G} = F^\prime({\tilde V}){\tilde F}^\prime({\tilde W})G$.
One may then introduce new conformally inertial coordinates which
diagonalize the metric
\begin{equation}
  T = \case12({\tilde W} + {\tilde V})\,, \quad
  X = \case12({\tilde W} - {\tilde V})\,.
\end{equation}

It turns out to be convenient to classify the geometries allowing
Killing tensors into three cases characterized by different conditions on
the form to which the conformal factor can be transformed.
These three cases can be characterized as follows \cite{ru:kt}:
\begin{eqnarray}\label{eq:threecases}
  (i) && \ (\hbox{null}) :
      \quad {\tilde G} = [A({\tilde W})V({\tilde V})
                           + B({\tilde W})][dV({\tilde V})/d{\tilde V}] \,,
\nonumber\\
  (ii) && \ (\hbox{nonnull H-J}) :
          \quad {\tilde G} = C(T) + D(X) \,,
\nonumber\\
  (iii) && \ (\hbox{nonnull harmonic}) :
\nonumber\\ && \
        {\tilde G}_{,{\tilde V}{\tilde V}}
                            + {\tilde G}_{,{\tilde W}{\tilde W}} =
             \case12({\tilde G}_{,TT} + {\tilde G}_{,XX}) = 0 \,.
\end{eqnarray} %
In the above expressions $A, B, C, D, V$ are arbitrary functions
and the variables $\tilde V, \tilde W$ or $T,X$ will be referred to
as symmetry-adapted variables.

The Killing tensor in case $(i)$
is characterized by having a degenerate eigenvalue corresponding to a single
null eigenvector.
The remaining cases $(ii)$ and $(iii)$
have a Killing tensor
characterized by nonnull eigenvectors.
The case $(iii)$ corresponds to a conformal factor
satisfying Laplace's equation which is therefore a harmonic function
of ${\tilde W}, {\tilde V}$ or equivalently of $T, X$.
Of the three Killing tensor cases it is only the case $(ii)$ that
correponds to Hamilton-Jacobi separability.
Cases $(i)$ and $(ii)$ are of special importance. The case
$(i)$ will be referred to as the ``null case'' while, for simplicity,
the case $(ii)$  will be referred to as the ``nonnull case''
unless explicitly qualified as the ``H-J case". The case $(iii)$ will
be referred to as the ``harmonic case."

The expression for the conformal factor in case $(i)$ can be simplified to
\begin{equation}\label{eq:ib}
     (ib) \quad  \tilde G = A(W)V + B(W)
\end{equation}
by changing the variables
to $V = \int V,_{\tilde V} d{\tilde V}$ and $\tilde W = W$.
However, the original expression is more convenient to use as starting point
for an analysis of some of the examples to be discussed below.
Note that the form $(i)$ or $(ib)$ is invariant under the
transformations $W \to W(\bar W)$, leading to an equivalence class of
symmetry-adapted dependent variables.

When a Killing vector exists rather than a nontrivial Killing tensor,
Eq.~(\ref{eq:threecases}) continues to hold with either $AB=0$
or $CD=0$. The condition $AB=0$ is associated with the existence of
a null Killing vector, leading to a flat Jacobi geometry. The condition
$CD=0$ corresponds to a nonnull Killing vector field, with
$C=0$ describing the timelike case and $D=0$ the spacelike case.

The case $(ii)$ is
also generalizable directly to a 2-dimensional subblock of the metric
in higher dimension or to the
case of complete separability where this factor has the form
\begin{equation}
    {\tilde G} = C(T) + \sum_i D_i(X^i) \,.
\end{equation}
The eigenvectors of the Killing tensor in the subblock
case are aligned with the conformally
inertial coordinate directions $T$ and $X$.
In the case of complete separability
one has a set of nonnull Killing tensors whose
eigenvectors are aligned with $T$ and $X^i$. Intermediate cases between
these two extreme cases are also possible.
In the special case of a Killing vector, one can choose symmetry-adapted
conformally inertial coordinates
such that the conformal factor is independent of one of them.

\subsection{How to use Killing tensor symmetries}

We are not just interested in obtaining a sufficient number of constants
of the motion to solve the problem implicitly. We want to obtain
{\it explicit\/} exact solutions. To do this one needs to decouple a
sufficient number of equations. In general
the Jacobi slicing gauge does not allow this even if one
has a sufficient number of Killing tensor symmetries.
However, there are other choices of the independent variable which
permits this if one chooses symmetry-adapted dependent variables.
Since there is some freedom in the choice of the dependent
variables, this freedom may be exploited to help simplify the problem further.
A specific choice of dependent variables often suggests which independent
variable to use to actually integrate the equations of motion.

To simplify the discussion
only 2-dimensional examples will be considered.
To find useful slicing gauges one must first
re-express the general-slicing-gauge Hamiltonian in terms
of the symmetry-adapted dependent variables and the special
conformal factor $\tilde G$ associated with the existence of one
of the three types of Killing symmetries. This then suggests a choice
of independent variable in terms of which
the equations of motion can be explicitly solved.

Suppose one has a Hamiltonian of the form
\begin{equation}
  H = \case12 x (- {\dot \alpha}{}^2 + {\dot \beta}{}^2)
           + x^{-1} U_{\rm(taub)} \,.
\end{equation}
By introducing the null variables
\begin{equation} \label{eq:nullvars}
   w=\alpha + \beta\,,\quad v= \alpha -\beta \,,
\end{equation}
one transforms the Hamiltonian to the form
\begin{equation}
  H = - \case12 x \, \dot w \, \dot v + x^{-1} U_{\rm(taub)} \,.
\end{equation}
This Hamiltonian
corresponds to a Jacobi geometry with the conformal factor
$G = |U_{\rm(taub)}|$ (recall that $x_J = 2|U_{\rm(taub)}|$).

Re-expressing the Hamiltonian in terms of a new set of null variables
$w = w({\bar W})$ and $v = v({\bar V})$ leads to
\begin{equation}\label{eq:barh}
  H = - \case12 x \, (dw/d{\bar W})(dv/d{\bar V}) \,
       \dot{\bar W} \, \dot{\bar V} + x^{-1} U_{\rm(taub)} \,.
\end{equation}
If the Jacobi geometry admits a Killing tensor and these new variables
are associated symmetry-adapted variables,
then one can make the identification
\begin{equation}
   \tilde G = (dw/d{\bar W})(dv/d{\bar V}) \, |U_{\rm(taub)}| \,.
\end{equation}
Inserting this relationship into the Hamiltonian leads to
\begin{equation}
  H = - \case12 x \,  |U_{\rm(taub)}|^{-1} \, \tilde G
       \dot{\bar W} \, \dot{\bar V} + x^{-1} U_{\rm(taub)} \,.
\end{equation}

It is convenient to introduce a new slicing gauge function $y$ defined by
\begin{equation}\label{eq:y}
  y = |U_{\rm(taub)}|{}^{-1} \, {\tilde G} \, x\,,
\end{equation}
when discussing symmetry-adapted slicing gauges.
Expressing the above Hamiltonian in terms of $\tilde G$ and $y$ yields
\begin{equation}\label{eq:hy}
  H = - \case12 y \, \dot{\bar W} \, \dot{\bar V}
  + y^{-1} \mathop{\rm sgn}\nolimits                 %\sgn
     (U_{\rm(taub)}) \, {\tilde G} \,.
\end{equation}
The two cases $(i)$ and $(ii)$ in Eq.~(\ref{eq:threecases})
will be treated separately when it comes to finding
slicing gauges which lead to decoupling and explicit solution of the
equations of motion.
The case $(iii)$ does not seem
to have the same importance as the first two Killing tensor cases and will
only be commented on briefly when it does occur.

\subsubsection{The null Killing tensor case}

The Lagrangian equations of motion corresponding to the previous
Hamiltonian (\ref{eq:hy}) are
\begin{eqnarray}
  && \ddot {\bar W} + y^{-1} (\partial y/\partial {\bar W}) \dot {\bar W}{}^2
\nonumber\\ &&\quad
 - 2 y^{-1} \partial (y^{-1} \mathop{\rm sgn}\nolimits %\sgn
                            (U_{\rm(taub)}){\tilde G}) /\partial {\bar V}
     = 0 \,,
\nonumber\\
  && \ddot {\bar V} + y^{-1} (\partial y/\partial {\bar V}) \dot {\bar V}{}^2
\nonumber\\ &&\quad
 - 2 y^{-1} \partial (y^{-1} \mathop{\rm sgn}\nolimits %\sgn
                             (U_{\rm(taub)}){\tilde G}) /\partial {\bar W}
     = 0 \,.
\end{eqnarray} %
Assume that one has chosen symmetry-adapted variables so that
the conformal factor takes the simpler form $(ib)$
\begin{equation}\label{eq:g}
    \mathop{\rm sgn}\nolimits %\sgn
     (U_{\rm(taub)}) \tilde G
          = {\bar A}({\bar W}){\bar V} + {\bar B}({\bar W}) \,.
\end{equation}
It is then possible to choose a
slicing gauge so that the equation of motion for the variable $\bar W$
decouples.
Those slicing gauges which allow this decoupling are characterized by the
condition
\begin{equation}
   y =  y(\bar W) \
\end{equation}
and will be referred to as decoupling slicing gauges.
A decoupling of the variable $\bar V$  occurs for the analogous conditions
with $\bar V$ and $\bar W$ interchanged.

For a given choice of $y(\bar W)$ one can choose new dependent
variables $W$ and $V$ defined by
\begin{equation}
   dW/d{\bar W} = y(\bar W)\,,\quad V = {\bar V}\,.
\end{equation}
This transformation
eliminates the quadratic first derivative term from the decoupled
equation and leads to the Hamiltonian
\begin{eqnarray}\label{eq:hnull}
  H &=& - \case12  \dot W \dot V +
  [d{\bar W}(W)/dW][{\bar A}({\bar W}(W)) V + B({\bar W}(W))]
\nonumber\\
    &\equiv& - \case12  \dot W \dot V +  A(W) V + B(W) \,,
\end{eqnarray}
which has its kinetic energy in the standard Minkowski null form.

A Hamiltonian of this final form leads to the decoupled equation
\begin{equation}
     \ddot W = 2 A(W) \,,
\end{equation}
which has the first integral
\begin{equation}\label{eq:Wfirstintegral}
   \dot W{}^2 = 4 \int A(W) \, dW + {\it const} \,.
\end{equation}
Rewriting this in the form
\FL
\begin{equation} \label{eq:Wgfe}
   \case12 \dot W{}^2 + {\cal U}(W) = E \,, \quad
    {\cal U}(W) = -2 \int A(W)\, dW  \,,
\end{equation}
where $E$ is a constant, allows ${\cal U}(W)$ to be interpreted as
a potential for this 1-dimensional problem.

By appropriately choosing the function $y(\bar W)$,
one can make the function
$A(W)$ take any desired form. However, there are preferred choices.
For example,
choosing
$A(W)$ to be proportional to one of the powers 0, 1, or $-3$  of $W$
or to a sum
of  terms involving the  powers 0 and 1 of $W$
leads to a generalized Friedmann problem with
elementary function solutions for $W$ as a function of the independent
variable. The first two single term choices lead to a linear and quadratic
potential, while in the last single term case the transformation
$\hat W = W^2$ leads to a linear potential. The combination of 0 and 1
power terms leads to a quadratic polynomial potential.
All of these potentials were discussed in section \ref{sec:gfp}.
After a choice of gauge, leading to a particular form for $A(W)$,
the remaining variable  $V$
is determined by first solving the Hamiltonian constraint for $\dot V$,
yielding
\begin{equation}
  \dot V = (2/\dot W) [A(W) V + B(W)] = {\ddot W} V/\dot W + 2B(W)/\dot W \,,
\end{equation}
and then integrating this linear first order equation, leading to
\begin{equation}
  V = {\dot W} \left( {\it const} + 2 \int B ({\dot W})^{-2}
            d\lambda \right) \,.
\end{equation}
Whether or not the integral can be
expressed in terms of elementary
functions depends on the specific case one is dealing with.

It follows from the above formulas that the expressions for $U_{\rm(taub)}$
and $x$ then become
\begin{eqnarray}\label{eq:nullseppot}
   U_{\rm(taub)} &=& [ dW(w)/dw ] \,  [ d V(v)/dv ] \,  [A(W) V + B(W)]
\nonumber\\
    &\equiv& [ g(w) V(v) + h(w) ][dV(v)/dv]
\nonumber\\
   x &=&  [ dW(w)/dw ] \,  [ d V(v)/dv ] \,.
\end{eqnarray} %
As noted above after Eq.~(\ref{eq:ib}),
the Jacobi geometry is flat if $g(w)=0$ or $h(w)=0$.
One can often see by inspection whether or not
a Taub potential is of the above form.
For such cases it is convenient
to start with the gauge function $x$ and derive Eq.~(\ref{eq:hnull}).
Those slicing gauges which allow decoupling are
characterized by the condition
\begin{equation}
   x =  x_1(w) \, dV(v)/dv \,,
\end{equation}
where $x_1(w)$ takes the role $y$ previously played.
Decoupling in the Taub slicing gauge itself can only occur if $V$ is a linear
function of $v$.
If $V$ is not a linear function of $v$ one makes a variable change from $v$
to $V$ to obtain decoupling. This leads to
\FL
\begin{equation}\label{eq:nullxgh}
  H = - \case12 x_1(w) \, \dot w \, \dot V
  +  x_1(w)^{-1} [ g(w) V + h(w) ] = 0\,,
\end{equation}
For a given choice of $x_1(w)$, the choice
\begin{equation}
   dW/dw = x_1(w) \,,
\end{equation}
of a new dependent variable $W$ leads to
\begin{eqnarray}
   x &=&  [ dW(w)/dw ] \,  [ d V(v)/dv ] \,,
\nonumber\\
   H &=&  - \case12  \dot W \dot V +  A(W) V + B(W) \,,
\end{eqnarray}
that is, the same expressions as the end result of taking the ``$y$
approach''.

Many Hamiltonians correspond to null Killing tensor cases,
thereby explaining the existence of a whole host of exact solutions.

\subsubsection{The nonnull Killing tensor case}

Making the transformation $T = \case12 ({\bar W} + {\bar V})$,
$X = \case12 ({\bar W} - {\bar V})$ transforms the Hamiltonian
in Eq.~(\ref{eq:hy}) to
\begin{equation}\label{eq:nonnully}
  H = \case12 y \, (-\dot{T}{}^2 +  \dot{X}{}^2)
  + y^{-1}  \mathop{\rm sgn}\nolimits %\sgn
            (U_{\rm(taub)}) \, {\tilde G} \,.
\end{equation}
If one has a nonnull Killing tensor case expressed in symmetry-adapted
variables so that
\begin{equation}
    \mathop{\rm sgn}\nolimits  %\sgn
     (U_{\rm(taub)}) \, {\tilde G} = C(T) + D(X) \,,
\end{equation}
then the choice
\begin{equation}
  y = 1\, ,\quad x = |U_{\rm(taub)}|{\tilde G}{}^{-1}\,,
\end{equation}
leads to the Hamiltonian
\begin{equation}\label{eq:hnonnull}
  H = \case12 (-\dot{T}{}^2 +  \dot{X}{}^2)
      + C(T) + D(X) \,,
\end{equation}
for which both the $T$ and $X$ equations decouple
\begin{equation}
  \ddot T = dC(T)/dT\, ,\quad \ddot X = - dD(X)/dX\,.
\end{equation}
These equations give the first integrals
\begin{equation}
  -\case12 {\dot T}{}^2 + C(T) = E_t\, ,\quad
   \case12 {\dot X}{}^2 + D(X) = E_x\,,
\end{equation}
where the integration constants have to satisfy
\begin{equation}
  E_t + E_x = 0\,,
\end{equation}
because of the Hamiltonian constraint.
Equation (\ref{eq:hnonnull}) corresponds to Hamilton-Jacobi separability and
is generalizable to any dimension.

\subsection{Power variables and power law slicing gauges}  \label{sec:power}

Power variables are often the simplest choice of dependent
variables adapted to Killing tensor symmetries.
Such  a variable is defined by an expression of the form
\begin{equation}
    u = u{}_{(0)} \prod_a R_a{}^{Q_a}
        = u{}_{(0)} e^{Q_a\beta^a}
        = u{}_{(0)} e^{Q_A\beta^A} \,.
\end{equation}
Similarly
power law slicing gauges are often the simplest
slicing gauges which lead to decoupling.
Such gauges are characterized by the following
form of the slicing gauge functions \cite{pll}
\begin{eqnarray}
 N &=& N_{(0)}\prod R_a{}^{Q_a}
    = N_{(0)} e^{Q_a \beta^{a}}
    = N_{(0)} e^{Q_A \beta^{A}} \,,\quad {\rm or}
\nonumber\\
    x^{-1} &=& [N_{(0)}/12] e^{(Q_A \beta^{A} - 3\beta^0)} \,.
\end{eqnarray} %

The conditions $Q_1=Q_2=Q_3$,
or equivalently $Q_\pm=0$ or $Q_\times=0$ as appropriate,
(and  $Q_4=0$ if a scalar field is present)
characterize the isotropic power law gauges
which can be useful for SH models
in the case of spatially isotropic
sources like an orthogonal perfect fluid or a cosmological constant term,
the Taub time gauge being an obvious example.
However, because of anisotropic spatial curvatures and sources, one needs
to exploit anisotropic dependence of the function $N$ on the individual
scale factors to make progress in obtaining solutions or simplifying the
equations.
The condition $Q_1+Q_2+Q_3=1$ or equivalently $Q_0=1$ leads to a
scale invariant independent variable.
The conformal gauge corresponds to both conditions holding, leading to
$N=N_{(0)} e^{\beta^0}$. Finally it is worth noticing that power variables
and power law slicing gauges are not only useful when it comes to finding
symmetry-adapted variables but also in the context of qualitative analysis.

%%%%%%%%%%%%%%%%%%%%%%%%%%%%%%%%%%%%%%%%%%%%%%%%%%%%%%%%%%%%%%%%%%%%%%%%%%%%%%

\subsection{Killing tensor symmetries in models characterized by
         SE-Hamiltonians}

\subsubsection{Null decoupling} \label{sec:null}
Consider a potential of the form
\begin{equation} \label{eq:nullsep}
   U_{\rm(taub)} =  [ g(w) V(v) + h(w) ][dV(v)/dv] \,.
\end{equation}
If $g(w)$, $h(w)$ and $V$ are sums of exponentials including possible
constant additive terms, or if $g(w) = 0$ and $V$ is linear in $v$,
then one has a SE-Hamiltonian corresponding to a null Killing tensor case.
Of particular interest is the case in which $V$ is a single exponential term
or linear in $v$ in the ``flat'' $g(w) = 0$ case.
The remainder of the subsection analyzes this situation.
The Taub potential for the cases
\begin{eqnarray}\label{eq:dvdv}
   V &=& e^{c_0v}/c_0 \,,
        \quad c_0\neq0 \,, \quad g(w)\neq0 \quad {\rm or} \quad g(w)=0 \,,
\nonumber\\
   V &=& v\,, \quad c_0=0 \,, \quad  g(w)=0 \,,
\end{eqnarray} %
can then be formally written as
\begin{equation} \label{eq:nutaub}
  U_{\rm(taub)} = [c_0 (\sum_{i = 0}^{p} A_i e^{a_i w})e^{c_0 v} +
             \sum_{i = 0}^{q} B_i e^{b_i w}] e^{c_0 v} \,.
\end{equation}
Note that the three separate cases
$c_0=0$, or $A_i=0$ or $B_i=0$ for all values of $i$ all
correspond to flat cases.

Choosing $x=x_1(w) e^{c_0v}$ and $V = e^{c_0v} / c_0$ as a new variable
in the $c_0 \neq 0$ case leads to the Hamiltonian
\begin{eqnarray}\label{eq:HwV}
  H &=& - \case12  x_1(w) \, \dot w \, \dot V
\\ &&\quad
      + x_1(w)^{-1} [c_0{}^2 (\sum_{i = 0}^{p} A_i e^{a_i w}) V +
                             \sum_{i = 0}^{q} B_i e^{b_i w}]\,.
\nonumber
\end{eqnarray}
In the case  $c_0 = 0$, $U_{\rm(taub)}$ is independent of $v$,
and the Hamiltonian takes this same form with
$V$ equal to $v$.

Usually the simplest class of choices for the remaining gauge function
$x_1(w)$ are power law slicing gauges. For such gauges this function
is proportional to an exponential, leading to
\begin{equation}
  x = {\hat B} e^{m w} e^{c_0 v} \,,
\end{equation}
and the Hamiltonian
\begin{eqnarray}
  H &=& - \case12  {\hat B} e^{m w} \, \dot w \, \dot V
\\ &&\quad
      + {\hat B}{}^{-1} e^{-m w}
           [c_0{}^2 (\sum_{i = 0}^{p} A_i e^{a_i w}) V +
                             \sum_{i = 0}^{q} B_i e^{b_i w}]\,.
\nonumber
\end{eqnarray}
Here one may choose $m$ to be either zero or nonzero.

\bigskip\paragraph{The gauge choice $ m=0$.}

Setting ${\hat B} = 1$ one then has
\begin{equation}
  H = - \case12   \dot w \, \dot V
      +  [c_0{}^2 (\sum_{i = 0}^{p} A_i e^{a_i w}) V +
             \sum_{i = 0}^{q} B_i e^{b_i w}]\,.
\end{equation}
This Hamiltonian leads to a decoupled second order differential equation
which in turn can be integrated to yield a first order differential equation
\begin{equation}
   {\dot w}{}^2 = 4 c_0{}^2 \sum_i (A_i e^{a_i w})/a_i + const
\end{equation}
if all the coefficients
$a_i$ are nonzero. If not, the single $a_i = 0$ term
gives rise to a term linear in $w$ which has to be added to this
expression in order to give the correct form for the first order
equation.

The decoupled first order equation for $w$
arising from the $a_i \neq 0$ case is simply the
generalized Friedmann equation,
while $V$ may be obtained formally by solving the Hamiltonian constraint
(which is linear in $\dot V$) for its derivative and integrating.
To simplify the process of integrating this final pair of equations
(abandoning further use of the Hamiltonian/Lagrangian approach),
one can reintroduce the gauge freedom
that was fixed above by a specific choice of $x_1(w)$ made
to obtain the decoupled first order equation.
However, usually this extra freedom is not needed to find
many simple explicit expressions for the solution of the system.

\bigskip\paragraph{The gauge choice $m\neq0$.}

If one instead chooses $m \neq 0$, and follows the procedure after
Eq.~(\ref{eq:nullxgh}) and uses
the power variable
\begin{equation}
      W =  e^{m w}
\end{equation}
and sets ${\hat B} = m$, then one obtains
\begin{eqnarray}\label{eq:HWV}
  H &=& - \case12  \dot W \, \dot V
\\ &&\quad
     +  m^{-1}[c_0{}^2 (\sum_{i = 0}^{p} A_i W^{a_i/m-1}) V +
             \sum_{i = 0}^{q} B_i W^{b_i/m-1}]\,.
\nonumber
\end{eqnarray}
In the case that all $a_i$ are nonzero,
the decoupled equation gives rise to the power form of
the generalized Friedmann equation.
An $a_i = 0$ term yields a logarithmic term in the
integrated expression of the decoupled second order differential equation.
As in the previous case, to integrate this final pair of equations, one can
reintroduce the slicing gauge freedom involving rescaling of
$\lambda$ derivatives
by functions of $W$ in order to simplify the problem.
Without resorting to this additional freedom, one can make the following
case by case analysis of some types of SE-Hamiltonians which often
arise.

\bigskip\paragraph{The case $A_i=0$ for all $i$ (the flat case).}

\subparagraph*{The gauge choice $m=0$.}
This leads to $\ddot w=0$,
so that $w$ is affinely related to the independent variable.
Solving the Hamiltonian
constraint for $\dot V$ leads to an easily integrated expression for $V$
(or $v$ if $c_0 = 0$)
which results in a linear combination of exponentials unless one of the
exponential coefficients $b_i$ vanishes, leading to a linear term.

\subparagraph*{The gauge choice $m\neq0$.}
This leads to
$\ddot W=0$ and $W$ is affinely related to $\lambda$.
Setting the additive constant to zero, again one can integrate the
expression for $\dot V$ to obtain $V$, this time resulting in a linear
combination of powers of the independent variable
\begin{equation}\label{eq:plsum}
  V = V_0 +\sum_i C_i \lambda^{b_i/m} \,,
\end{equation}
unless one of the coefficients $b_i$ vanishes, leading to a logarithmic
term. (Note that if $c_0 = 0$ all the above expressions for the $A_i = 0$
case remain valid with $V$ replaced with $v$.)

\bigskip\paragraph{The case $c_0 \neq 0, A_0 \neq 0, A_i = 0$ ($i \geq 1$).}

\subparagraph*{Subcase $a_0=0$.}
Here the choice
$m=0$ is appropriate, leading to $\ddot w = {\it const}$ and quadratic
solutions for $w$.

\subparagraph*{Subcase $a_0 \neq 0$.}
Here the three choices $m= a_0,
\pm\case12 a_0$ lead to elementary function solutions for $W$ as described
after equation (\ref{eq:Wgfe}).

\bigskip\paragraph{Two examples.}

\subparagraph*{One null exponential potential term in any dimension.}
When a problem is characterized by a single null potential term,
one first uses a Lorentz transformation from the original conformally
inertial coordinates to a new set $\bar\beta{}^0, \bar\beta,\bar\beta{}^S$
so that the Hamiltonian takes the form
\begin{equation}
  H = \case12 x [ - \dot{\bar \beta}{}^{0\,2} + \dot{\bar \beta}{}^2
                  + \sum_S \dot{\bar\beta}{}^{S\,2} ]
       + \case12 x^{-1} K e^{{\bar C}({\bar \beta{}^0} + {\bar \beta})}\,.
\end{equation}
Then one introduces the null variables
\begin{equation}
   {\bar w} = {\bar \beta{}^0} + {\bar \beta}\,,\quad
   {\bar v} = {\bar \beta{}^0} - {\bar \beta}\,.
\end{equation}
In these variables the Hamiltonian then takes the form
\begin{equation}
   H = - \case12 x {\dot {\bar v}} \, {\dot {\bar w}} +
          \case12 x^{-1} [ \Theta{}^2 + K e^{\bar C {\bar w}} ] \,,
\end{equation}
where $\Theta{}^2 = \sum_S {\bar p_S}{}^2$
arises from nonnull cyclic variables $\bar\beta{}^S$
and $S$ denotes the index
labeling the nonnull cyclic variables, assuming that $x$ is chosen to
be independent of them.
The variables ${\bar \beta{}^S}$ can be evaluated in terms of
quadratures arising from the equations
\begin{equation}
   {\dot {\bar \beta{}^S}} = x^{-1} {\bar p_S}\,.
\end{equation}

The full problem can therefore be decomposed into a set of quadratures
and a 2-dimensional problem coming from the above
reduced Hamiltonian for $w$ and $v$.
This is a flat null Killing tensor case, $A_i = 0$, which has been
discussed above.
One need only identify the constants
$c_0 = 0$, $b_0 = 0$, $b_1 = {\bar C}$,
$B_0 = \case12 \Theta{}^2$, $B_1 = \case12 K$ in
equation (\ref{eq:HwV}).

\subparagraph*{One exponential potential term in two dimensions.}
The one potential term case corresponds to a flat Jacobi geometry.
The single potential term can be identified in two ways with the
explicit terms in equation (\ref{eq:nutaub}).
Either $A_0 c_0 \neq 0$ or $B_0 \neq 0$.
These two different identifications lead to different natural
slicing gauges.

\medskip%\typeout{>>>> medskip here!******************}

If $A_0 = 0$, then $B_0 \neq 0$, then one can choose $m=0$ and a translation
in $\lambda$ leading to $w\sim \lambda$. Integration of the Hamiltonian
constraint yields an exponential expression for $V$ unless $b_0=0$ in
which case $V\sim \lambda + const$.
In this latter case the Taub slicing gauge potential is null,
i.e., it depends only on the null variable $v$.
If one chooses instead $m \neq 0$, then one can make a translation in
$\lambda$ such that $W\sim \lambda$ and
$V\sim \lambda^{b_0/m} + const$ if $b_0 \neq 0$
or $V\sim \ln \lambda + const$ if $b_0=0$.
In the first case the Taub potential is nonnull and the choice
$m=b_0$ makes $V$ affinely related to the independent variable.
When both variables are affinely related to $\lambda$, the choice of slicing
gauge makes the single potential term in the Hamiltonian a constant, i.e., the
gauge is the Jacobi slicing gauge.

\medskip%\typeout{>>>> medskip here!******************}

If $A_0 c_0 \neq 0$, then either $a_0=0$ or $a_0 \neq 0$.  If $a_0=0$, then
with $m=0$, the variable $w$ is quadratic in $\lambda$ and the expression
for `$\ln V$' is  a standard integral.  If $a_0 \neq 0$ then the three choices
$m=a_0, \pm \case12 a_0$ are relevant according to the
discussion following equation (\ref{eq:Wfirstintegral}).
The choice $m=a_0$ leads to a quadratic solution for $W$. Choosing the zero of
the independent variable
to eliminate the linear coefficient leads to power law solutions for
$V$, i.e., $V\sim \lambda^\sigma$, where $\sigma$ is determined by the
quadratic coefficient and $B$.
For the choice $m=\case12 a_0$, $W$ is a hyperbolic/trigonometric sine or
cosine and the expression for `$\ln V$' is a standard integral.
For the choice $m=-\case12 a_0$, $W^2$ is quadratic in $\lambda$ leading to
`$\ln V$' being a quartic expression.
An entirely parallel discussion holds with $w$ and $v$ interchanged
corresponding to those cases where $v$ decouples. This will lead to
additional slicing gauges.

\bigskip\paragraph{Linear decoupling.}

It is not necessary to choose null variables to solve problems admitting
null Killing tensors. The only requirement is that a variable and
its derivative occur linearly in the Hamiltonian.
As an example,
the Hamiltonian problem corresponding to the
potential of equation (\ref{eq:nutaub})  can be treated as follows.
Making a transformation
\begin{equation}
    v = (\ln g + w)/ c_0  \leftrightarrow  g = e^{c_0 v - w} \,,
\end{equation}
and choosing a new gauge function $ y = x/g$ leads to
\begin{eqnarray} \label{eq:Hwg}
      H &=& - y (\frac{1}{2c_0}) ( \dot w \dot g + g \dot w{}^2 )
\\ &&\quad
          + y^{-1} [c_0 (\sum_{i = 0}^{p} A_i e^{(a_i+1) w})g +
             \sum_{i = 0}^{q} B_i e^{b_i w}] e^w \,.
\nonumber
\end{eqnarray}
This Hamiltonian leads to the decoupled equation
\FL
\begin{equation}
   \ddot{w} + ( y^{-1} \frac{dy}{dw}   -1 ) \dot w{}^2
      - 2c_0{}^2 y^{-2} \sum_i A_i e^{(a_i + 2)w} = 0   \,.
\end{equation}
For the choice $y=D e^{dw}$ this leads to
\begin{equation}
   \ddot{w} + ( d-1 ) \dot w{}^2
      - \frac{2c_0{}^2}{D^2} \sum_i A_i e^{(a_i -2d + 2)w} = 0   \,,
\end{equation}
which is just the second order form for the generalized Friedmann
equation discussed in section \ref{sec:gfp}.

Note that this discussion starting from Eq.~(\ref{eq:Hwg})
easily generalizes to include all real values of
$g$.
The identification $y=-4 c_0 e^{2w} z$ shows that the Hamiltonian
(\ref{eq:tnutm}) of the Taub-NUT-M spacetime is of this form.

%\input dz5039tre

%%%%%%%%%%%%%%%%%%%%%%%%%%%%%%%%%%%%%%%%%%%%%%%%%%%%%%%%%%%%%%%%%%%

\subsubsection{Nonnull decoupling} \label{sec:nonnull}

A particular type of nonnull decoupling occurs
for SE-Hamiltonians in any dimension
when after a Lorentz transformation
to new conformally inertial coordinates $\bar\beta{}^A$ (where $\bar\beta{}^0$
denotes the timelike variable),
each exponential
potential term involves only a single new (nonnull) variable.
In other words the Taub potential takes the form
\begin{equation}\label{eq:nonnull}
    U_{\rm(taub)} = \sum_A \sum_i B_{iA} e^{\bar b{}^{iA} \bar\beta{}^A} \,.
\end{equation}
Then the total Hamiltonian is just the sum of independent Hamiltonians $H_A$
for each new variable, constrained only by the Hamiltonian constraint
on the sum of the individual energies
\begin{eqnarray}
      H &=& \sum_A H_A \,, \quad \sum_A E_A = 0\,,
\nonumber\\
      H_A &=& \case12\eta_{AB}({\dot {\bar \beta}}{}^B)^2 +
      \sum_i B_{iA} e^{\bar b{}^{iA} \bar\beta{}^A} = E_A\,,
\end{eqnarray}
Each equation $H_A=E_A$ is a generalized Friedmann equation.

The simplest example of this occurs for a single (nonnull) potential term.
The Hamiltonian expressed in terms of the new $\bar \beta{}^A$ variables
then has the form
\begin{equation}
   H = \case12 x \eta_{AB} \dot{\bar \beta{}^A} \dot{\bar \beta{}^B}
   +  \case12 x^{-1} K e^{{\bar C^P}{\bar \beta{}^P}} = 0 \quad
   ({\rm no\ sum\ over\ } P)\,.
\end{equation}
In the timelike case the index $P$ will assume the value 0, while in the
spacelike case it will assume one of the remaining allowed values.

Letting $x$ only depend on ${\bar \beta{}^P}$ results in the
equations of motion
$(x \dot{\bar \beta}{}^Q) \,{\dot {}} = 0$
for the remaining cyclic variables ${\bar \beta{}^Q}$, $Q\neq P$.
These lead to constant
values of the momenta ${\bar p_Q} = x \eta_{QR}{\dot {\bar \beta{}^R}}\,.$
Thus the cyclic variables are determined by the equations
\begin{equation}
   {\dot {\bar \beta{}^Q}} = x^{-1} \eta^{QR} {\bar p_R}\,.
\end{equation}
The noncyclic variable may be solved for using
the Hamiltonian constraint, which may
conveniently be rewritten in the standard form of a generalized Friedmann
problem with two exponential terms
\begin{equation}
   ({\dot {\bar \beta{}^P}})^2 =
   \cases{x^{-2}[\Sigma{}^2 + Ke^{{\bar C^P}{\bar \beta{}^P}}] \,,
         &$P = 0 \leftrightarrow$ timelike case\,,\cr
          x^{-2}[\Gamma - Ke^{{\bar C^P}{\bar \beta{}^P}}] \,,
         &$P\neq0 \leftrightarrow$ spacelike case\,,\cr}
\end{equation}
where $\Sigma{}^2 = \sum_Q {\bar p_Q}{}^2$ and
$\Gamma = - \eta^{QR}{\bar p_Q}{\bar p_R}$.
This problem
is easily solved in various choices of slicing gauge and power variables
as discussed in section \ref{sec:gfp}.

There are 2-dimensional SE-Hamiltonians corresponding to
nonnull Killing tensor cases
which require nontrivial conformal transformations as well.
If one starts with the Hamiltonian in Eq.~(\ref{eq:nonnully}) and assumes
that the conformal factor is a polynomial in $T$ and $X$ of the following
form
\begin{equation}\label{eq:TX}
   \mathop{\rm sgn}\nolimits %\sgn
   (U_{\rm(taub)}) \, {\tilde G} = \sum_{i= 0}^n (A_i T^i + B_i X^i) \,,
\end{equation}
which corresponds to a simply solved separable problem,
then the variable transformation
\begin{equation}
  T = e^{aw} + e^{bv} \, ,\quad X = e^{aw} - e^{bv} \,,
\end{equation}
where $a$ and $b$ are arbitrary nonzero constants, and the relation
(following from the definition (\ref{eq:y}))
\begin{equation}
   y = (4abe^{aw + bv})^{-1} x
\end{equation}
leads to the Hamiltonian
\begin{eqnarray}
  H &=& - \case12 x \, \dot w \, \dot v
      + x^{-1}(4ab)\sum_{i = 0}^n \sum_{k = 0}^i(A_i + (-1)^k B_i)
                          {i\choose k}
\nonumber\\ &&\quad
   \times%\typeout{continued product on next line: is this correct?}
        e^{(i - k + 1)aw + (k + 1)bv}  \,.
\end{eqnarray}
Thus the identification
\begin{eqnarray}
  && U_{\rm(taub)} =
\\ &&
      (4ab)\sum_{i = 0}^n \sum_{k = 0}^i(A_i + (-1)^k B_i){i\choose k}
           e^{(i - k + 1)aw + (k + 1)bv} \,,
\nonumber
\end{eqnarray}
can be made. There are several solvable cases appearing in the literature
corresponding to polynomials of low degree in Eq.~(\ref{eq:TX}).

\subsubsection{Nonnull-null decoupling} \label{sec:nnn}

For those cases whose Jacobi geometry allows both a null Killing tensor and
a nonnull Killing tensor, one may take yet another approach in solving
the field equations.
For example, if the Taub potential of a 2-dimensional problem
is both of the form
(\ref{eq:nutaub}), corresponding to $w$ decoupling,
and (\ref{eq:nonnull}),
then the Hamiltonian must take the form
\begin{eqnarray}
   H &=& - \case12 x \dot w \dot v + x^{-1} U_{\rm(taub)}  \,,
\nonumber\\
   U_{\rm(taub)} &=& D_1 e^{2(bw+cv)} + D_2 e^{bw+cv}
\nonumber\\ &&\quad
               + D_3 e^{2(-bw+cv)}  + D_4 e^{-bw+cv} \,.
\end{eqnarray} %
A similar expession holds for $v$ decoupling with $w$ and $v$ interchanged.

One may then boost $(w,v)=(k\bar w, k^{-1} \bar v)$ to
$(\bar w, \bar v) = (\bar\alpha + \bar\beta,\bar\alpha - \bar\beta)$,
where $k>0$ is determined so that each exponential term in the potential
depends on only one of the new conformally inertial coordinates
$\bar\alpha$ or $\bar\beta$, namely by the condition
\begin{equation}
          \zeta \equiv 2b k = \pm 2 c k^{-1}\,,
     \quad \rightarrow \quad        \  k= |c/b|^{1/2} \,.
\end{equation}
Then if $ \mathop{\rm sgn}\nolimits %\sgn
bc =1$, the potential becomes
\begin{equation}
        U_{\rm(taub)} = D_1 e^{2\zeta\bar\alpha} + D_2 e^{\zeta\bar\alpha}
                 + D_3 e^{-2\zeta\bar\beta} +  D_4 e^{-\zeta\bar\beta} \,,
\end{equation}
while if $ \mathop{\rm sgn}\nolimits %\sgn
bc = -1$, $\bar\alpha$ and $\bar\beta$ are interchanged
in the potential.

Assuming $ \mathop{\rm sgn}\nolimits  %\sgn
bc =1$, then two power law
slicing gauge choices lead to mutual decoupling of a pair of variables,
the null variable $\bar w$ and one of the new nonnull inertial
coordinates. These correspond to making the $D_2$ and $D_4$ potential
terms constant respectively. These choices are
\begin{eqnarray}
  {\rm case\ } (a):\quad  x &=& e^{\zeta\alpha} \ ;\
                         \hbox{$\bar w$ and $\bar\alpha$ decouple,}
\nonumber\\
  {\rm case\ } (b):\quad  x &=&  e^{-\zeta\beta} \ ;\
                         \hbox{$\bar w$ and $\bar\beta$ decouple,}
\end{eqnarray} %
with the respective Hamiltonians
\begin{eqnarray}
  {\rm case\ } (a):\quad
     H &=& - \case12 e^{ \zeta\bar\alpha} \dot{\bar w}
                     ( 2 \dot{\bar\alpha} - \dot{\bar w} )
                 + D_1 e^{\zeta\bar\alpha} + D_2
\nonumber\\ &&\quad
             + D_3 e^{\zeta(\bar\alpha - 2\bar w)} + D_4 e^{-\zeta\bar w}\,,
\nonumber\\
  {\rm case\ } (b):\quad
     H &=& - \case12 e^{-\zeta\bar\beta} \dot{\bar w}
                   ( \dot{\bar w} - 2 \dot{\bar\beta} )
                 + D_1 e^{\zeta(2 \bar w - \bar\beta} + D_2 e^{\zeta\bar w}
\nonumber\\ &&\quad
                 + D_3 e^{-\zeta\bar\beta} + D_4 \,.
\end{eqnarray}
Similar results hold with $\bar\alpha$ and $\bar\beta$ interchanged
for $ \mathop{\rm sgn}\nolimits %\sgn
bc = -1$.

The decoupled equations for case $(a)$ are
\begin{eqnarray}
      0 &=& \delta L /\delta \bar\alpha
        = e^{\zeta\bar\alpha} [ - \ddot{\bar w} - \case12\zeta \dot w{}^2
               + \zeta D_1 + \zeta D_3 e^{-2\zeta\bar w} ] \,,
\nonumber\\
      0 &=& \delta L /\delta \bar w +  \delta L /\delta \bar\alpha
              + \zeta H
\nonumber\\
    &=& e^{\zeta\bar\alpha} [ - \ddot{\bar \alpha} -\zeta\dot{\bar\alpha}{}^2
               + 2\zeta D_1 + \zeta D_2 e^{-\zeta \bar\alpha} ] \,,
\end{eqnarray} %
while those for case $(b)$ are
\begin{eqnarray}
      0 &=& \delta L /\delta \bar\beta
        = - e^{-\zeta\bar\beta} [ - \ddot{\bar w} + \case12\zeta \dot w{}^2
               + \zeta D_1 + \zeta D_3 e^{2\zeta\bar w} ] \,,
\nonumber\\
      0 &=& \delta L /\delta \bar w +  \delta L /\delta \bar\beta
              - \zeta H
\nonumber\\
 &=& - e^{-\zeta\bar\alpha} [ - \ddot{\bar\beta} + \zeta \dot{\bar\beta}{}^2
               + 2\zeta D_3 + \zeta D_4 e^{\zeta \bar\beta} ] \,.
\end{eqnarray} %

Each of these four decoupled equations is equivalent to the second order form
(\ref{eq:diffgfe})
of a generalized Friedmann equation (\ref{eq:diffgfeint}).
They have the following respective values of the
potential term $f(e^\theta)$ and the constant $\delta$ appearing in the
latter equations
\begin{equation}
    f(e^\theta) =
    \cases{
           - \zeta D_1 - \zeta D_3 e^{-2\zeta \bar w} \,,
                & $\delta = \case12 \zeta\,,\ \theta= \bar w$ \,,
  \strut\cr
           2 \zeta D_1 + \zeta D_2 e^{-\zeta \bar \alpha}  \,,
                & $\delta = \zeta\,,\  \theta= \bar\alpha$ \,,
  \strut\cr
           - \zeta D_1 - \zeta D_3 e^{2\zeta \bar w} \,,
                & $\delta = -\case12 \zeta\,,\ \theta= \bar w$ \,,
  \strut\cr
           -2 \zeta D_3 - \zeta D_4 e^{\zeta \bar\beta} \,,
                & $\delta = - \zeta\,,\ \theta=\bar\beta $\,.
  \strut\cr
              }
\end{equation}
The equivalent generalized Friedmann equations are then
respectively
\begin{equation}
    \dot\theta{}^2 = h(e^\theta) + {\cal E} e^{-2\delta\theta}
   = \cases{
        - D_1 + D_3 e^{-2\zeta \bar w} + {\cal E} e^{-\zeta \bar w} \,,
                \strut & \cr
       D_1 + D_2 e^{-\zeta \bar \alpha} + {\cal E} e^{-2\zeta \bar\alpha}\,,
                \strut & \cr
          D_1 - D_3 e^{2\zeta \bar w} + {\cal E} e^{\zeta \bar w} \,,
                \strut & \cr
       D_3 + D_4 e^{\zeta \bar\beta} + {\cal E} e^{2\zeta \bar\alpha} \,.
                \strut & \cr
              }
\end{equation}
Finally the following power variables convert these equations into
quadratic potential problems
\begin{equation}
    \zeta^{-2} \dot u{}^2 = \cases{
        - D_1 u^2 + D_3 + {\cal E} u \,,
                \strut & $u=e^{\zeta \bar w}$ \,,\cr
          D_1 u^2 + D_2 u + {\cal E} \,,
                \strut & $ u=e^{\zeta \bar\alpha}$ \,,\cr
          D_1 u^2 - D_3 + {\cal E} u \,,
                \strut & $ u=e^{-\zeta\bar w}$ \,,\cr
          D_3 u^2 + D_4 u + {\cal E} \,.
                \strut & $ u=e^{-\zeta\bar\beta}$ \,.\cr
               }
\end{equation}
The solutions of these equations describe 1-dimensional motion
in a quadratic potential and lead to solutions for the
dependent variable $u$ which are
affinely related to exponential or trigonometric or hyperbolic sines
and cosines of an argument affinely related to the independent variable.
Of course this method can also be used to treat the single potential term
case, but the previous two methods are simpler.

\subsubsection{Power variables, power law slicing gauges and SE-Hamiltonians}

If one looks at the literature on exact solutions one almost always
find the solution expressed in power variables and power law slicing gauges.
Why is this the case? The answer is that practically all solvable
problems are described by SE-Hamiltonians with a relatively few number
of potential terms. As seen above, the simplest symmetry-adapted variables
and slicing gauges are usually power variables and power slicing gauges
directly related to these exponential terms.
On the other hand, if one has many exponential
terms one might be forced to use non-power law variables.
For example, this happens when the function $V$,
in the null Killing tensor case $(i)$, is a sum of
exponential terms.

\subsection{The intrinsic approach to null Killing tensor problems}
\label{sec:inull}

Apart from the general case of the Jacobi slicing gauge,
the only specific slicing gauges which have been considered here
are the power
law gauges. This subsection will show how another important class of
slicing gauges arises in a natural way for the null Killing tensor cases.
These gauges are the so-called ``intrinsic" slicing gauges.

Suppose one chooses symmetry-adapted variables and a
symmetry compatible slicing gauge in
the null Killing tensor case
so that one obtains the Hamiltonian (\ref{eq:HWV})
\begin{equation}\label{eq:ham}
   H = -\case12 {\dot W}{\dot V} + A(W)V + B(W) = 0\,.
\end{equation}
Then one obtains the first order decoupled equation for $W$
\begin{eqnarray}\label{eq:dec}
   {\dot W}^2 &=& 2 [E - {\cal U}(W)] = F(W)\,,
\nonumber\\ &&\quad {\rm where}
         \quad {\cal U}(W) = -2\int A(W) dW \,.
\end{eqnarray}
For most functions $A(W)$,
this equation does not admit solutions expressible in terms of
elementary functions.
For example, consider a function $A(W)$
which consists of more than one term without being linear in $W$ (if it is
linear then it is
integrable in terms of elementary functions as already discussed).
For some cases of this type it is possible to find the solution
in terms of elementary functions by use of an intrinsic slicing gauge.
As discussed in section \ref{sec:gfpi},
such a slicing gauge is characterized by choosing
some simple function
of the metric components as the independent variable.
In the present problem, one can
reintroduce the gauge freedom in equations (\ref{eq:ham}) and  (\ref{eq:dec})
by introducing
a new independent variable $\bar\lambda$  such that
$N = N_{\rm(taub)} x^{-1} z(W)^{-1}$.
This choice leads to the decoupled equation
\begin{equation}
   {\dot W}^2 = z(W)^{-2}F(W) \,,
\end{equation}
where the dot refers to the new independent
variable $\bar\lambda$.
Choosing $z = F^{1/2}$ leads to  $W=\bar\lambda$
as the independent variable (setting
the constant of integration to
zero),
so that the slicing gauge is clearly an intrinsic one.
Inserting
$W = \bar\lambda$ into the Hamiltonian constraint and expressing this
in the new slicing gauge yields
\begin{equation}
   {dV\over d\bar\lambda}
       =  2 F(\bar\lambda)^{-1}[ A(\bar\lambda) V + B(\bar\lambda)]\,,
\end{equation}
which is easily solved. However, whether or not
the solution can be expressed in terms of elementary
functions depends on the explicit expressions for $A, B$ and $F$.

To be more explicit, consider the interior Schwarzschild case
where the usual Schwarzschild radial coordinate is related to the metric
scale factors by the intrinsic slicing condition
$r=R_1=R_2=e^w$.
Using the power variables $W = e^w$, $V = \case23 e^{\frac23 v}$ and the
gauge choice $x = e^{w + \frac32 v}$ leads to
\FL
\begin{eqnarray}
 &&  H = -\case12 {\dot W}{\dot V} + 24[(1 - \kappa\rho_{(0)} W^2)V
      + \kappa(\rho_{(0)} + p_{(0)})W^\frac52] \,,
\nonumber\\
  && {\dot W}^2 = 4[W - \frac13 \kappa\rho_{(0)} W^3 + const] = F\,.
\end{eqnarray} %
For the particular case of nonsingular solutions,
smoothness conditions at the center (where $W= r = 0$)
(see e.g. \cite{stephani}) require that
the integration
constant occurring in the expression for the decoupled variable must be
zero. Reintroducing the gauge freedom and choosing
$z = F^{1/2} = [96(W - \frac13 \kappa\rho_{(0)} W^3)]^{1/2}$ leads to the
usual Schwarzschild gauge $r= W = \bar\lambda$
\begin{eqnarray}
  {dV \over dr}
     &=& \case12[(1 - \kappa\rho_{(0)} r^2)V
\nonumber\\ &&\quad
     + \kappa(\rho_{(0)}
                  + p_{(0)})r^\frac52]/[r - \frac13 \kappa\rho_{(0)} r^3]\,.
\end{eqnarray}
This equation is easily integrated and going back to the original metric
variables one finds the simple standard
expression for the interior Schwarzschild solution \cite{stephani}.
The general case with a nonzero constant has more complicated solutions.

%%%%%%%%%%%%%%%%%%%%%%%%%%%%%%%%%%%%%%%%%%%%%%%%%%%%%%%%%%%%%%%%%%%%%%%

\subsection{Killing tensor symmetries for a subclass of 2-dimensional models}
\label{sec:HKT}

By reducing a given problem, either by exploiting symmetries or by
specializing to a subcase, one often ends up with a reduced system
having only a few degrees of freedom. Apart from the trivial case when
there is only a single degree of freedom left, the simplest reduced
systems have two degrees of freedom. Many problems can be
described by a reduced 2-dimensional Hamiltonian of the special form
\begin{equation}
     H = \case12 (-\dot\alpha{}^2 + \dot\beta{}^2) + U
       = - \case12 \dot w \dot v + U \,,
\end{equation}
where $w,v$ are the standard null variables of Eq.~(\ref{eq:nullvars}),
and
\begin{equation}\label{eq:hom_pot}
     U  = e^{2c\alpha} F(\beta) \,, \quad {\rm or} \quad
         U = e^{2c\beta} F(\alpha) \,,
\end{equation}
where $c$ is a constant. Although $c$ (if nonzero)
can always be normalized to unity
by a suitable rescaling of $\alpha$ and $\beta$ we choose not to do so
here in order to facilitate comparison with the table
below. However, the translational freedom in $\beta$ will be used to
simplify formulas.

When $c\neq0$ the potential form
(\ref{eq:hom_pot}) corresponds exactly to the case when the associated
Jacobi metric
\begin{equation}\label{eq:homjac}
                        J_{AB} = 2| U | \eta_{AB} \,,
\end{equation}
admits a homothetic symmetry generated by $\xi = \partial/ \partial
\alpha$ (timelike homothetic Killing vector (HKV) case) or $\xi =
\partial/ \partial \beta$ (spacelike HKV case) respectively
\begin{equation}
    \hbox{\it\char'44}_{\hbox{$\xi$}}  J_{AB} = 2c J_{AB} \,.  % \pounds
\end{equation}
In the
case $c=0$ the potential depends only on a single variable
and $\xi$ reduces to
a Killing vector symmetry. It follows from the form of $\xi$ that the
above variables are adapted to this symmetry. The problem of
classifying the function $F(\beta)$ [or $F(\alpha)$] for which the
Jacobi metric (\ref{eq:homjac}) admits Killing tensor symmetries has
been analyzed in \cite{ru:kt}. As explained in that reference it is
sufficient to consider the timelike HKV case. The spacelike HKV case
can then easily be obtained by an appropriate transformation.

In this subsection all potentials will be given for 2-dimensional
models which admit a second rank Killing tensor of a given weight
under the homothetic symmetry, subject to the
assumption that the Killing tensor $K_{AB}$ is characterized by a
homothetic weight $2b$ through the equation
$\hbox{\it\char'44}_{\hbox{$\xi$}} K_{AB} = 2bc K_{AB}$.     % \pounds
This includes some cases which were not stated explicitly
in \cite{ru:kt}.

The classification of potentials admitting such Killing
tensors depends on two parameters describing
properties of the Killing tensor. The first
parameter is the sign of the determinant of the conformal part of
the Killing tensor, $\Sigma = \mathop{\rm sgn}\nolimits %\sgn
\det(P_{AB})$, where $P_{AB} =
K_{AB}-\frac12 K J_{AB}$ and $K=K^A{}_A$.
The Killing tensor type is related to $\Sigma$ according to
\begin{equation}
    \Sigma = \cases{
             0 \,, \strut & (null) \cr
             1 \,, \strut & (nonnull H-J) \cr
            -1 \,, \strut & (nonnull harmonic) \,,\cr}
\end{equation}
corresponding respectively to the three cases
$(i)$, $(ii)$, and $(iii)$ of Eq.~(\ref{eq:threecases}).
The second parameter is the homothetic weight factor $b$.
The cases $b=1$ or $b=0$ require special treatment compared to $b\neq 0,1$.
With the three values of $\Sigma$, this leads
to nine different cases altogether.

We now enumerate the potentials of the form (\ref{eq:hom_pot})
admitting Killing tensors corresponding to these cases.
The three cases corresponding to a null Killing
tensor are collectively given by
\begin{eqnarray}\label{eq:NHKV_pot}
&& (A)\quad (b \neq 1; \Sigma=0) :
\nonumber\\ &&\quad
   U  = \hbox{[two exponential term case I in Table 3.]} \,,
\nonumber\\
&& (B) \quad(b=1; \Sigma=0) :
   U =    [C_1 c(w-v) + C_2] e^{2cw}  \,,
\end{eqnarray}
where case A also includes the case $b = 0 = \Sigma$.
When $b=1$, the spacelike HKV case is obtained by interchanging
$w$ and $v$ in the corresponding expression in (\ref{eq:NHKV_pot}).

For nonnull Killing tensors, the function $F(\beta)$ is given by one of the
following expressions (modulo a translation of $\beta$)
\begin{eqnarray}\label{eq:NNHKV_pot}
&&
(C)\quad (b\neq0,1;\ \Sigma=1) :
\nonumber\\ && \qquad
     C_1 \cosh^s [2c\beta/(s+2)]
                     + C_2 \sinh^s [2c\beta/(s+2)] \,,
\nonumber\\ &&
(D)\quad (b\neq0,1;\ \Sigma=-1) :
\nonumber\\ && \qquad
           {\Re e}\left\{D[e^{2c\beta/(s+2)}
                               + ie^{-2c\beta(s+2)}]^s\right\} \,,
\nonumber\\ &&
(E)\quad (b=1;\ \Sigma=1) :
\nonumber\\ && \qquad
         \hbox{[two exponential term case III in Table 3]} \,,
\nonumber\\ &&
(F)\quad (b=1;\ \Sigma=-1) :
\nonumber\\ && \qquad
         e^{2c k  \beta}     \{ C_1  \cos[2c(1- k ^2)^{1/2} \beta]
\nonumber\\ && \qquad\qquad\qquad
           + C_2 \sin[2c(1- k ^2)^{1/2} \beta]   \} \,,
\nonumber\\ &&
(G) \quad (b=0; \Sigma=1) :
\nonumber\\ && \qquad
         C_1 \ln\coth (c\beta) + C_2  \,,
\nonumber\\ &&
(H)\quad (b=0; \Sigma=-1) :
\nonumber\\ && \qquad
         C_1 \arctan e^{2c\beta} + C_2  \,,
\end{eqnarray}
where $s = -2b/(b-1)$ and $ k $ ($| k |<1$) are real parameters while $D$
is a complex parameter.
In cases $(C)$ and $(D)$ the potential can be expressed explicitly as a
sum of exponential terms if $s$ is an integer. The subcases with
exactly two exponential terms are given explicitly in section \ref{sec:KT2}.
Case $(D)$ can also be expressed explicitly as a real
function in the form
\begin{eqnarray}
&&  F = \cosh^{s/2} [4c\beta/(s+2)]
             \{ C_1 \cos[s \arctan e^{4c\beta/(s+2)}]
\nonumber\\ &&\quad
             + C_2 \sin[s \arctan e^{4c\beta/(s+2)}]  \} \,.
\end{eqnarray}
Using multiple angle formulas the trigonometric expression inside the
curly brackets can be converted to algebraic form provided that $s$ is
a rational number $m/n$. However, since this involves solving a
polynomial equation of degree $|n|$, explicit algebraic expressions can
only be guaranteed for $|n| \leq 4$.
Some of the above Killing tensor cases admit special Killing vector cases,
e.g., setting $C_1=0$ or $C_2=0$ in expression $(C)$ leads to such a case.

%%%%%%%%%%%%%%%%%%%%%%%%%%%%%%%%%%%%%%%%%%%%%%%%%%%%%%%%%%%%%%%%%%%%%%%%%%%%
\subsection{Killing tensor cases for Taub potentials
with two exponential terms} \label{sec:KT2}

An important special case of 2-dimensional systems occurs when the
potential is a sum of two exponential terms
\begin{eqnarray}\label{eq:KT2null}
       U &=& C_1 e^{p_1 w + q_1 v} + C_2 e^{p_2 w + q_2 v}
\nonumber\\
       &=& C_1 e^{c_1\alpha + d_1\beta} + C_2 e^{c_2\alpha + d_2\beta} \,.
\end{eqnarray}
Extracting all the two exponential term potential cases from the various
types of the previous section and adding the flat case with a null HKV
and the nonnull Killing vector case corresponding to $c = 0$ in
Eq.~(\ref{eq:hom_pot})
leads to Table 3 for the corresponding parameter values.
In the case (V) of this table, the type of nonnull Killing tensor
depends on the relative sign $Z = \mathop{\rm sgn}\nolimits %\sgn
(C_1C_2)$ of the two terms
in the potential. For $Z=1$ one has a nonnull H-J Killing tensor
case ($\Sigma = 1$) while for
$Z=-1$ one has a nonnull harmonic Killing tensor case ($\Sigma = -1$).

%\typeout{**********************************Table 3:}

The null (I) and flat (II) cases are easily treated using results from the
null decoupling section \ref{sec:null}. (Case I admits additional nonnull
Killing tensor cases for certain parameter values, for such cases one may
choose nonnull solution techniques.) Referring to that section, the flat
case corresponds to $A_i = 0$, while the null cases correspond to $A_i \neq
0$. In case (III), decoupling can be achieved by an appropriate Lorentz
transformation leading to two Friedmann equations.
The case (IV) is a nonnull Killing vector
case where decoupling can also be accomplished
by a Lorentz transformation leading
directly to a single generalized Friedmann equation.
In the remaining
nonnull cases (V), one can introduce power variables leading to an easily
solved problem with a potential which is a quadratic form in the new
variables. An example of such a case has been dealt with in
\cite{exact}.

%%%%%%%%%%%%%%%%%%%%%%%%%%%%%%%%%%%%%%%%%%%%%%%%%%%%%%%%%%%%%%%%%%%%%%%%%%%%%%
% section 5
%%%%%%%%%%%%%%%%%%%%%%%%%%%%%%%%%%%%%%%%%%%%%%%%%%%%%%%%%%%%%%%%%%%%%%%%%%%%%%

\section{Invariant Submanifolds and How to Obtain Them} \label{sec:inv}

Solving the Einstein field equations in general seems to be impossible,
particularly in view of recent results that the only
generalized local symmetries of these equations
are due to scale invariance and the diffeomorphism group \cite{torgen},
and these symmetries are insufficient to lead to a general solution.
To find special solutions one imposes space-time symmetries and/or
other restrictions on the dependent variables so that one obtains
a more tractable consistent subsystem of differential equations.
In other words one tries to find
``invariant submanifolds" of the original system of field equations.
Even imposing enough space-time symmetries to
reduce the field equations to ordinary differential
equations as one does to obtain the HH models
still does not lead to such tractable subsystems in general.
One must impose further conditions to be able to actually find
exact solutions.
There is no general systematic method of discovering invariant submanifolds.
It is here that creativity and imagination and even plain luck
play a role in rooting out these hidden structures.
There are many particular ways in which invariant submanifolds
have been found, but few of these successes involve a systematic method.
Many methods require an arbitrary function like an unspecified equation
of state or an unspecified scalar field potential to produce solutions.
In this brief section one systematic
method will be presented which does not rely on the existence
of arbitrary functions and is
relevant to many though not all of the
known invariant submanifolds.
In particular for Hamiltonian problems
this method also yields the class of
exact power law (EPL) solutions. EPL solutions have
been studied in \cite{wai,epl}.

Hamiltonians which are reducible to the following form
play a crucial role in the discussion of HH models
\begin{equation}
    H = \case12 \chi \eta_{\mu\nu} \dot y{}^\mu \dot y{}^\nu
          + \chi^{-1} {\cal U}\,,
\end{equation}
where $\chi$ and ${\cal U}$ are analogous to the previous slicing gauge
function
$x$ and the Taub potential.
Choosing $\chi=1$ leads to the equations
\begin{equation}
      \ddot y{}^\mu = - \eta^{\mu\nu} \partial {\cal U} / \partial y^\nu \,.
\end{equation}

If $\partial {\cal U}/ \partial y^\mu =0$
holds for some value $y_{(0)}^\mu$ of a particular coordinate
$y^\mu$ independent of the values of the remaining
coordinates,
and if this condition is compatible with the Hamiltonian constraint,
then $y^\mu = y_{(0)}^\mu$ describes an invariant submanifold.
The equations for the remaining variables
on this submanifold are given by the above Hamiltonian
after having inserted the conditions
$y^\mu = y_{(0)}^\mu$ and $\dot y{}^\mu = 0$ (for that coordinate alone).
Invariant submanifolds within invariant submanifolds are also possible.
Reduction down to one dimension automatically leads to a solution
since the Hamiltonian constraint only involves a single variable,
thus leading to a quadrature.

Lorentz transformations of the $\beta^A$ variables
are symmetry transformations of the
Minkowski metric appearing in the expression for the kinetic energy function.
They play a crucial role in finding many
invariant submanifolds.
All HH models of the previous section
have Hamiltonians or reduced Hamiltonians with
Taub potentials which can be put into the following form
by a Lorentz transformation from the variables $\beta^A$ to new ones
$\bar\beta{}^A$
\begin{equation} \label{eq:inv}
     U_{\rm(taub)} = \sum_i e^{c_i\bar\beta{}^0} F_i(\bar\beta{}^P)\,,
          \quad    P\neq 0 \,,
\end{equation}
If $\partial F_i/ \partial \bar\beta{}^P =0$ holds for all values of $i$
for some particular value $\bar\beta{}^P_{(0)}$ of a particular coordinate
$\bar\beta{}^P$ independent of the values of the remaining
coordinates,
and if this condition is compatible with the Hamiltonian constraint,
then $\bar\beta{}^P= \bar\beta{}^P_{(0)}$
describes an invariant submanifold.
In many HH cases the index value $0$ and some definite value $P\neq0$
can be interchanged in this discussion,
but the Hamiltonian constraint
seems to prevent the existence of invariant submanifolds of this type.
1-dimensional invariant submanifolds with one exponential term lead
directly to EPL solutions.

%%%%%%%%%%%%%%%%%%%%%%%%%%%%%%%%%%%%%%%%%%%%%%%%%%%%%%%%%%%%%%%%%%%%%%%%%%%%%%
% section 6
%%%%%%%%%%%%%%%%%%%%%%%%%%%%%%%%%%%%%%%%%%%%%%%%%%%%%%%%%%%%%%%%%%%%%%%%%%%%%%

\section{Problems Leading to Exact Solutions and How to Solve Them}

This section will survey the cases which lead to exact solutions.  The method
of solution which works in each case will be specified by referring to
previous sections, without going through the mechanical details of obtaining
and presenting the solution explicitly. In fact, as has been shown, there are
often several ways one can solve a given problem and hence more than one
representation of the solution exists.  Specific examples of how to use the
methods of this article to produce the actual spacetime metrics which
correspond to these solutions
are given by Uggla \cite{exact} and Uggla and Rosquist \cite{exactii}.

Except for a few special class B cases and Bianchi type VI$_0$,
all exact solutions arise from spacetimes which admit either additional
continuous spacetime symmetries (Killing vectors and/or homothetic Killing
vectors) or additional continuous intrinsic symmetries (Killing vectors).
The latter are symmetries of the intrinsic geometry of the individual
homogeneous hypersurfaces which are not necessarily
spacetime symmetries.
As in section 2, the diagonal and nondiagonal models are treated
separately, but the diagonal models are collected according
to the dimension of the intrinsic symmetry group.
Unless otherwise stated, the only perfect fluid solutions being considered
here are those for which $p = (\gamma-1)\rho$.

%%%%%%%%%%%%%%%%%%%%%%%%%%%%
\subsection{Diagonal models}

The possible dimensions
of the intrinsic symmetry group of the geometry of the HH hypersurfaces
are 6, 4, and 3.
Beginning with dimension 6, models are considered with
only $\beta^0$-dependent sources and possible scalar fields.
This class of models includes the
Bianchi type I and V models
and the SH constant spatial curvature type IX models
(there are no static models of this latter type).
The SH models belonging to this class are intrinsically isotropic.

Next diagonal models with a 4-dimensional intrinsic symmetry group
and with only $\beta^0,\beta^+$-dependent sources are treated.
These models are all intrinsically LRS and
include the Bianchi type I, II, and V models,
the LRS Bianchi type III, VIII, and IX models,
the SH Kantowski-Sachs models,
and the static spherically symmetric models.
Note that apart from the SH Bianchi type IX FRW perfect fluid
solutions and the SH LRS Bianchi type VIII and IX stiff
perfect fluid solutions,
there are no other known exact perfect fluid solutions
for these two Bianchi types.

Finally the diagonal SH Bianchi type VI vacuum and perfect fluid
models are considered.

\subsubsection{Models with a 6-dimensional intrinsic symmetry group}
\label{sec:6d}

\paragraph{Sources not including a scalar field.}

The Hamiltonian can be written as \cite{hhscalar}
\begin{eqnarray}
      H &=& \case12 x ( -\dot{\beta}{}^{0\,2} + \dot{\beta}{}^{+\,2}
                                 + \dot{\beta}{}^{-\,2} )
\nonumber\\ &&\quad
                 + x^{-1} [ -72 k e^{4\beta^0}
                                + U_{\rm(taubs)}(\beta^0)  ]\,,
\end{eqnarray}
where $U_{\rm(taubs)}$ is the source potential expressed in the Taub
slicing gauge. The variables $\beta^\pm$ are equal to zero for the
type IX models while $\beta^+$ is equal to zero for the type V models.
One can choose 1-forms so that the parameter $k$ has the values 0 for
Bianchi type I, 1 for type IX
(correponding to the choice $n^{(1)} = n^{(2)} = n^{(3)} = 2$),
and $-1$ for type V (corresponding to the choice $a=1$).

Letting $x$ only depend on $\beta^0$ results in the
equations of motion
$(x \dot{\beta}{}^\pm) \,{\dot {}} = 0$,
which lead to constant
values of the momenta $ p_\pm = x \dot{\beta}{}^\pm$.
Thus $\beta^\pm$ are determined by the equations
\begin{equation}
   \dot{\beta}{}^\pm = x^{-1} p_\pm \,.
\end{equation}
One may solve for $\beta^0$ using
the Hamiltonian constraint
\begin{equation}
  ( \dot{\beta}{}^0 )^2 =
      x^{-2}[\Sigma{}^2 - 72 k e^{4\beta^0} + U_{\rm(taubs)} ]   \,,
\end{equation}
where $\Sigma{}^2 = p_+{}^2 + p_-{}^2$.
This equation immediately gives a quadrature for $\beta^0$.
The most interesting case is when $U_{\rm(taubs)}$ is a sum
of exponential terms and this problem reduces to the
the generalized Friedmann problem.
References to some of
the literature on the most notable SH solutions with a 6-dimensional
intrinsic symmetry group are given in Table 4.
The SH vacuum type I solution is usually associated with Kasner who
found the corresponding static solution \cite{kas}.
The isotropic vacuum type V solution is just the Milne
form of Minkowski spacetime. Useful references for FRW and FRW-$\Lambda$
models
are Harrison \cite{harr}, Vajk \cite{vajk}, Anderson \cite{and} and
Misner, Thorne and Wheeler \cite{mtw}. The book by
Kramer et al \cite{kraetal} is also useful in this context as well as
for further references on models in Table 4 with symmetry groups of
dimension 3 and 4.

%\typeout{**********************************Table 4:}

\bigskip\paragraph{Sources including a scalar field.}

The Hamiltonian is
\begin{eqnarray}
      H &=& \case12 x ( -\dot{\beta}{}^{0\,2} + \dot{\beta}{}^{+\,2}
                                 + \dot{\beta}{}^{-\,2}
                                 + \dot{\beta}{}^{\dagger\,2} )
\nonumber\\ &&\quad
              - 24 x^{-1} [ 3 k e^{4\beta^0}
              + \epsilon \kappa e^{6\beta^0} V_{\rm(sc)}(\beta^\dagger) ] \,.
\end{eqnarray}

\subparagraph*{Solvable cases.}
For most scalar potentials this is not a solvable problem.
However, if $V_{\rm(sc)}$ is the sum of exponential terms, then
some solutions do exist.
An interesting example is the case of
scalar field models with a single exponential
potential $V_{\rm(sc)} = e^{-2c \beta^\dagger}$ \cite{hall}.
For such models with $\beta^\pm=0$, which includes
the isotropic models, the Taub potential is given by
\begin{equation}
    U_{\rm(taub)} = -24  [ 3 k e^{2w +2v}
                    + \epsilon \kappa e^{(3-c) w + (3+c)v } ] \,,
\end{equation}
where $w= \beta^0 + \beta^\dagger$
and $v= \beta^0 - \beta^\dagger$.
If $k=0$ this is a simply solvable one-exponential-term problem.
When $k\neq0$ there are two solvable cases. The first
case $c=1$ corresponds to the $A_i=0$ flat null decoupling case
of section \ref{sec:null}. The second case
$c=2$ corresponds to the nonflat null Killing tensor case \cite{ru:kt}.
The Jacobi metric of the case $k=0$ and $\beta^\pm = 0$ with an arbitrary
scalar potential, $V_{\rm(sc)}$, admits a timelike HKV.
Therefore the Killing tensor cases $(A)$ through $(H)$
of section \ref{sec:HKT} apply and lead to exact solutions.
The particular case $(C)$ with $s=2$ leads to the solutions found in
\cite{der1,der2}. However, one can easily produce many other solutions
of comparable physical interest.

\subparagraph*{Invariant submanifolds.}
If $V_{\rm(sc)}$ has relative extrema, then one has an invariant submanifold
corresponding to the corresponding fixed value of the scalar field.
The resulting problem yields a generalized Friedmann equation where the
scalar potential reduces to an effective cosmological constant.

There are other more interesting invariant submanifolds obtained by a
different method \cite{hhscalar,ellmad,goe,goenoe,noe}.
These correspond
to exact solutions describing inflationary models in cosmology
as well as static domain walls in an astrophysical context.

\subsubsection{Models with a 4-dimensional intrinsic symmetry group}
\label{sec:4d}

For the family of intrinsically LRS class A models, which can be chosen to
satisfy $n^{(1)} = n^{(2)}$, it is convenient to introduce the notation
$\sigma = n^{(1)}n^{(3)}$. One must set $n^{(3)}=0$ to obtain the
remaining LRS models, for which
the curvature parameter $\sigma$ continues to have its previous meaning.
The sources considered in this subsection may include a cosmological constant,
electro-magnetic fields and perfect fluids.

The Hamiltonian for this family of spacetimes is given by
\begin{eqnarray}\label{eq:hamtwoplus}
   H &=& \case12 x \eta_{AB} \dot\beta{}^A \dot\beta{}^B
   + 24 x^{-1} [\case14  n^{(3)\,2} e^{-4(2\beta^+ - \beta^0)}
\nonumber\\ &&\quad
   + \epsilon \sigma e^{2(2\beta^0 - \beta^+)}
   - \epsilon e^2 e^{-2(2\beta^+ - \beta^0)}
\nonumber\\ &&\quad
   - \epsilon\Lambda e^{6\beta^0}] +
   U_{\rm(fluid)}\,,
\nonumber\\ &&\quad
   [A,B = 0, +, -]\,,
\end{eqnarray} %
where $\beta^-$ must vanish except for the Bianchi types I and II
where it is a cyclic variable, provided that $x$ is assumed to be
independent of this variable.
The system associated with this
Hamiltonian provides several interesting examples of
null and/or nonnull Killing tensor cases which in turn give rise to
many exact solutions.

\bigskip\paragraph{Vacuum, $\Lambda$, EM field, SH stiff perfect fluid.}

\subparagraph*{The case $\Lambda=0$.}
The Hamiltonian for this case can be nicely expressed
in terms of a new pair of conformally inertial coordinates obtained by
the Lorentz transformation \cite{uni}
\begin{eqnarray}\label{eq:boost}
    (\overline{\beta}{}^0,\overline{\beta}{}^+) &=&
              3^{-1/2} (2\beta^0-\beta^+, -\beta^0 +2\beta^+) \,,
\nonumber\\
    (\beta^0,\beta^+) &=&
              3^{-1/2} (2\overline{\beta}{}^0+\overline{\beta}{}^+,
                 -\overline{\beta}{}^0 -2\overline{\beta}{}^+) \,,
\end{eqnarray} %
in terms of which the Hamiltonian takes the following form
\begin{eqnarray}
  H &=& \case12 x(-\dot{\overline{\beta}}{}^{0\, 2}
   +\dot{\overline{\beta}}{}^{+\, 2} +\dot{\beta}{}^{-\,2})
   + 24 x^{-1} [\case14 n^{(3)\,2}e^{-4\sqrt3 \overline{\beta}{}^+}
\nonumber\\ &&\quad
   + \epsilon\sigma e^{2\sqrt3 \overline{\beta}{}^0 }
  - \epsilon e^2 e^{-2\sqrt3 \overline{\beta}{}^+} + \kappa \rho_{(0)} ] \,.
\end{eqnarray}

It is easy to see that this Hamiltonian is
a nonnull H-J Killing tensor case.
Furthermore in the Taub slicing gauge $x=1$,
one has a completely decoupled Hamiltonian
\begin{eqnarray}
  H_{\rm(taub)} &=& -H_{\overline{\beta}{}^0 } +H_{\overline{\beta}{}^+}
                    +H_{\beta^-} + H_{\rm(stiff)} \,,
\nonumber\\
 H_{\overline{\beta}{}^0 } &=& -\case12 \dot{\overline{\beta}{}^0 }\,^2
    + 24 \epsilon\sigma e^{2\sqrt3 \overline{\beta}{}^0 }
   =E_{\overline{\beta}{}^0 }\,,
\nonumber\\
 H_{\overline{\beta}{}^+} &=& \case12 \dot{\overline{\beta}{}}{}^+\,^2
 +\case14 n^{(3)\,2}  e^{-4\sqrt3 \overline{\beta}{}^+}
 - \epsilon {\rm e}^2 e^{-2\sqrt3 \overline{\beta}{}^+}
           =E_{\overline{\beta}{}^+}\,,
\nonumber\\
 H_{\beta^-} &=& \case12 \dot{\beta}{}^-\,^2=E_{\beta^-}\,,
\nonumber\\
 H_{\rm(stiff)} &=& 24\kappa \rho_{(0)}
                 = E_{\rm(stiff)} \quad (\epsilon = -1) \,.
\end{eqnarray} %
The full problem therefore consists of three generalized Friedmann
problems interpretable as 1-dimensional problems with exponential
potentials and constant energies, restricted only by the constraint
\cite{ecs}
\begin{equation}
  E_{\overline{\beta}{}^0 } + E_{\overline{\beta}{}^+} + E_{\beta^-}
        + E_{\rm(stiff)}  = 0 \,.
\end{equation}
Each of the 1-dimensional motion problems is governed by a generalized
Friedmann equation,
with a potential for
the variables $\beta^-$ (when nonzero), $\overline{\beta}{}^0$, and
$\overline{\beta}{}^+$ having in general one, two and three exponential terms
respectively, the latter occurring as an ``equally spaced" exponential
coefficient case (see section \ref{sec:gfp}), all of which are equivalent to
generalized Friedmann problems with a quadratic potential.

Apart from the trivial cyclic variable $\beta^-$ which is present for
Bianchi types I and II,
the natural variables which lead to quadratic potentials
for the other two degrees of freedom are
\begin{eqnarray}
 U_{\overline{\beta}{}^0 }
           &=& e^{-\sqrt3 \overline{\beta}{}^0 } \,,\quad \sigma\neq0\,,
\nonumber\\
 U_{\overline{\beta}{}^+}
     &=&  \cases{e^{2\sqrt3 \overline{\beta}{}^+} \,, & $n^{(3)}\neq0$ \,,\cr
                   e^{\sqrt3 \overline{\beta}{}^+} \,,  & $n^{(3)}=0$\,.\cr}
\end{eqnarray} %

\subparagraph*{The case $\beta^- =0= U_{\rm(fluid)}$.}
These models may be
re-examined as an example of a 2-dimensional null Killing tensor case
treated in section \ref{sec:null}.
Letting
\begin{equation}
     w= \beta^0 + \beta^+\,, \quad v = \beta^0 - \beta^+ \,,
\end{equation}
the Taub time gauge potential takes the form
\begin{eqnarray}\label{eq:utaubfour}
  U_{\rm(taub)} &=& 24[\case14 n^{(3)\,2} e^{-2w+6v}
              + \epsilon \sigma e^{w + 3v}
\nonumber\\ &&\quad
              - \epsilon e^2 e^{-w +3v} -\epsilon \Lambda e^{3(w+v)}] \,.
\end{eqnarray}
For the vacuum case
this expression can be identified with
equation (\ref{eq:nutaub}), with
$c_0 = 3$ and $A_0 = 2(n^{(3)})^2/3 $, allowing $w$ to decouple.
This permits a nonzero cosmological constant term since only a single
null variable is required to decouple, in contrast with the  previous
nonnull discussion where the cosmological constant had to be zero.

The natural power variables and slicing gauge function are
\begin{equation}
      V=e^{3v} \,,\quad W= e^{mw} \,, \quad x= 3m V W \,,
\end{equation}
leading to
\begin{eqnarray}
  H &=& - \case12 \dot W \dot V + \case8m    [
      \case14 n^{(3)\,2} W^{-\case2m -1} V
       + \epsilon \sigma W^{\case1m -1}
\nonumber\\ &&\quad
       - \epsilon e^2 W^{-\case1m -1}
       -\epsilon \Lambda W^{\case3m -1}     ] \,.
\end{eqnarray}
Since $a_0= -2 \neq0$, the three choices $m = -2, \pm 1 $ lead to
to elementary function solutions for $W$.
The choice $m=1$, which makes the $\epsilon$ term a constant,
was first introduced in the SH context by Misner and Taub \cite{tnutm}.
The choice $m=-1$, which makes the $e^2$ term a constant,
was first introduced by Brill \cite{brill} in the SH case.
A third new slicing gauge arises for the choice $m=-2$ which
makes the $n^{(3)}$ term proportional to $V$.
Note that $A_0=0$ for the spherically symmetric models. Thus the choice
$m=1$ leads directly to the standard expression for the Reissner-Nordstr\"om
solution with cosmological constant.

If one also sets $\Lambda=0$ then one has a nonnull-null case
which may be solved as in section \ref{sec:nnn}.

If there are several terms equal to zero there are even more slicing gauges
and dependent variables one can choose to solve the problem. As an example,
consider the LRS Bianchi type II, III and the KS vacuum models
which correspond to a 2-dimensional problem with a
Taub potential consisting of a single exponential term.
This term is a nonnull exponential,
and the corresponding problem is
easily solved using the methods of sections \ref{sec:null}
or \ref{sec:nonnull}.

References to some of the literature on the more prominent
solutions
are given in Table 5. Apart from the solutions indicated in this table, it's
worth noting that the general LRS solution with an electromagnetic field
and a cosmological constant have been given by Cahen and Defrise \cite{cahdef}.
A useful reference and guide to the literature on solutions with
electromagnetic fields is the work by MacCallum \cite{macem}.
MacCallum, together with Siklos, has also made a thorough
investigation of HH vacuum models with a cosmological constant \cite{macsik}.
For a discussion on LRS models see \cite{ve}.

%\typeout{**********************************Table 5:}

\bigskip\paragraph{Static perfect fluids.}

The most interesting static models are the astrophysically relevant
spherically symmetric ones. The Bianchi type I models
are also of some interest as
cylindrically or plane symmetric ($\beta^- = 0$) models.
Other static Bianchi models do not seem to be particularly
interesting physically and will not be considered here.

\subparagraph*{Spherically symmetric models.}
For the astrophysical
spherically symmetric models, various equations of state have been
considered.

\smallskip%\typeout{subsubparagraph}
{The case $p=(\gamma-1)\rho\,$:}
For the usual equation of state with $1<\gamma<2$
one has the Taub potential
\begin{equation}
    U_{\rm(taub)} = 24 [ e^{4\beta^0-2\beta^+}
              + \kappa p_{(0)} e^{ (6-\eta)\beta^0 + 2\eta \beta^+} ] \,,
\end{equation}
where $\eta=\gamma/(\gamma-1)$.

Making the boost
\begin{eqnarray}\label{eq:genboost}
   \beta^0 &=& \Gamma ( \overline{\beta}{}^0   + v \overline{\beta}{}^+ ) \,,
\nonumber\\
   \beta^+ &=& \Gamma ( v \overline{\beta}{}^0 + \overline{\beta}{}^+   ) \,,
            \quad \hbox{where } \Gamma = (1-v^2)^{-1/2}\,,
\end{eqnarray} %
with the value
$v=\case12 (\eta-2)/(\eta+1) = \case12 (2-\gamma)/(2\gamma -1)$
of the boost parameter,
leads to
\begin{eqnarray}
&&  U_{\rm(taub)} = 24  e^{A\overline{\beta}{}^0} [ e^{B\overline{\beta}{}^+}
                    + \kappa p_{(0)} e^{ C \overline{\beta}{}^+} ]  \,,
\nonumber\\
&&   A  = \case{3 \Gamma (\eta+2)}{\eta+1} \,,\quad
     B  = -\case{6 \Gamma}{\eta+1} \,,\quad
     C  = \case{3 \Gamma (\eta-2)^2}{2(\eta+1)} \,.
\end{eqnarray} %
This potential is of the same form as Eq.~(\ref{eq:inv}) with a single term
and has a nonzero minimum value. A 1-dimensional invariant submanifold
corresponds to this minimum value, leading to a generalized Friedmann
equation with one potential term easily solved using the methods of
section \ref{sec:gfp}. This solution is a special case of solutions
found by Tolman \cite{tolman}. The solution with $\gamma = 4/3$
has also been found by Klein \cite{klein}.

\smallskip%\typeout{subsubparagraph}
{The case $\rho= \rho_{(0)}$:}
This case has been treated in section \ref{sec:inull}.
The nonsingular interior solution was
first found by Schwarzschild \cite{sch}.
Solutions which have a singularity are also null cases and have been
investigated by Volkoff \cite{volkoff} and Wyman \cite{wyman}.

\smallskip%\typeout{subsubparagraph}
{The case of an unspecified equation of state:}
The Taub potential is
\begin{equation}
    U_{\rm(taub)} = 24 [ e^{4\beta^0-2\beta^+}
              + \kappa e^{6\beta^0}  p(\beta^0 - 2\beta^+) ] \,.
\end{equation}
Making the boost (\ref{eq:boost}) leads to
\begin{equation}
    U_{\rm(taub)} = 24 [ e^{2\sqrt3 \overline{\beta}{}^0}
      + \kappa e^{4\sqrt3 \overline{\beta}{}^0} f(\overline{\beta}{}^+) ] \,,
\end{equation}
where $f= e^{2\sqrt3\overline{\beta}{}^+} p(\overline{\beta}{}^+)$.
If $f$ has a minimum for some
value of $\overline{\beta}{}^+$ then one obtains an invariant submanifold
as discussed in section \ref{sec:inv}.
Unfortunately such a minimum leads to an unphysical equation of state
$p=-\rho/3$.

However, this problem has the same form as the scalar field problem
with an unspecified
scalar potential mentioned above and dealt with in \cite{hhscalar}.
Thus the same
method can be applied to the present problem
and will produce invariant submanifolds and corresponding
exact solutions.
Alternatively one can specify
$f(\overline{\beta}{}^+)$ or $p(\overline{\beta}{}^+)$
to be some function so that one obtains a problem for which one might find an
invariant submanifold or a
Killing tensor and thus exact solutions.
Once a solution is found the equation of state can be derived.
Unfortunately, the general solution to the Killing tensor problem is
not available at present. However, all of the solvable cases found
in the literature
can be recovered by a certain ansatz for a conformal transformation relating
the standard null variables $(w,v)$ to a set of null variables $(W,V)$ which
are adapted to the symmetry \cite{rosquist:star}.

The starting point is to write down the Jacobi metric in standard null
variables $w = \beta^0 + \beta^+$, $v = \beta^0 - \beta^-$ leading to (modulo
a constant factor)
\begin{equation}\label{eq:jacobi_metric_ss}
    ds_J{}^2 = -2\left\{
       e^{w+3v} + \kappa e^{3(w+v)} p({\textstyle\frac{-w+3v}{2}}) \right\}
                    dw dv \,.
\end{equation}
The ansatz we use for the conformal transformation is
\begin{equation}
    e^w = W^r \,,\quad e^v = V^s \,,
\end{equation}
where $r$ and $s$ are constants to be determined. Applying this
transformation to the Jacobi metric (\ref{eq:jacobi_metric_ss}) yields
\begin{eqnarray}\label{eq:G_adapted}
     ds_J{}^2 &=& -2{\tilde G}(W,V) dW dV \,,
\nonumber\\
    {\tilde G} &=& rs \left[ W^{r-1}V^{3s-1}
              + \kappa W^{3r-1}V^{3s-1}h(Y) \right] \,,
\end{eqnarray}
where $h(Y) = p(\log Y)$ and
$Y = e^{\beta^3} =W^{-r}V^{3s/2}$
(recall that $\beta^3 = \beta^0 - 2\beta^+$).
We next look for conditions on $r$, $s$
and $h(Y)$ which make the variables $W$ and $V$ symmetry adapted with respect
to a Killing vector or a Killing tensor. We do this by inserting the
expression for $\tilde G$ in (\ref{eq:G_adapted}) into the equations
(\ref{eq:threecases}). Analysis of the resulting set of equations leads to
the solvable cases given in Table 6
(for how one explicitly solves these cases see \cite{rosquist:star}).

The cases for which the equation of state is of physical interest are the
null case with $s=2/3$ (Schwarzschild's interior solution), the
Hamilton-Jacobi case with $r=1$, $s=1/3$, (the Killing vector case, where
$a=b$, corresponds to Buchdahl's generalized polytropic solution of index
five \cite{buchdahl:poly_five}) and finally the Hamilton-Jacobi case with
$r=2$, $s=2/3$ (setting $c_+=0$ gives Buchdahl's generalized polytrope of
index one \cite{buchdahl:poly_one,buchdahl:lectures}, while the general
$c_-\neq0$ case was recently given by Simon \cite{simon:sspf}).

%\typeout{**********************************Table 6:}

\subparagraph*{Bianchi type I models.}
Again several equations of state are of interest.

\smallskip%\typeout{subsubparagraph}
{The case $p=(\gamma-1)\rho$:}
These models correspond to a problem with a
single exponential potential term
\begin{equation}
    U_{\rm(taub)} = 24\kappa p_{(0)} e^{ (6-\eta)\beta^0 + 2\eta \beta^+} \,,
\end{equation}
which is nonnull for $1< \gamma < 2$ and null for $\gamma=2$ since
$\eta=\gamma/(\gamma-1)$. These two types of problems were dealt with in
section  \ref{sec:null} and \ref{sec:nonnull}.

\smallskip%\typeout{subsubparagraph}
  {The case $\rho= \rho_{(0)} + (\eta-1) p$:} The Taub potential is
\begin{equation}
    U_{\rm(taub)} = 24\kappa
         [ (\rho_{(0)} + p_{(0)}) e^{ (6-\eta)\beta^0 + 2\eta \beta^+}
                  - \rho_{(0)} e^{6\beta^0} ] \,.
\end{equation}
For $\eta=6$ (or $\gamma = 6/5$) it is easily seen that one has a nonnull
Killing tensor case, thus easily solved with the methods of section
\ref{sec:nonnull}. The solution was first found by Evans \cite{eva}.

\bigskip\paragraph{Spatially homogeneous Bianchi type II nonstiff
               perfect fluid models.}

These models have the Hamiltonian
\begin{eqnarray}
   H &=& \case12 x \eta_{AB} \dot\beta{}^A \dot\beta{}^B
\nonumber\\ &&\quad
   + 24 x^{-1} [\case14   e^{-4(2\beta^+ - \beta^0)}
   + \kappa \rho_{(0)} e^{3(2-\gamma)\beta^0}  ] \,,
\end{eqnarray}
where
$ [A,B = 0, +, -]\,$.
Making the boost (\ref{eq:genboost})
with
\begin{equation}
      v = \case18 ( 3 \gamma -2 )
\end{equation}
leads to the Taub potential
\begin{eqnarray}
   U_{\rm(taub)} &=& 6 e^{3\Gamma (2-\gamma)\overline{\beta}{}^0}
       [ e^{-3\Gamma (6-\gamma)\overline{\beta}{}^+/2}
\nonumber\\ &&\quad
   + 4 \kappa \rho_{(0)}
      e^{e\Gamma (2-\gamma) (3\gamma-2)\overline{\beta}{}^+/8} ] \,.
\end{eqnarray}
The $\overline{\beta}{}^+$-dependent factor in this potential has a minimum.
As discussed in section \ref{sec:inv} this
corresponds to a 2-dimensional invariant submanifold,
leading to a
2-dimensional problem with one nonnull exponential term,
easily solved using sections  \ref{sec:nonnull} and \ref{sec:null}.
When $\overline{\beta}{}^-\neq0$ this  yields the Collins solution \cite{col},
while for $\overline{\beta}{}^- =0$ it gives an LRS EPL solution
\cite{col,colstu,coletal}.

\bigskip\paragraph{Spatially homogeneous KS and Bianchi type III nonstiff
               perfect fluid models.}

Referring to sections \ref{sec:mtm} and \ref{sec:classBm},
the orthogonal perfect fluid models
have a Hamiltonian of the form
\begin{eqnarray}
    H &=& \case12 x ( - \dot{\beta}{}^{0\,2} + \dot{\beta}{}^{+\,2} )
              + 24 x^{-1} [ \sigma e^{4\beta^0 - 2\beta^+}
\nonumber\\ &&\quad
              + \kappa \rho_{(0)} e^{3(2-\gamma)\beta^0} ]\,.
\end{eqnarray}

Transforming to the null variables
$w=\beta^0 + \beta^+$ and $v=\beta^0 - \beta^+$
allows the potential to be identified with (\ref{eq:KT2null})
of the two term case of section \ref{sec:KT2}, with the following
correspondence between the parameters
\begin{eqnarray}
    C_1 &=& 24\sigma\,, \quad C_2 = 24 \kappa \rho_{(0)}\,,
\nonumber\\
    p_1 &=& 1\,,
 q_1 = 3\,, \quad
    p_2 = q_2 = 3(2-\gamma)/2 \,.
\end{eqnarray} %
Solvable cases occur for the following parameter values:

\subparagraph*{The flat null $A_i =0$ case.}
Condition $(II)$ in  Table 3
corresponds to the radiation value $\gamma = 4/3$.

\subparagraph*{The null $A_i \neq 0$ case.}
Condition $(I)$ yields physical solutions for the
dust value $\gamma=1$ and the value $\gamma=\case53$.

The above solutions can be found in
\cite{ru:kt,doro,kansac,col,komche,exactii}.

\subsubsection{Spatially homogeneous Bianchi type VI models} \label{sec:6}

\paragraph{Vacuum models.}

All known type VI vacuum models are Taub symmetric ($\beta^- = 0$) and
correspond to a 2-dimensional problem with a
Taub potential which is the single exponential term
in equation (\ref{eq:Bvac}),
which is nonnull except for Bianchi type VI$_0$ where it is null.
These are easily solved using the methods of sections
\ref{sec:null} and \ref{sec:nonnull}.
These solutions were first found by
\cite{ellmac,eva,col,mac}.

\bigskip\paragraph{Solvable perfect fluid models.}

Referring to sections \ref{sec:mtm} and \ref{sec:classBm},
the orthogonal perfect fluid models
have a Hamiltonian of the form
\begin{eqnarray}
    H &=& \case12 x ( - \dot{\beta}{}^{0\,2} + \dot{\beta}{}^{\times\,2} )
\nonumber\\ &&\quad
              + 24 x^{-1} [ c^{-2} e^{4(\beta^0 - cq \beta^\times)}
                            + \kappa \rho_{(0)} e^{3(2-\gamma)\beta^0} ]\,,
\end{eqnarray}
where $c^{-2}= q^2 + 3 a^2$.

Transforming to the null variables
$w=\beta^0 + \beta^\times$ and $v=\beta^0 - \beta^\times$
allows the potential to be identified with (\ref{eq:KT2null})
of the two term case of section \ref{sec:KT2}, with the following
correspondence between the parameters
\begin{eqnarray}
    C_1 &=& 24c^{-2}\,, \quad C_2 = 24 \kappa \rho_{(0)}\,,
    p_1 = 2(1 - cq)\,,
\nonumber\\
    q_1 &=& 2(1 + cq)\,, \quad
    p_2 = q_2 = 3(2-\gamma)/2 \,.
\end{eqnarray} %
The physical cases correspond to
\begin{equation}
       0\le cq \le 1\,, \quad  1\le \gamma \le 2\,.
\end{equation}
Solvable cases occur for the following parameter values:

\subparagraph*{The flat null $A_i =0$ case.}
Condition $(I)$ in Table 3
yields $cq = \case14(3\gamma-2)$.
The resulting solution was first found by Collins \cite{col}
and has been presented by Wainwright \cite{wai} corresponding to the
form given in equation (\ref{eq:plsum}) with $m=\case32$.

\subparagraph*{The null $A_i \neq 0$ case.}
Conditions $(I)$ of Table 3
with $a_0 \neq 0$ yield the conditions
$cq=\pm \case12 (4 - 3\gamma)$ or $cq= \case18 (3\gamma + 2)$, respectively,
and solutions found by
Uggla \cite{exact} and Uggla and Rosquist \cite{exactii}.

\subparagraph*{The nonnull case.}
There also exist two examples of the nonstiff perfect fluid solutions
corresponding to nonnull cases in which
the potential may be reduced to a quadratic expression in  the two
natural power variables (case (V) in Table 3.
These correspond to the values
$cq = \case45$ and $\gamma = \case65$, leading to the
solution given by Uggla \cite{exact}.
These exact solutions together with the above null type VI$_h$ ones
were found using the methods developed in the present article
and are the first new orthogonal SH non-EPL perfect fluid solutions
found in several decades.
The type VI$_h$ ($h \neq 0$) stiff perfect fluid solution can be obtained
by making a boost that leads to a generalized Friedmann problem
(the type VI$_{h= 0}$ stiff perfect fluid case is contained in the flat
$A_i = 0$ case discussed above).

\subparagraph*{Invariant submanifold perfect fluid models.}
Making the boost (\ref{eq:genboost}) with value $v= (3\gamma-2)/(4cq) < 1$
leads to the Taub potential
\begin{eqnarray}
     U_{\rm(taub)} &=& 24 e^{3\Gamma (2-\gamma) \overline{\beta}{}^0}
   [ c^{-2} e^{ [3\gamma-2 - 4(cq)^2 ]\overline{\beta}{}^\times / (cq) }
\nonumber\\ &&\quad
          + \kappa \rho_{(0)}
        e^{ 3\Gamma(2-\gamma)(3\gamma-2)\overline{\beta}{}^\times/(4cq)} ]\,.
\end{eqnarray}
For the type VI$_h$ models,
the $\overline{\beta}{}^\times$-dependent factor of this potential
has a minimum provided that $(3\gamma-2)/4 < (cq)^2$,
yielding an EPL solution.
This solution was discussed by Hsu and Wainwright \cite{hsuwai}.

\subparagraph*{Bianchi type VI$_0$ models.}
For nonstiff perfect fluid models
the boost (\ref{eq:genboost}) in the $\beta^+$-direction with
the value $v=-\case14 (3\gamma-2)$ leads to the Taub potential
\begin{eqnarray}
  &&  U_{\rm(taub)} = 6 e^{ 3\Gamma(2-\gamma)\overline{\beta}{}^0}
      [ e^{A \overline{\beta}{}^+} h_-{}^2
           + 4\kappa \rho_{(0)} e^{B \overline{\beta}{}^+} ] \,,
\nonumber\\
  &&  A = 3 \Gamma (2-\gamma) \,,\quad
     B  = - \case34 \Gamma (2-\gamma) (3\gamma-2) \,.
\end{eqnarray} %
The $\overline{\beta}{}^+,\beta^-$-dependent factor has a minimum, yielding
a 1-dimensional invariant submanifold which corresponds to the same
type VI$_0$ EPL solution just discussed.
However, these new dependent variables are useful for a qualitative
discussion of the dynamics for this class of models \cite{latestage}.

%%%%%%%%%%%%%%%%%%%%%%%%%%%%%%%%%%%
\subsection{Nondiagonal models}

\subsubsection{Stationary cylindrically symmetric models}

The stationary cylindrically symmetric vacuum models
have the
reduced Hamiltonian (\ref{eq:Hcyl}) which corresponds to the single nonnull
exponential potential term case and is therefore solvable as discussed
in section \ref{sec:nonnull}.
For an explicit representation of this solution see \cite{kraetal}.

\subsubsection{Spatially homogeneous Bianchi type VI$_{-1/9}$ models}
\label{sec:619}

\paragraph{Vacuum models.}

The boost (\ref{eq:genboost}) with the value
$v= 2/(3\sqrt3)$ transforms the Taub potential from
(\ref{eq:Hsixn}) to
\FL
\begin{equation}
    U_{\rm(taub)} = e^{8 \Gamma \overline{\beta}{}^0/3}
      [ \case32 p_\phi{}^2 e^{4\sqrt3 \Gamma \overline{\beta}{}^+}
         + 32 e^{-10\sqrt3 \Gamma \overline{\beta}{}^+/9} ] \,.
\end{equation}
The $\overline{\beta}{}^+$-dependent factor has a minimum,
which corresponds to a
1-dimensional invariant submanifold, leading to an EPL solution
\cite{wai,robtra}.

\bigskip\paragraph{Perfect fluid models.}

The same boost as in the previous vacuum case
for $\gamma=10/9$ transforms the Taub potential from
(\ref{eq:Hsixn}) to
\begin{eqnarray}
    U_{\rm(taub)} &=& e^{8\Gamma \overline{\beta}{}^0/3}
      [ \case32 p_\phi{}^2 e^{4\sqrt3 \Gamma \overline{\beta}{}^+}
         + 32 e^{-10\sqrt3 \Gamma \overline{\beta}{}^+/9}
\nonumber\\ &&\quad
         + 24 \ell^{10/9} e^{8\Gamma \overline{\beta}{}^+/3} ] \,.
\end{eqnarray}
The $\overline{\beta}{}^+$-dependent factor has a minimum,
which corresponds to a
1-dimensional invariant submanifold, leading to an EPL solution
\cite{wai}. No known exact solutions exist for other values of $\gamma$.

\subsubsection{Spatially homogeneous class A models belonging to the
            symmetric case} \label{sec:nondiagA}

\paragraph{Bianchi type II perfect fluid models.}

Here we are going to show how one can obtain tilted type II EPL solutions by
finding invariant submanifolds without explicitly knowing the function $Y$
occuring in the fluid potential. Tilted models are quite complicated and
therefore the manipulations become rather cumbersome. However, by using
computor algebra they can be done.

Equations (\ref{eq:vstar}), (\ref{eq:yfluid}) and (\ref{eq:centrpot})
yield the total potential
\begin{eqnarray}
 &&   U_{\rm(taub)} = U_{\rm(c)} + U_{\rm(G)} + U_{\rm(fluid)}
\\
  &&   = 24\kappa^2\ell^2 (v_3)^2 (n^{(1)})^{-2} [
       e^{-4\sqrt3\beta^-}
\nonumber\\ &&\
       + B e^{4(\beta^0 + \beta^+ + \sqrt3\beta^-)}
       + C e^{3(2-\gamma)\beta^0} Y^{-\gamma/2}
          (\gamma Y - \gamma + 1) ] \,,
\nonumber
\end{eqnarray} %
where $B$ and $C$ are constants defined by
\begin{eqnarray}
     B &=& \frac14\kappa^{-2} (n^{(1)})^4\ell^{-2}(v_3)^{-2} \,,
\nonumber\\
     C &=& \kappa^{-1}(n^{(1)})^2 \ell^{-(2-\gamma)}(v_3)^{-2} \,.
\end{eqnarray} %
It can be simplified by first performing a boost with velocity $v=
(4-3\gamma)/2$ (excluding the stiff fluid case $\gamma=2$) in the $\beta^+$
direction
\begin{equation}
      \beta^0 = \Gamma_v (  \tilde\beta^0 + v \tilde\beta^+) \,,\
      \beta^+ = \Gamma_v (v \tilde\beta^0 +   \tilde\beta^+) \,,\
      \beta^- =  \tilde\beta^- \,,
\end{equation}
where
\begin{equation}
    \Gamma_v = (1-v^2)^{-1/2}
             = 2 [3(3\gamma-2)(2-\gamma)]^{-1/2} \,.
\end{equation}
Then the relation (\ref{eq:F}) simplifies to
\begin{equation}
     F = A e^{(4/\Gamma_v)\tilde\beta^+} \,,
\end{equation}
where $A$ is a constant given by
\begin{equation}
     A = \gamma^{-2}\ell^{-2(\gamma-1)}(v_3)^2 \,,
\end{equation}
thus leading to $Y=Y(\tilde\beta{}^+)$ since $FY^{\gamma - 1} - Y + 1 = 0$.
Expressed in these variables the potential takes the form
\FL
\begin{eqnarray}
 &&  U_{\rm(taub)} =
    24\kappa^2\ell^2 (v_3)^2 (n^{(1)})^{-2}
       [ e^{-4\sqrt3\tilde\beta^-}
\\ &&\quad
    + B e^{6(2-\gamma) \Gamma_v (\tilde\beta^0+\tilde\beta^+)
                             + 4\sqrt3 \tilde\beta^-}
\nonumber\\ &&\quad
   + C e^{(3/2)(2-\gamma) \Gamma_v
          [2\tilde\beta^0 - (3\gamma-4)\tilde\beta^+]} Y^{-\gamma/2}
                               (\gamma Y - \gamma + 1) ] \,.
\nonumber
\end{eqnarray} %
A further boost with velocity $w=-(\sqrt3/4)\Gamma_v(2-\gamma)
=-\sqrt{2-\gamma}/[2\sqrt{3\gamma-2}]$ in the $\tilde\beta{}^-$ direction
\begin{eqnarray}
     \tilde\beta^0 &=& \Gamma_w (  \bar\beta{}^0 + w \bar\beta{}^-) \,,\quad
     \tilde\beta^+ = \bar\beta{}^+ \,,
\nonumber\\
     \tilde\beta^- &=& \Gamma_w (w \bar\beta{}^0 +   \bar\beta{}^-) \,,
\end{eqnarray}
where
\begin{equation}
     \Gamma_w = (1-w^2)^{-1/2}
              = \frac{2\sqrt{3\gamma-2}}{\sqrt{13\gamma-10}} \,,
\end{equation}
leads to
\begin{equation}
    U_{\rm(taub)} = 24\kappa^2\ell^2 (v_3)^2 (n^{(1)})^{-2}
        e^{3(2-\gamma)\Gamma_w\bar\beta{}^0}
                     \Phi(\bar\beta{}^+,\bar\beta{}^-)\,,
\end{equation}
where
\begin{eqnarray}
 && \Phi(\bar\beta{}^+,\bar\beta{}^-) =  e^{-4\sqrt3\Gamma_w\bar\beta{}^-}
       + B e^{6(2-\gamma)\Gamma_v\bar\beta{}^+
              - 4\sqrt3\Gamma_w(1+2\alpha)\bar\beta{}^-}
\nonumber\\ &&
       + C e^{(3/2)(2-\gamma)(4-3\gamma)\bar\beta{}^+
              - 4\sqrt3\Gamma_w(1+\alpha)\bar\beta{}^-}
        Y^{-\gamma/2} (\gamma Y - \gamma + 1) \,.
\end{eqnarray} %
and $\alpha = -(\Gamma_w)^{-2}$. This form of the potential displays
explicitly its homothetic character in the sense that $\partial/\partial
\bar\beta{}^0$ is a homothetic vector of the associated Jacobi geometry.

To locate possible extremal values of the
function $\Phi$ it is convenient to express it as
\begin{eqnarray}
     \Phi &=& Z^k + B A^{-2b-1} Z^{k+2} Y^{2a+1}(Y-1)^{2b+1}
\nonumber\\ &&\quad
            + C A^{-b} Z^{k+1} Y^a (Y-1)^b (\gamma Y-\gamma+1) \,,
\end{eqnarray}
where we have introduced the variable
$Z = e^{(4\sqrt3/\Gamma_w)\bar\beta{}^-}$
and the constants
\begin{eqnarray}
     a &=& -\frac{5\gamma-4}{2(3\gamma-2)} \,,\quad
     b = \frac{4-3\gamma}{2(3\gamma-2)}  \,,
\nonumber\\
    k &=& \alpha^{-1} = -(\Gamma_w)^2 = -\frac{4(3\gamma-2)}{13\gamma-10} \,.
\end{eqnarray}
The potential can be further simplified by making a suitable translation in
$\bar\beta{}^-$ or equivalently rescaling $Z$ by $Z = \delta\bar Z{}$ where
$\delta = A^{b+1/2}B^{-1/2}$. This gives finally $\Phi = \delta^k \bar\Phi$
where
\begin{equation}
     \bar\Phi = Z^k + D(Y) Z^{k+2} + E(Y) Z^{k+1} \,.
\end{equation}
where
\begin{eqnarray}
      D(Y) &=& Y^{2a+1} (Y-1)^{2b+1} \,,
\nonumber\\
      E(Y) &=& 2 Y^a (Y-1)^b (Y-1+\gamma^{-1}) \,.
\end{eqnarray} %

To find possible extremal points of $\bar\Phi$ we calculate the derivatives
\begin{eqnarray}
   \bar\Phi_{,Y}      &=& \bar Z{}^{k+1} [D'(Y)\bar Z{}+E'(Y)] \,,
\nonumber\\
  \bar\Phi_{,\bar Z{}} &=& \bar Z{}^{k-1} [k+D(Y)\bar Z{}^2+E(Y)\bar Z{}] \,,
\end{eqnarray} %
where
\begin{eqnarray}
        D'(Y) &=& Y^{2a}(Y-1)^{2b}[2(a+b+1)Y-2a-1] \,,
\\
        E'(Y) &=& 2 Y^{a-1} Y^{b-1} \{ (a+b+1)Y^2
\nonumber\\ &&\quad
     + [-2a+b-1+(a-b)\gamma^{-1}]Y + a(1-\gamma^{-1})   \} \,.
\nonumber
\end{eqnarray} %
Equating the first of these expressions to zero while noting that $D'(Y)>0$
gives $\bar Z{} = -E'(Y)/D'(Y)$. Inserting
this result in the second equation yields
an equation for $Y$ having a single solution given by
\begin{equation}
     Y - 1 = \frac{(3\gamma-4)(7\gamma-10)}{2\gamma(17\gamma-18)} \,,
\end{equation}
leading to
\begin{equation}
     \bar Z{} = Y^{-a}(Y-1)^{-b}
               \frac{4\gamma(17\gamma-18)}{(11\gamma-10)(7\gamma-10)} \,.
\end{equation}
The conditions $Z>0$ and $Y>1$ can only be fulfilled if $\gamma>10/7$.

We conclude that there exists an exact power law solution with values of $\bar
\beta^\pm$ corresponding to the minimum values of $Y$ and $\bar Z{}$ through
the relations
\begin{eqnarray}
    e^{(4/\Gamma_v)\bar\beta{}^+} &=& A^{-1}F = A^{-1}Y^{1-\gamma}(Y-1) \,,
\nonumber\\
    e^{(4\sqrt3/\Gamma_w)\bar\beta{}^-} &=& A^{b+1/2}B^{-1/2}\bar Z{} \,.
\end{eqnarray} %
This is exactly Hewitt's EPL solution for the equation of state parameter in
the range $10/7<\gamma<2$ \cite{hew}. (Note that the solution with
$\gamma=10/7$ occurring in \cite{hew} is not a tilted one.)
Since the Jacobi metric admits a timelike HKV, one can apply
the methods developed in \cite{latestage} to obtain a complete picture of the
behaviour at late times of these models for $10/7<\gamma<2$.

\bigskip\paragraph{Bianchi type VI$_0$ stiff perfect fluid models.}

For stiff perfect fluids and
the usual values of the structure constants $n^{(1)} =  -n^{(2)} = 1$,
the Taub potential takes the form
\begin{equation}
      U_{\rm(taub)}= 6(4 + \kappa v_3{}^2) e^{4(\beta^0 + \beta^+)}
              + 24 \kappa \ell^2\,.
\end{equation}
Transforming to null variables this is seen to correspond to the
easily solved flat null case $c_0=0$ of section \ref{sec:null}.
This solution was first found by Wainwright et al \cite{wainstiff}.

\subsection{Summary of the spatially homogeneous perfect fluid models}

All of the exact SH perfect fluid solutions arising from
Hamiltonian models are now collected in Table 7.
For the sake of completeness the remaining known ``non-Hamiltonian''
exact tilted perfect fluid solutions are listed in Table 8.
Whether or not some of the
models corresponding to these cases admit a Hamiltonian of the ``standard''
form $H = T + U$ is  not clear. We have not been able to find such a
formulation for any of the models using the present approach.
Perhaps the ``comoving'' approach of MacCallum and Taub \cite{mactau}
could be used to produce Hamiltonians in at least the tilted type V and type
VI$_0$ cases.

%\typeout{**********************************Table 7:}

%\typeout{**********************************Table 8:}

%%%%%%%%%%%%%%%%%%%%%%%%%%%%%%%%%%%%%%%%%%%%%%%%%%%%%%%%%%%%%%%%%%%%%%%%%%%%%
% section 7
%%%%%%%%%%%%%%%%%%%%%%%%%%%%%%%%%%%%%%%%%%%%%%%%%%%%%%%%%%%%%%%%%%%%%%%%%%%%%%

\section{Beyond 4-dimensional General Relativity}
The techniques presented in this article are valuable even outside the
realm of 4-dimensional general relativity.
The only requirement needed to apply them in this larger context
is a Hamiltonian theory for which the Hamiltonian
has a quadratic form kinetic energy function.

\subsection{Higher dimensionsional theories}

With the emergence of unified field theories constructed using
higher-dimensional spacetimes, the door opened to the search for
higher-dimensional analogs of many of the 4-dimensional HH solutions
of Einstein's equations or related field equations.
These occur as solutions of Einstein's equations in higher dimensions and
its numerous generalizations---Kaluza Klein and supersymmetric variations
of Einstein's equations, with coupling to various matter sources---both
in the context of static HH spacetimes
and SH cosmological spacetimes.

Although details change, the same general picture applies and again
one sees in the literature on this topic the same kinds of similarities
that characterize those HH solutions listed by Kramer et al
in the 4-dimensional case. The simplest models for which solutions can
be found are again diagonal, most characterized by the natural generalizations
of the 4-dimensional Hamiltonians, with the same Lorentz structure of the
kinetic part playing a key role in the properties of the dynamics.
One may take the mathematical discussion of the present article
and apply it with slight modifications directly to the higher-dimensional
case \cite{j:high}.

\subsection{Nonminimally coupled scalar fields}

Nonminimally coupled scalar field models are
described by an action of the form
\begin{eqnarray}\label{eq:action}
     S  &=& \frac12 \int d^4x
            \sqrt{-{}^{(4)}g} \,\bigl[\kappa^{-1} A(\phi) \,{}^{(4)}R
\nonumber\\ &&\qquad \qquad
         - B(\phi) g^{\alpha\beta} \phi_{,\alpha}\phi_{,\beta}
         - 2 V(\phi) \bigr] \,,
\end{eqnarray}
where ${}^{(4)}g = \det(g_{\alpha\beta})$. If one is so inclined, one
may also add other matter fields to this action, and below we will
add a perfect fluid contribution.  The function $B(\phi)$ can
be set equal to 1 without loss of generality by using the freedom to redefine
the scalar field by a transformation
$\hat\phi = f(\phi)$ together with an accompanying
redefinition of $A$.
However, it is convenient to retain a general expression $B(\phi)$
for easier comparison with the literature. The coupling between gravity and
matter is said to be minimal if $A(\phi)$ is constant. One can then set
$A=1$, $B=1$. All other couplings are referred to as nonminimal. Often used
nonminimal couplings include conformal coupling characterized by $A(\phi) =
1-\frac16\phi^2$, $B(\phi) =1$, and Brans-Dicke couplingfor which $A(\phi) =
\kappa\phi$, $B(\phi) = \omega/\phi$, and $V(\phi) =0$. Some typical scalar
potentials are
\begin{equation}\label{eq:Vcases}
  V(\phi) =
   \cases{
       \Lambda/\kappa             & cosmological constant, \cr
       \frac12 m^2 \phi^2         & mass term, \cr
       \frac1{4!}\lambda \phi^4   & $\lambda \phi^4$ theory, \cr
       m^2 \mu^{-2} e^{\mu \phi}  & exponential potential, \cr }
\end{equation}
where $m$, $\lambda$ and $\mu$ are the physical parameters
characterizing the various cases.

Varying the action (\ref{eq:action}) with respect to the metric and the
scalar field yields
\begin{eqnarray}
&&     \frac{\delta S}{\delta g^{\alpha\beta}}
      =  \case12 \sqrt{-{}^{(4)}g} (
          \kappa^{-1} A(\phi) G_{\alpha\beta}
        + \kappa^{-1} (\mathop{\rlap{\hbox{$\sqcap$}}\sqcup}\nolimits % \dal
                         A(\phi) ) g_{\alpha\beta}
\nonumber\\ &&\
        - \kappa^{-1} A(\phi)_{;\alpha\beta}
        - B(\phi) \phi_{,\alpha}\phi_{,\beta}
        + [\case12 B(\phi) (\nabla\phi)^2  + V(\phi)] g_{\alpha\beta}) \,,
\nonumber\\
&&     \frac{\delta S}{\delta\phi}
     = \sqrt{-{}^{(4)}g} [
           B(\phi) \mathop{\rlap{\hbox{$\sqcap$}}\sqcup}\nolimits \phi % \dal
           + (2\kappa)^{-1} A'(\phi) {}^{(4)}R
\nonumber\\ &&\
           - V'(\phi) - \case12 B'(\phi) (\nabla\phi)^2   ] \,,
\end{eqnarray} %
where
$(\nabla f)^2 = g^{\alpha\beta} f_{,\alpha} f_{,\beta}$ and
$\mathop{\rlap{\hbox{$\sqcap$}}\sqcup}\nolimits
          f = g^{\alpha\beta} f_{;\alpha\beta}$.      % \dal

\subsubsection{Diagonal spatially homogeneous scalar field models}

We will assume that the scalar field is SH and therefore a function of
$t$ only. Adding an orthogonal perfect fluid
with an equation of state $p = (\gamma-1)\rho$ to the action
(\ref{eq:action}) for the diagonal SH models,
leads to the Hamiltonian
\begin{eqnarray}
&& H = - 4\kappa N n^\alpha n^\beta \frac{\delta S}{\delta g^{\alpha\beta}}
         + 2\kappa N g^{1/2} n^\alpha n^\beta T_{{\rm(pf)}\alpha\beta} \\
\nonumber\\ &&
     = 6 N^{-1} e^{3\beta^0} \left[
           A(\phi)\eta_{AB}\dot\beta^A \dot\beta^B
           -  A'(\phi) \dot\phi \dot\beta^0
           + \frac{\kappa}{6} B(\phi) \dot\phi{}^2 \right]
\nonumber\\ &&\
           + N \left[ e^{\beta^0} V^\ast A(\phi)
           + 2\kappa e^{3\beta^0} V(\phi)
           + 2\kappa \rho_{(0)} e^{-3(\gamma-1)\beta^0} \right] \,,
\nonumber
\end{eqnarray} %
where $T_{{\rm(pf)}\alpha\beta}$ is the energy-momentum tensor of the perfect
fluid.
Here $V^\ast$ is given by (\ref{eq:vstar}) for the class A models
for which $A,B = 0,+,-$, by
$V^\ast = 2c^{-2}e^{\beta^0 - 4cq\beta^\times}$, $A,B = 0,\times$ for the
type V and VI$_h$ models, by
$V^\ast = -2\sigma e^{\beta^0 - 2\beta^+}$, $A,B = 0,+$ for the
MT models, and
by $V^\ast = - 6 k$ for the FRW models, which are described by the line
element
\begin{equation}
     ds^2 = - N^2 dt^2 + R^2 \left[ \frac{dr^2}{1 - k r^2}
              + r^2 (d\theta^2 + \sin^2 \theta d\phi^2) \right] \,,
\end{equation}
with $k$ taking the standard values
$1,0,-1$ for the closed, flat, and open models respectively and for which
$\ln R = \beta^0\,, \beta^\pm = 0$.
(The isotropic case can of course be obtained from the other models. However,
since this case is by far the most discussed when it comes to nonminimally
coupled models, it is given explicitly here.)
The kinetic energy of a minimally coupled
model can be transformed to a manifestly conformally flat form by the choice
$B(\phi) = constant$ corresponding to a redefinition of $\phi$. For
nonminimally coupled models we make the field redefinitions ({\it cf.\/}
\cite{page:scalar})
\begin{eqnarray} \label{eq:redefscalar}
    \tilde\beta{}^0 &=& \beta^0 + \ln A^{1/2} \,,\quad
    \tilde\beta{}^P = \beta^P\,,
\\
    \tilde\beta{}^\dagger  &=& 12^{-1/2} \int d\phi
    [A(\phi)]^{-1}\sqrt{3[A'(\phi)]^2+2\kappa A(\phi)B(\phi)} \,,
\nonumber
\end{eqnarray} %
where $P$ takes appropiate values, e.g., $\pm$ for the class A models.
This transformation leads to a manifestly conformally flat kinetic energy
\begin{eqnarray}
  T
 &=& 6 N^{-1} e^{3\tilde\beta{}^0} [\tilde A({\tilde\beta}{}^\dagger)]^{-1/2}
           \bigl(\eta_{AB}\dot{\tilde\beta}{}^A \dot{\tilde\beta}{}^B
                 + \dot{\tilde\beta}{}^\dagger{}\,^2 \bigr)
\nonumber\\
 &=& \case12 x_{(nm)}
           \bigl(\eta_{AB}\dot{\tilde\beta}{}^A \dot{\tilde\beta}{}^B
                 + \dot{\tilde\beta}{}^\dagger{}\,^2 \bigr) \,,
\end{eqnarray}
where $\tilde A({\tilde\beta}^\dagger) = A(\phi)$ and
$x_{(nm)} = 12N^{-1} e^{3{\tilde\beta}^0}
[\tilde A({\tilde\beta}^\dagger)]^{-1/2}$
$= A(\phi) x$.
Note also that the redefined scalar field
is consistent with the definition made in section
\ref{sec:HA} so that ${\tilde\beta}^\dagger = \beta^\dagger$
in the minimal coupling limit $A \rightarrow 1$, $B
\rightarrow 1$. With the above variable choice the potential takes the form
\begin{eqnarray}\label{eq:nmpot}
     U &=& 24 x_{(nm)}^{-1}       [
         \case12 V^\ast e^{4{{\tilde\beta}^0}}
  + \kappa e^{6{{\tilde\beta}^0}} [\tilde A({\tilde\beta}^\dagger)]^{-2}
    \tilde V({\tilde\beta}^\dagger)
\nonumber\\ &&\quad
        + \kappa \rho_{(0)} e^{3(2-\gamma){{\tilde\beta}^0}}
        [\tilde A({\tilde\beta}^\dagger)]^{-(4-3\gamma)/2}      ] \,,
\end{eqnarray}
where $\tilde V({\tilde\beta}^\dagger) = V(\phi)$.
Note that if the potential is given by $V = {\tilde c} A^2$ and
if $\gamma = 4/3$ (i.e., radiation), then the above Hamiltonian coincides
with the one in general relativity describing a massless scalar
field and a cosmological constant $\Lambda = \kappa {\tilde c}$.
Thus intrinsically isotropic models
(i.e., the isotropic models and the type I and V models,
collectively characterized by $V^\ast = -6k$ when choosing
$a=1$ for the type V models)
are solvable if $x_{(nm)}$ is chosen to depend only on ${\tilde\beta}^0$
since this leads to a 1-dimensional generalized Friedmann problem.
Intrinsically isotropic stiff fluid models
with the same
scalar field potential, $V = {\tilde c} A^2$, are also solvable since they
lead to a separable potential.
The corresponding Jacobi metric
therefore admits a second rank Killing tensor.

\bigskip\paragraph{The conformally coupled case.}

Conformal coupling corresponds to the choice $A(\phi) = 1-\phi^2/6$,
$B(\phi) = 1$.
Using Eq.~(\ref{eq:redefscalar}) leads to a redefinition
of the scalar field given by the relation
$\phi = \sqrt6 \tanh {\tilde\beta}^\dagger$ implying
$\tilde A({\tilde\beta}^\dagger) = \cosh^{-2} {\tilde\beta}^\dagger$.
For a model with
a cosmological constant, a mass term, and a quartic term the total potential
can be written in terms of the redefined fields as
\begin{eqnarray}
     U &=& 24x_{(nm)}^{-1}    [
          \case12 V^\ast e^{4{{\tilde\beta}^0}}
          +  e^{6{{\tilde\beta}^0}}
        ( \Lambda \cosh^4 {\tilde\beta}^\dagger
\nonumber\\ &&\quad
            + 3\kappa m^2 \cosh^2 {\tilde\beta}^\dagger
           \sinh^2 {\tilde\beta}^\dagger
      + 3\kappa \lambda \sinh^4 {\tilde\beta}^\dagger    )  ] \,.
\end{eqnarray}
Obviously this is a SE-Hamiltonian. Apart from the ``trivially''
solvable intrinsically isotropic case with
$m=0$, $\lambda =0$, $\Lambda = 0$ (correponding to setting ${\tilde c} = 0$
in the previous general discussion), there are also a number of other
sets of values of the parameters for which the model is solvable
\cite{br:conf}.
The isotropic case
with $\lambda=0$, $\Lambda =0$ was shown to be chaotic by Calzetta and El
Hasi \cite{ceh:scchaos}.

\bigskip\paragraph{A solvable case with nonconformal quadratic coupling.}

Consider a nonminimally quadratically coupled model with
$A(\phi) = 1- \xi \phi^2$.
For these models it follows that the
$V(\phi) = {\tilde c} A^2 = \kappa^{-1} \Lambda A^2$ case,
discussed above, corresponds
to a model with an arbitrary cosmological constant
$\Lambda$, mass $m = 2 \sqrt{ -\kappa^{-1} \Lambda \xi}$, and a quartic term
with $\lambda = 24\kappa^{-1} \Lambda \xi^2$. To have a physically reasonable
mass term we must have $\Lambda\xi < 0$.

\bigskip\paragraph{Brans-Dicke models.}

In this case the scalar field coupling is defined by the relations
$A(\phi) = \kappa \phi\,, B(\phi) = \omega/\phi$, and $V(\phi) = 0$.
Redefining the scalar field by
$\phi = \kappa^{-1} e^{2\nu {{\tilde\beta}^\dagger}} =
\kappa^{-1} \tilde A({\tilde\beta}^\dagger)$ where
$\nu = \sqrt{3/(3 + 2\omega)}$ leads to the
Hamiltonian
\begin{eqnarray}\label{eq:hambd}
     H &=& \case12 x_{(nm)} (-\dot{\tilde\beta}{}^0\,^2
                 + \dot{\tilde\beta}{}^+\,^2 + \dot{\tilde\beta}{}^-\,^2
                 + \dot{\tilde\beta}{}^\dagger\,^2)
\\ &&\quad
        + 24 x_{(nm)}^{-1} \left[
              \case12 V^\ast e^{4{{\tilde\beta}^0}}
              + \kappa \rho_{(0)}
    e^{3(2-\gamma){{\tilde\beta}^0} - (4-3\gamma)\nu{{\tilde\beta}^\dagger}}
      \right] \,.
\nonumber
\end{eqnarray}
Note that this is a SE-Hamiltonian and that for radiation ($\gamma=4/3$),
it is equivalent to the general relativistic Hamiltonian with
two noninteracting perfect fluids,
one stiff and the other radiation
(provided one chooses $x_{(nm)}$ to be independent of
${\tilde\beta}^\dagger$). Furthermore, if the fluid term in the above
Hamiltonian is zero then there
is a 1--1 correpondence between solvable stiff fluid models in general
relativity and vacuum solutions in Brans-Dicke theory.

As seen from the above discussion there is a
close mathematical relationship between the nonminimally
coupled scalar field Hamiltonians (and particularly Brans-Dicke theory)
and the SE-Hamiltonians occuring in general relativity.
This explains the numerous exact solutions one has obtained in these
theories and the equally numerous number of articles describing them in the
literature
(for Brans-Dicke theory see e.g., \cite{bd1??,bd2??}).
Moreover, the above discussion shows how one
easily can find new ones if one is so inclined.

\subsection{A note on quantum cosmology}

The results of this article may be used as a first step in quantizing
SH models.
Ashtekar et. al. \cite{atu} have quantized the intrinsically multiply
transitive vacuum models.
As can be seen from sections \ref{sec:6d} and \ref{sec:4d},
these all have nonnull decoupling in the Taub slicing gauge.
For any nonnull decoupling case
the Hamiltonian takes the form
\begin{equation}
     H = \sum_\mu  H_\mu (y^\mu,p_\mu)
\end{equation}
when expressed in symmetry adapted dependent variables and in a
slicing gauge leading to decoupling.
As done in \cite{atu},
one can make a canonical transformation such that each decoupled
Hamiltonian is reduced to the square of a momentum $P_\mu$,
leading to
\begin{equation}
       H = \case12  \eta^{\mu\nu} P_\mu P_\nu  \,.
\end{equation}
The only trace of the original potential is to be found in the
ranges of the values of the new variables.
At this stage one has a complete set of observables (constants of the
motion) and one can follow the
quantization procedure used in \cite{atu} to quantize these models.
However, even for nonnull solvable models one obtains a complete set
of observables, and the
quantization procedure discussed by Torre \cite{tor} can be used
to quantize them.

%%%%%%%%%%%%%%%%%%%%%%%%%%%%%%%%%%%%%%%%%%%%%%%%%%%%%%%%%%%%%%%%%%%%%%%%%%%%%
% section 8
%%%%%%%%%%%%%%%%%%%%%%%%%%%%%%%%%%%%%%%%%%%%%%%%%%%%%%%%%%%%%%%%%%%%%%%%%%%%%

\section{Concluding Remarks}

There are other methods than the ones presented in this article which
exist for producing exact solutions. Their relationship
to the present ones is discussed below, but the relationship
is not completely understood and deserves further attention.
The section is concluded with a general discussion on a number of
different issues.

\subsection{Relationship to other solution techniques}

\subsubsection{Comparison with solution generating techniques}

Various solution generating techniques have been developed for
vacuum, electromagnetic or stiff perfect fluid spacetimes with one or two
commuting Killing vectors (see e.g., \cite{kraetal,ger1,ger2}).
These techniques rely on the existence of symmetries  which allow one
to find new solutions from a given solution within the infinite-dimensional
space of solutions being considered.
For the  finite-dimensional
Hamiltonian problems studied here, one can also use the
Killing tensor symmetries to generate new solutions from a particular one,
but in practice this is a mute point since one finds the entire family
at once.

There are also solution generating methods which produce new solutions
from a particular one but with different source or symmetry
characteristics \cite{kraetal,wainstiff,vop}.
Although the present approach analyzes separate Hamiltonian problems,
one could also use the  variation of parameters idea  of \cite{vop}
to establish relationships between different Hamiltonian problems.

For models with an infinite number of degrees of freedom, one can impose
conditions on various geometric quantities or on the functional form
of the line element and still obtain a nontrivial problem corresponding
to an invariant submanifold. This is in stark contrast to the
situation for the finite number of degrees of freedom of the
HH models where such conditions usually result in inconsistencies
(except in the case when one has an unspecified function, like an arbitrary
scalar field potential).
Thus it is critical to have systematic methods for finding invariant
submanifolds for such models.

\subsubsection{Comparison with the exact solution method
of Maartens and Wolfaardt}

Maartens and Wolfhaardt consider a certain class of systems of second order
differential equations and find a constant of the motion linear in the
first derivatives \cite{maawol}.
For Hamiltonian systems of this type it therefore
seems reasonable that this symmetry corresponds to a Killing vector symmetry
since the latter is associated with a constant of the motion which is
linear in the momenta. However, their method is also applicable to
non-Hamiltonian problems.

They apply their analysis to diagonal
SH models. They rederive the Bianchi type I solutions with either
a cosmological constant or a perfect fluid and the orthogonal
Bianchi type II stiff perfect fluid solutions.
These Hamiltonian models do indeed admit Killing vector symmetries.
However, they also apply their method to a non-Hamiltonian
tilted Bianchi type V stiff model and thereby obtain an exact
perfect fluid solution \cite{wainstiff,vop}.

\subsubsection{Comparison with Hewitt's exact solution method}

For polynomial systems of ordinary differential equations one can search
for algebraic invariant curves, which then lead to exact solutions.
Hewitt has applied such a method to 2-dimensional systems arising
from the Einstein field equations for certain
cosmological models \cite{hew:alg}, looking for linear and quadratic
algebraic invariant curves.
For 2-dimensional systems
the existence of a sufficient number of
such curves not only produces the corresponding exact
solutions but also makes it possible to solve the full system.
The search for these invariant curves is quite complicated and relies
on algebraic computing, making it difficult to
extend the approach to higher dimensions
or to invariant curves of higher degree.
Another consideration is the degree of the polynomials occurring
in the system of equations, which must be sufficiently low for practical
use.

Such 2-dimensional polynomial systems
can be derived if, for example, the problem is 2-dimensional and the
Taub potential has at most two exponential terms.
All the models of this type which admit Killing
tensors are given in section \ref{sec:KT2}.
The Killing tensors give rise to constants of the motion
which are linear or quadratic in the momenta and involve exponential
factors.

Sometimes these constants of the motion lead to
linear or quadratic curves in the polynomial system
for certain values of those constants, but not always.
Thus there is an overlap with Hewitt's method
but it is not clear how large it is.
So far all cases which have been found by that method correspond to the
existence of Killing tensors although there are many Killing tensor cases
which don't lead to linear or quadratic algebraic curves.
On the other hand there may exist Hewitt cases
which do not correspond to Killing tensors.

The present approach has the advantage that one can immediately see whether
or not exact solutions occur by inspection of a
{\it single function\/}, the Taub potential, by hand,
without attacking a {\it whole system\/} of differential equations using
algebraic computing. Furthermore it is easier to obtain an SE-Hamiltonian
in ```standard'' form than by rewriting the field equations in polynomial
form. It also has the advantage that higher dimensional problems and those
with more complicated Taub potentials are easily handled. On the other hand
Hewitt's method can be applied to 2-dimensional
problems which do not arise from Hamiltonian systems of the usual
type.

\subsection{Discussion}

The common practice of obtaining exact HH solutions in gravitational physics
is to examine each new scenario as a new problem in isolation without
considering its mathematical relationship to other such problems.
However, if a problem admits a Hamiltonian formulation,
where the kinetic part of
the Hamiltonian can be put in the ``conformally flat" form,
then the present approach can be applied.
Since this approach
has made it possible to unify, extend and bring
order to many apparently unrelated special results of the existing literature,
it should prove to be a useful tool in future studies.

There are many models not explicitly treated in the present article
which could be investigated with these methods. Among these are
a variety of static \cite{kraetal} and self-similar
models \cite{ear}.
Some timelike self-similar models have already
been treated in this way and some new solutions found \cite{u:tss}.
Other sources or combinations of sources may also be considered
leading to an abundance of models.

All of these examples lie within conventional general relativity.
However, the most likely applications will arise in exploring
alternative gravitational theories. For example, of the numerous
articles which regularly appear in this area,
a randomly chosen one \cite{pim} analyzes a Bianchi type I supergravity
model, which can be completely explained in terms of the present analysis.
This is not untypical.
We are not aware of any solvable case in the literature on HH models
which cannot be explained by the existence of rank two Killing tensor and
Killing vector symmetries.
It would be interesting
to find an explicit solvable case solution not admitting such Killing
symmetries.

Finally
the present framework is not just valuable for the goal of searching
for exact solutions but may serve as the starting point for a
qualitative analysis of the more general behavior of the field
equations. There are various kinds of Hamiltonian symmetries which
may not be sufficient to lead to exact solutions. Nevertheless by
adapting the variables to these symmetries, one obtains a simpler
qualitative description. An example of such a symmetry is the
homothetic Killing vector symmetry which many models exhibit \cite{geom}.
This was exploited to develop an
intuitive qualitative picture of the dynamics of a number of models
in \cite{latestage}. This follows in the footsteps of the
well known picture of the Mixmaster dynamics of the diagonal
Bianchi type IX models when expressed in Hamiltonian form in terms of
the Misner parametrization \cite{mis:mini,mis:qc,mis:cqd} and generalized as
much as possible to the general case for all Bianchi types in
\cite{uni}.
One can also use adapted variables to attempt a
so-called regularization of the field equations \cite{unireg,u:IX}.
Thus it seems clear that the tools presented here may prove
useful in many applications involving the rich dynamics of HH models.

%%%%%%%%%%%%%%%%%%%%%%%%%%%%%%%%%%%%%%%%%%%%%%%%%%%%%%%%%%%%%%%%%%%%%%%%%%%%%%
%%%%%%%%%%%%%%%%%%%%%%%%%%%%%%%%%%%%%%%%%%%%%%%%%%%%%%%%%%%%%%%%%%%%%%%%%%%%%%

%%%%%%%%%%%%%%%%%%%%%%%%%%%%%%%%%%%%%%%%%%%%%%%%%%%%%%%%%%%%%%%%%%%%%%%%%%%%%%
%%%%%%%%%%%%%%%%%%%%%%%%%%%%%%%%%%%%%%%%%%%%%%%%%%%%%%%%%%%%%%%%%%%%%%%%%%%%%%
%\typeout{************}
%\typeout{For some reason the caption command is not working correctly!}
%\typeout{And the horizontal space on the narrow tables is bad.}
%\typeout{************}

\onecolumn\typeout{************}
\typeout{8 Tables at end of file}
\typeout{************}

\renewcommand{\arraystretch}{2}

%%%%%%%%%%%%%%%%%%%%%%%%%%%%%%%%%%%%%%%%%%%%%%%%%%%%%%%%%%%%%%%%%%%%%%%%%%%%
% TABLE 1
%%%%%%%%%%%%%%%%%%%%%%%%%%%%%%%%%%%%%%%%%%%%%%%%%%%%%%%%%%%%%%%%%%%%%%%%%%%%

\mediumtext

\begin{table}
\begin{center}
\begin{tabular}{ccccccc}  \hline %\baselineskip=18pt
\strut   & I &II&VI$_0$& VII$_0$ & VIII & IX
    \\ \hline\hline
\strut$n^{(1)}$ & 0 & 0 &  1 & 1 & 1  & 1 \\ \hline
\strut$n^{(2)}$ & 0 & 0 &$-1$& 1 & 1  & 1 \\ \hline
\strut$n^{(3)}$ & 0 & 1 &  0 & 0 &$-1$& 1 \\ \hline
\end{tabular}
\end{center}
\caption{}
%Table 1.
Canonical choices of the symmetry parameters $n^{(a)}$.
For nondiagonal type II models the choice
$(n^{(1)},n^{(2)},n^{(3)}) = (1,0,0)$ is more convenient than the one
in the table.
%}
\end{table}

\vspace*{5mm}
%%%%%%%%%%%%%%%%%%%%%%%%%%%%%%%%%%%%%%%%%%%%%%%%%%%%%%%%%%%%%%%%%%%%%%%%%%%%%%
% TABLE 2
%%%%%%%%%%%%%%%%%%%%%%%%%%%%%%%%%%%%%%%%%%%%%%%%%%%%%%%%%%%%%%%%%%%%%%%%%%%

\narrowtext

\begin{table}
\begin{center}
\begin{tabular}{ccc} \hline
\strut $(r_1,r_2)$ &$\Delta$ & $\delta$
    \\ \hline\hline
\strut $(0,1)$ & $2 q_2 -q_1$ & $q_2 - q_1$
    \\ \hline
\strut $(1,0)$ & $2 q_1 -q_2$ &$q_1 - q_2$
    \\ \hline
\strut $(0,2)$ & $q_2$ & $\case12 (q_2 - q_1)$
    \\ \hline
\strut $(2,0)$ & $q_1$ & $\case12 (q_1 - q_2)$
    \\ \hline
\strut $(1,2)$ & $q_2$ & $q_2 - q_1$
    \\ \hline
\strut $(2,1)$ & $q_1$ & $q_1 - q_2$
     \\ \hline
\end{tabular}
\end{center}
\caption{}
%Table 2.
The choices of $(\Delta,\delta)$ for the two-term potential case which
lead to a linear or quadratic potential. $(r_1,r_2)$ are the new powers.
%}
\end{table}

%%%%%%%%%%%%%%%%%%%%%%%%%%%%%%%%%%%%%%%%%%%%%%%%%%%%%%%%%%%%%%%%%%%%%%%%%%%
% TABLE 3
%%%%%%%%%%%%%%%%%%%%%%%%%%%%%%%%%%%%%%%%%%%%%%%%%%%%%%%%%%%%%%%%%%%%%%%%%%%%

\widetext

\begin{table}
 \begin{center}
  \begin{tabular}{cccccc} \hline
     & $b$ & $\Sigma$ & $(p_i,q_j)$  & $(c_i,d_j)$ & HKV KT case
       \\ \hline\hline
    I & $b\neq 1$ & $0$ & \begin{tabular}{c}
                    $p_i=2p_j$ or\\
                    $q_i=2q_j$ \\
                    \end{tabular}
         & \begin{tabular}{c}
            $c_i + d_i = 2(c_j + d_j)$ or\\
            $c_i - d_i = 2(c_j - d_j)$ \\
          \end{tabular}
         & $(A)$
         \\ \hline
   II & $-$ & $-$ & \begin{tabular}{c}
                     $p_1=p_2$ or $q_1=q_2$ \\
                    \end{tabular}
         & \begin{tabular}{c}
            $c_1-c_2 = \pm (d_1-d_2)$  \\
           \end{tabular}
         & --
         \\ \hline
   III & $1$ & $1$ & $p_1q_2+p_2q_1=0$  & $c_1c_2 - d_1d_2 = 0$ & $(E)$
         \\  \hline
   IV & $-$ & $-$ & $p_1q_2-p_2q_1=0$  & $c_1d_2 - c_2d_1 = 0$ & --
         \\ \hline
   V & $\case12$ & $Z=\pm1$ & \begin{tabular}{c}
                          $p_i=3p_j$ and \\$q_j=3q_i$
                         \end{tabular}
         & \begin{tabular}{c}
             $c_i+d_i=3(c_j+d_j)$ and\\ $c_j-d_j=3(c_i-d_i)$
           \end{tabular}
         & $Z = \cases{ 1 & $(C)$\cr
                       -1 & $(D)$\cr}$
         \\ \hline
  \end{tabular}
 \end{center}
\caption{}
\label{tab:KT2}
%Table 3.
Null and nonnull parameters for 2-dimensional two exponential terms
models admitting Killing tensors up to second rank. The index pair
$(i,j)$ is a permutation of $(1,2)$ where appropriate in the table.
The parameters $b$ and $\Sigma$ are undefined in cases II and IV.
The last column relates the different cases to the general HKV Killing
tensor cases.
%}
\end{table}

\vspace*{5mm}
%%%%%%%%%%%%%%%%%%%%%%%%%%%%%%%%%%%%%%%%%%%%%%%%%%%%%%%%%%%%%%%%%%%%%%%%%%%%
% TABLE 4
%%%%%%%%%%%%%%%%%%%%%%%%%%%%%%%%%%%%%%%%%%%%%%%%%%%%%%%%%%%%%%%%%%%%%%%%%%%%

\begin{table}
\begin{center}
\begin{tabular}{ccllll}  \hline\hline
\vtop{\halign{\strut #\hfil\cr Bianchi\cr type\cr}} & dim &
vacuum & $\Lambda$-term &
\vtop{\halign{#\hfil\cr perfect\cr fluid\cr}} &
\vtop{\halign{#\hfil\cr $\Lambda$-term plus\cr perfect fluid\cr}}
\\ \hline\hline
I & 6 & Mink         & deS\cite{des} & FRW    &  FRW-$\Lambda$\\ \hline
I & 4 & Kasner\cite{kas}, Mink
          & Saunders\cite{sau} &
       \vtop{\halign{\strut#\hfil\cr
             Jacobs\cite{jac} \cr
             Robinson\cite{rob} \cr
             Raychaudhuri\cite{ray} \cr
             Doroshkevich\cite{doro} \cr
             Ste-Ell\cite{steell} \cr
             Vai-Elt\cite{ve} \cr}}
                & Saunders\cite{sau}  \\ \hline
I & 3 & Kasner\cite{kas}
          & Saunders\cite{sau} &
                \vtop{\halign{\strut#\hfil\cr
                     Jacobs\cite{jac} \cr
                     Robinson\cite{rob} \cr
                     Raychaudhuri\cite{ray} \cr}}
                   & Saunders\cite{sau}  \\ \hline
V & 6 & Mink(Milne\cite{hsuwai})
           & deS\cite{des}, anti-deS
                   & FRW  & FRW-$\Lambda$ \\ \hline
V & 3 & Joseph\cite{jos}
          &    & Ell-Mac\cite{ellmac}
                 &   \\ \hline
IX& 6 & --- & deS\cite{des} & FRW  & FRW-$\Lambda$ (Ein\cite{ein}) \\ \hline
\end{tabular}
\end{center}
\caption{}
\label{tab:hh611}
%Table 4.
SH models with 6-dimensional intrinsic symmetry group not including
a scalar field. The abbreviations Mink, deS, anti-deS, Vai-Elt, Ste-Ell,
Ell-Mac and Ein stand respectively for Minkowski, de Sitter, anti-de Sitter,
Vajk-Eltgroth, Stewart-Ellis, Ellis-MacCallum and Einstein. The dimension
column in this and all subsequent tables refers to the dimension
of the spacetime symmetry group.
%}
\end{table}

%%%%%%%%%%%%%%%%%%%%%%%%%%%%%%%%%%%%%%%%%%%%%%%%%%%%%%%%%%%%%%%%%%%%%%%%%%%%%%
% TABLE 5
%%%%%%%%%%%%%%%%%%%%%%%%%%%%%%%%%%%%%%%%%%%%%%%%%%%%%%%%%%%%%%%%%%%%%%%%%%%%

\begin{table}
\begin{center}
\begin{tabular}{llllll}  \hline\hline
\vtop{\halign{\strut#\hfil\cr
sym\cr type\cr}} & dim &
vacuum & $\Lambda$-term & em-term &
\vtop{\halign{\strut#\hfil\cr
$\Lambda$-term $+$ \cr em-term\cr}}
\\ \hline\hline
B I   & 4 & Kasner\cite{kas} ($\epsilon=\pm 1$) &
      \vtop{\halign{\strut#\hfil\cr
            Saunders\cite{sau} ($\epsilon=-1$)\cr
            N-H\cite{novhor} ($\epsilon=1$)\cr}}    &
      \vtop{\halign{\strut#\hfil\cr
            Rosen\cite{rosen} ($\epsilon=-1$)\cr
            Kar\cite{kar} ($\epsilon=1$)\cr
            McVittie \cite{mcv} ($\epsilon=1$)\cr}} &
                                                                    \\ \hline
B I   & 3 & Kasner\cite{kas} ($\epsilon=\pm 1$) &
            Saunders\cite{sau} ($\epsilon=-1$) &
      \vtop{\halign{\strut#\hfil\cr
            Datta\cite{dat} ($\epsilon=-1$)\cr
            Bonnor\cite{bon} ($\epsilon=1$)\cr
            Jacobs\cite{jac:em} ($\epsilon=-1$)\cr}} &
                                                                    \\ \hline
B II  & 4 & Taub\cite{tau} ($\epsilon=-1$) & &
            \vtop{\halign{\strut#\hfil\cr
                  Ruban\cite{rub} ($\epsilon=-1$)\cr
                  Barnes\cite{barnes} ($\epsilon=-1$)\cr}} & \\ \hline
B II  & 3 & Taub\cite{tau} ($\epsilon=-1$) & & & \\ \hline
B III & 4 & K-S\cite{kansac} ($\epsilon=-1$) & &
            Datta\cite{datiii} ($\epsilon=1$)  & \\ \hline
KS    & 4 & K-S\cite{kansac} ($\epsilon=-1$) & & & \\ \hline
SSS   & 4 & Schwar.\cite{sch} ($\epsilon=1$) &
            Kottler\cite{kot} ($\epsilon=1$) &
            Reissner\cite{rei} ($\epsilon=1$)  &
            Nord.\cite{nor} ($\epsilon=1$)  \\ \hline
B VIII & 4 & Taub\cite{tau} ($\epsilon=-1$) & & & \\ \hline
B IX  & 4 & \vtop{\halign{\strut#\hfil\cr
                  Taub\cite{tau} ($\epsilon=-1$)\cr
                  NUT\cite{nut} ($\epsilon=1$)\cr}} &
            Brill\cite{brill} ($\epsilon=-1$) & & \\ \hline
\end{tabular}
\end{center}
\caption{}
\label{tab:hh612}
%Table 5.
SH models with 4-dimensional intrinsic symmetry group
not including a scalar field. The abbreviations B, SSS,
N-H, Schwar., Nord. and NUT stand respectively for Bianchi,
static spherically symmetric,
Novotn\'y-Horsk\'y, Schwartzschild, Nordstr\"om and
Newman-Unti-Tamburino.
%}
\end{table}

%%%%%%%%%%%%%%%%%%%%%%%%%%%%%%%%%%%%%%%%%%%%%%%%%%%%%%%%%%%%%%%%%%%%%%%%%%%%%
% TABLE 6
%%%%%%%%%%%%%%%%%%%%%%%%%%%%%%%%%%%%%%%%%%%%%%%%%%%%%%%%%%%%%%%%%%%%%%%%%%%%

\begin{table}
 \begin{center}
  \begin{tabular}{ccccc} \hline
     $r$ & $s$ & $p(\beta^3)$ & equation of state
                                       & Killing tensor type \\ \hline\hline
   1 & - & $-aY^4+bY^2$ & $a(5p+\rho)^2 = 2b^2(3p+\rho)$ & null \\ \hline
     2 & - & $-aY^5+bY^4$ & $a^4(6p+\rho)^5 = b^5(5p+\rho)^4$ & null \\ \hline
     - & $1/3$ & $-aY^2+b$ & $3p+\rho=2b$ & null \\ \hline
     - & $2/3$ & $-a+bY^{-1}$ & $\rho=a$ & null \\ \hline
            1 & $1/3$ & $aY^4-2bY^2+\Sigma a$
       & $\displaystyle\frac{(5p+\rho-4\Sigma a)^2}{2\Sigma a-3p-\rho}
                                                   =\frac{8b^2}{a}$
%               & $\cases{ \Sigma= 1 & H-J     \cr
%                          \Sigma=-1 & harm. \cr}$ \\ \hline
                & \begin{tabular}[c]{c}
                    H-J     $(\Sigma= 1)$ \\
                    harm. $(\Sigma=-1)$ \\
                  \end{tabular}  \\ \hline
     2 & $2/3$ & \begin{tabular}[c]{c}
                   $(6Y)^{-1}\left[-c_+(1+Y)^6+c_-(1-Y)^6\right]$ \\
                   $(6Y)^{-1} \,{\Re e}[c_0(1+iY)^6]$ \\
                 \end{tabular}
               & \begin{tabular}[c]{c}
                   $\rho = c_+(1+Y)^5+c_-(1-Y)^5$ \\
                   $\rho = {\Im m}[c_0(1+iY)^5]$ \\
                 \end{tabular}
               & \begin{tabular}[c]{c}
                   H-J \\
                   harm.
                 \end{tabular}
               \\ \hline
  \end{tabular}
 \end{center}
\caption{}
\label{tab:hh612b}
%Table 6.
Perfect fluids corresponding to Killing tensor cases. In the first two
coulumns a ``-" means that the value of $r$ or $s$ is arbitrary. The
parameters $a$, $b$, $c_\pm$ are real constants while $c_0$ is a
complex constant. The shorthand notation $Y=e^{\beta^3}$
is also used. For
the last case, ($r=2$, $s=2/3$), it is not possible to write down an
equation of state in closed form. Instead the expression for
$\rho(\beta^3)$
is given in the fourth column. The abbreviations ``H-J" and ``harm."
stand for Hamilton-Jacobi and harmonic respectively.
The Hamilton-Jacobi Killing tensor case with
$(r,s)=(1,1/3)$ reduces to a Killing vector case when $a=b$.
%}
\end{table}

%%%%%%%%%%%%%%%%%%%%%%%%%%%%%%%%%%%%%%%%%%%%%%%%%%%%%%%%%%%%%%%%%%%%%%%%%%%
% TABLE 7
%%%%%%%%%%%%%%%%%%%%%%%%%%%%%%%%%%%%%%%%%%%%%%%%%%%%%%%%%%%%%%%%%%%%%%%%%%%%

\begin{table}
\begin{center}
\begin{tabular}{cclllll}  \hline\hline
\vtop{\halign{#\hfil\cr Bianchi\cr type\cr}} & dim & $1 \leq \gamma \leq 2$ &
solution & EPL & Orth. & relevant sections
\\ \hline\hline
I   & 6 & $\gamma$ & flat FRW
    & yes & yes & \ref{sec:6d},\ref{sec:gfp} \\ \hline
I   & 4 & \vtop{\halign{\strut#\hfil\cr
           $1$\cr $4/3$ \cr $\gamma$\cr}} &
    \vtop{\halign{\strut#\hfil\cr
          Robinson\cite{rob} \cr Dor.\cite{doro} \cr
          Jacobs\cite{jac}, S-E\cite{steell}\cr}}
                   & \vtop{\halign{\strut#\hfil\cr
                      no \cr no \cr no\cr}}
                   & \vtop{\halign{\strut#\hfil\cr
                             yes \cr yes \cr yes\cr}} &
    \vtop{\halign{\strut#\hfil\cr
          \ref{sec:6d}, \ref{sec:nonnull}, \ref{sec:gfp}, \ref{sec:null}\cr
          \ref{sec:6d}, \ref{sec:nonnull}, \ref{sec:gfp}, \ref{sec:null}\cr
          \ref{sec:6d}, \ref{sec:nonnull}, \ref{sec:gfp}, \ref{sec:null}\cr}}
    \\ \hline
I   & 3 & \vtop{\halign{\strut#\hfil\cr
                $1$\cr $\gamma$ \cr}} &
    \vtop{\halign{\strut#\hfil\cr
                  Robinson\cite{rob} \cr Jacobs\cite{jac}\cr}}
                   & \vtop{\halign{\strut#\hfil\cr
                                    no \cr no\cr}}
                   & \vtop{\halign{\strut#\hfil\cr
                                   yes \cr yes\cr}} &
    \vtop{\halign{\strut#\hfil\cr
          \ref{sec:6d}, \ref{sec:nonnull}, \ref{sec:gfp}\cr
          \ref{sec:6d}, \ref{sec:nonnull}, \ref{sec:gfp}\cr}}
    \\ \hline
II  & 4 & \vtop{\halign{\strut#\hfil\cr
                        $\gamma<2$\cr 2\cr}}
          & \vtop{\halign{\strut#\hfil\cr
                  C-S\cite{colstu} \cr
                  Collins\cite{col}  \cr}}
    & \vtop{\halign{\strut#\hfil\cr
                    yes\cr no\cr}}
    & \vtop{\halign{\strut#\hfil\cr
                    yes\cr yes\cr}}
    & \vtop{\halign{\strut#\hfil\cr
            \ref{sec:4d}, \ref{sec:inv}, \ref{sec:gfp} \cr
            \ref{sec:4d}, \ref{sec:nonnull}, \ref{sec:gfp} \cr}}
    \\ \hline
II  & 3 & \vtop{\halign{\strut#\hfil\cr
                        $\gamma$\cr 2 \cr $10/7<\gamma<2$\cr}}
    &  \vtop{\halign{\strut#\hfil\cr
             Collins\cite{col}  \cr
             Collins\cite{col}  \cr
             Hewitt\cite{hew}  \cr}}
    & \vtop{\halign{\strut#\hfil\cr
                    no\cr no\cr yes\cr}}
    & \vtop{\halign{\strut#\hfil\cr
                    yes\cr yes\cr no\cr}}
    & \vtop{\halign{\strut#\hfil\cr
            \ref{sec:4d}, \ref{sec:inv}, \ref{sec:null} \cr
            \ref{sec:4d}, \ref{sec:nonnull}, \ref{sec:gfp} \cr
            \ref{sec:nondiagA}, \ref{sec:inv}, \ref{sec:gfp} \cr}}
    \\ \hline
III & 4 & \vtop{\halign{\strut#\hfil\cr
                        1,4/3 \cr 5/3\cr 2\cr}}
    & \vtop{\halign{\strut#\hfil\cr
      K-C\cite{komche} \cr U-R\cite{exactii} \cr
      K-S\cite{kansac} \cr}}
    & \vtop{\halign{\strut#\hfil\cr
                    no\cr no\cr no\cr}}
    & \vtop{\halign{\strut#\hfil\cr
                    yes\cr yes\cr yes\cr}}
    & \vtop{\halign{\strut#\hfil\cr
            \ref{sec:4d}, \ref{sec:null}\cr
            \ref{sec:4d}, \ref{sec:null}\cr
            \ref{sec:4d}, \ref{sec:nonnull}\cr}}
    \\ \hline
KS  & 4 & \vtop{\halign{\strut#\hfil\cr
                        1 \cr 4/3\cr 5/3\cr 2\cr}}
    & \vtop{\halign{\strut#\hfil\cr
            Dor.\cite{doro}\cr K-C\cite{komche} \cr
            R-U\cite{ru:kt}\cr K-S\cite{kansac} \cr}}
    & \vtop{\halign{\strut#\hfil\cr
                    no\cr no\cr no\cr no\cr}}
    & \vtop{\halign{\strut#\hfil\cr
                    yes\cr yes\cr yes\cr yes\cr}}
    & \vtop{\halign{\strut#\hfil\cr
            \ref{sec:4d}, \ref{sec:null}\cr
            \ref{sec:4d}, \ref{sec:null}\cr
            \ref{sec:4d}, \ref{sec:null}\cr
            \ref{sec:4d}, \ref{sec:nonnull}\cr}}
    \\ \hline
V   & 6 & $\gamma$ & open FRW & no & yes & \ref{sec:6d}, \ref{sec:gfp}
    \\ \hline
V   & 3 & $\gamma$ & E-M\cite{ellmac}
    & no & yes & \ref{sec:6d}, \ref{sec:nonnull}, \ref{sec:gfp}
    \\ \hline
VI$_h$ & 3 & \vtop{\halign{\strut#\hfil\cr
                   $2(2cq+1)/3$\cr
                   $\gamma<2$\cr
                   2\cr
                   2\cr
                   6/5\cr
                   $2(2\pm cq)/3$\cr
                   $2(4cq-1)/3$\cr
                   $10/9$\cr
                   2\cr}}
    & \vtop{\halign{\strut#\hfil\cr
            Collins\cite{col} \cr
            Collins\cite{col} \cr
            E-M\cite{ellmac} ($h=0$)\cr
            Collins\cite{col} ($h\neq 0$)\cr
            Uggla\cite{exact} ($h=-3/16$)\cr
            U-R\cite{exactii} \cr
            U-R\cite{exactii} \cr
            Wain.\cite{wai19} ($h=-1/9$) \cr
            Wain. et al\cite{wainstiff}($h=0$) \cr}}
    & \vtop{\halign{\strut#\hfil\cr
                    no\cr yes\cr no\cr no\cr no\cr no\cr no\cr yes\cr no\cr}}
    & \vtop{\halign{\strut#\hfil\cr
              yes\cr yes\cr yes\cr yes\cr yes\cr yes\cr yes\cr yes\cr no\cr}}
    & \vtop{\halign{\strut#\hfil\cr
            \ref{sec:6}, \ref{sec:null}\cr
            \ref{sec:6}, \ref{sec:inv}, \ref{sec:gfp}\cr
            \ref{sec:6}, \ref{sec:null}\cr
            \ref{sec:6}, \ref{sec:nonnull}, \ref{sec:gfp}\cr
            \ref{sec:6}, \ref{sec:nonnull}\cr
            \ref{sec:6}, \ref{sec:null}\cr
            \ref{sec:6}, \ref{sec:null}\cr
            \ref{sec:619}, \ref{sec:inv}, \ref{sec:gfp}\cr
            \ref{sec:nondiagA}, \ref{sec:null}\cr}}
    \\ \hline
VIII & 4 & 2 & Jantzen\cite{vop} & no & yes &
               \ref{sec:4d}, \ref{sec:nonnull}, \ref{sec:gfp}
    \\ \hline
IX  & 6 & $\gamma$ & closed FRW & no & yes & \ref{sec:6d}, \ref{sec:gfp}
    \\ \hline
IX  & 4 & 2 & Barrow\cite{bar} & no & yes &
              \ref{sec:4d}, \ref{sec:nonnull}, \ref{sec:gfp}
    \\ \hline
\end{tabular}
\end{center}
\caption{}
\label{tab:hht63}
%Table 7.
Hamiltonian spatially homogeneous perfect fluid models.
The abbreviations Orth.,
S-E, C-S, K-C, ``Dor.", K-S, U-R, R-U, E-M and ``Wain." stand for
Orthogonal, Stewart-Ellis, Collins-Stewart, Kompanets-Chernov, Doroshkevich,
Kantowski-Sachs, Rosquist-Uggla,
Uggla-Rosquist, Ellis-MacCallum and Wainwright. One can set $q = 1$ in
the expression $cq = q/\sqrt{q^2 + 3a^2}$ for type VI$_h$. Note that when
$cq = 1/2$, then VI$_{h=-1}$=III. A yes in the ``Orth." column implies that
the model is orthogonal while a no implies a tilted model.
%}
\end{table}

%%%%%%%%%%%%%%%%%%%%%%%%%%%%%%%%%%%%%%%%%%%%%%%%%%%%%%%%%%%%%%%%%%%%%%%%%%%%%%
% TABLE 8
%%%%%%%%%%%%%%%%%%%%%%%%%%%%%%%%%%%%%%%%%%%%%%%%%%%%%%%%%%%%%%%%%%%%%%%%%%%%

\begin{table}
\begin{center}
\begin{tabular}{cccccc}  \hline\hline
\vtop{\halign{#\hfil\cr Bianchi\cr type\cr}} &
dim & $1 \leq \gamma \leq 2$ & solution & EPL & Orth.
\\ \hline\hline
V   & 4 & \vtop{\halign{\strut#\hfil\cr
                        1\cr 2\cr}} &
          \vtop{\halign{\strut#\hfil\cr
                 Farnsworth\cite{farnsw}\cr Maartens and Nel\cite{manel}\cr}}
        & \vtop{\halign{\strut#\hfil\cr
                        no\cr no\cr}}
        & \vtop{\halign{\strut#\hfil\cr
                        no\cr no\cr}}
    \\ \hline
V   & 3 & 2 &
           Maartens and Wolfaardt\cite{mawol} & no & no
    \\ \hline
VI$_0$  & 3 & 4/3, 1.0411 $\leq \gamma \leq$ 1.7169 &
              Rosquist\cite{ros}, Rosquist and Jantzen\cite{rosjan}
              & yes, yes
              & no, no
    \\ \hline
VI$_h$  & 3 & 2 & Wainwright et al\cite{wainstiff} & no & no
    \\ \hline
VII$_h$  & 3 & \vtop{\halign{\strut#\hfil\cr
                             2\cr 2\cr}}
    & \vtop{\halign{\strut#\hfil\cr
               Barrow\cite{bar}\cr Wainwright et al\cite{wainstiff}\cr}} &
      \vtop{\halign{\strut#\hfil\cr
                    no\cr no\cr}} &
      \vtop{\halign{\strut#\hfil\cr
                    yes\cr no\cr}}
    \\ \hline
\end{tabular}
\end{center}
\caption{}
\label{tab:hht63b}
%Table 8.
Non-Hamiltonian spatially homogeneous perfect fluid models.
The numerical values 1.0411 and 1.7169 are numerical approximations
calculated by Rosquist and Jantzen.
%}
\end{table}
%%%%%%%%%%%%%%%%%%%%%%%%%%%%%%%%%%%%%%%%%%%%%%%%%%%%%%%%%%%%%%%%%%%%%%%%%%
\end{document}